\documentclass[twocolumn,tightened,twocolappendix]{aastex63}

\usepackage{mathtools}
\usepackage{chngcntr}
\usepackage{amssymb}
\usepackage{color}
\usepackage{rotating,graphicx}

\def\hin{{{\rm H}\,{\sc i}}}
\def\wise{{\it WISE}~}
\def\planck{{\it Planck}~}
\def\act{{\it Atacama Cosmology Telescope}}
\def\jvla{{\it Jansky Very Large Array}}
\def\spt{{\it South Pole Telescope}~}
\def\msun{{M$_{\odot}$}}

\accepted{May 19, 2023}

\defcitealias{Madhavacheril2020}{M20}
\defcitealias{Zu2015}{Z15}
\defcitealias{Arnaud2010}{A10}
\defcitealias{Kim2022}{K22}

\shorttitle{tSZ effect in the CGM}
\shortauthors{Das, Chiang \& Mathur}

\begin{document}
\title{Detection of thermal Sunyaev{\textendash}Zel{\textquoteright}dovich Effect in the circumgalactic medium of low-mass galaxies - a surprising pattern in self-similarity and baryon sufficiency}

\correspondingauthor{Sanskriti Das}
\email{snskriti@stanford.edu}

\author[0000-0002-9069-7061]{Sanskriti Das}
\affil{Kavli Institute for Particle Astrophysics and Cosmology, Stanford University, 452 Lomita Mall, Stanford, CA\,94305, USA}
\affiliation{Department of Astronomy, The Ohio State University, 140 West 18th Avenue, Columbus, OH 43210, USA}
\author{Yi-Kuan Chiang}
\affiliation{Institute of Astronomy and Astrophysics, Academia Sinica, Astronomy-Mathematics Building, No.\,1, Section\,4, Roosevelt Road, Taipei 10617, Taiwan}
\affil{Center for Cosmology and Astroparticle Physics, 191 West Woodruff Avenue, Columbus, OH 43210, USA}
\author{Smita Mathur}
\affiliation{Department of Astronomy, The Ohio State University, 140 West 18th Avenue, Columbus, OH 43210, USA}
\affil{Center for Cosmology and Astroparticle Physics, 191 West Woodruff Avenue, Columbus, OH 43210, USA}

\begin{abstract}
We report on the measurement of the thermal Sunyaev{\textendash}Zel{\textquoteright}dovich (tSZ) Effect in the circumgalactic medium (CGM) of 641,923 galaxies with $\rm M_\star$=$\rm 10^{9.8-11.3}M_\odot$ at $z<$0.5, pushing the exploration of tSZ Effect to lower-mass galaxies compared to previous studies. We cross-correlate the galaxy catalog of \wise and SuperCosmos with the Compton-$y$ maps derived from the combined data of \act~and \planck\!. We improve on the data analysis methods (correcting for cosmic infrared background and Galactic dust, masking galaxy clusters and radio sources, stacking, aperture photometry), as well as modeling (taking into account beam smearing, ``two-halo" term, zero-point offset). We have constrained the thermal pressure in the CGM of $\rm M_\star$=$\rm 10^{10.6-11.3}M_\odot$ galaxies for a generalized NFW profile and provided upper limits for $\rm M_\star$=$\rm 10^{9.8-10.6}M_\odot$ galaxies. The relation between $\rm M_{500}$ ({obtained from an empirical $\rm M_\star$--$\rm M_{200}$ relation and a concentration factor}) and $\rm \tilde Y^{sph}_{R500}$ ({a measure of the} thermal energy within R$_{500}$) is {$>$2$\sigma$} steeper than the self-similarity and the deviation from the same that has been reported previously in higher mass halos. {We calculate the baryon fraction of the galaxies, $f_b$, assuming the CGM to be at the virial temperature that is derived from $\rm M_{200}$.} $f_b$ exhibits a non-monotonic trend with mass, with $\rm M_\star$=$\rm 10^{10.9-11.2}$\msun~galaxies being baryon sufficient.
\end{abstract}

\keywords{Sunyaev-Zeldovich effect --- Millimeter astronomy --- Extragalactic astronomy --- Observational Cosmology --- Circumgalactic medium --- Hot Intergalactic medium --- Hot ionized medium --- Galaxy evolution --- Galaxy environments --- Galactic winds --- Galaxy processes --- Cosmic microwave background radiation}

\section{Introduction}\label{sec:intro}
The circumgalactic medium (CGM) is the multiphase halo of gas and dust surrounding the stellar disk and the interstellar medium (ISM) of a galaxy \citep{Putman2012,Tumlinson2017,Mathur2022,Faucher-Giguere2023}. As a nexus of the accretion from the intergalactic medium (IGM), galactic outflow, and recycling precipitation, the CGM plays a crucial role in the formation and evolution of a galaxy and might harbor the missing \textit{galactic} baryons \citep{Cen1999}. The most massive and volume-filling phase of the CGM is predicted to be $\geqslant 10^6$\,K hot and fully ionized \citep[e.g.,][]{Spitzer1956,Schaye2015,Hopkins2018,Nelson2018,LiMiao2020}. This hot gas can be observed in the emission and absorption lines of highly ionized (He-like, H-like) metal ions and free-free continuum emission in X-ray \citep{Mathur2022}, and the Sunyaev-Zel'dovich (SZ) effect \citep{Sunyaev1969,Mroczkowski2019}. 

The SZ effect is a distortion in the spectrum of the cosmic microwave background (CMB) due to the inverse Compton scattering of the CMB photons with the free electrons in the intervening ionized media \citep{Sunyaev1969}. Thermal SZ (tSZ) effect, the strongest of different SZ effects, is characterized by the Compton-$y$ parameter: $y = (\sigma_{\rm T}/m_e c^2) \int {\rm P_e} dl$\footnote{$\sigma_{\rm T}, m_e$ and c are Thompson's scattering cross section, rest mass of the electron, and the speed of light in vacuum, respectively.}. It is a measure of the thermal pressure (${\rm P_e} = n_e \rm k_B T_e$) or thermal energy density of the free electrons of the relevant medium integrated along the line-of-sight. If the physical properties of the halo gas are driven by the gravitational potential, the thermal energy of the halo gas ($\propto \int{\rm P_e dV} \propto \int y \rm dA$) would solely depend on the virial mass of the halo; referred to as self-similarity. However, simulations predict that the self-similarity breaks in low-mass halos as the role of galactic feedback becomes increasingly important compared to the gravitation \citep[e.g.,][]{Lim2021,Kim2022,Pop2022}. 

The hot CGM of the Milky Way has been extensively observed in X-ray emission \citep[e.g.,][]{Snowden2000,Henley2010,Das2019c,Kaaret2020,Gupta2021,Bluem2022,Gupta2023,Bhatt2023} and X-ray absorption \citep[e.g.,][]{Williams2005,Gupta2012,Nicastro2016b,Gatuzz2018,Das2019a,Das2021b,Lara2023}. It is incredibly challenging to probe the hot CGM of external galaxies in emission due to the spatially and temporally variable foreground dominating the total X-ray emission. So far, the extended emission from the hot CGM has been detected around a handful of individual massive ($\rm M_\star >$10$^{11.3}$M$_\odot$) super-L$^\star$ galaxies \citep[e.g.,][]{Anderson2016,Bogdan2017,Li2017} and one external L$^\star$ galaxy \citep{Das2019b,Das2020a}. There have also been stacking efforts to detect the CGM in emission by cross-correlating the eROSITA data with galaxy catalogs at $z<$0.1 \citep{Chadayammuri2022,Comparat2022}. 

However, X-ray emission is biased toward gas at higher density and higher emissivity, and thus it is primarily sensitive to the inner region of the CGM. X-ray absorption is not affected by the bias mentioned above, but it has sparse spatial sampling due to pencil beams. Also, X-ray absorption entirely depends on metal ions which makes the results dependent on assumed abundance ratios of metals and metal-to-hydrogen ratio. In addition to that, the lack of sufficient X-ray bright QSO (quasi-stellar object) sightlines passing through the CGM of external galaxies limits the scope to probe the hot CGM of external galaxies in absorption \citep[e.g.,][]{Mathur2021,Nicastro2023}. 

The SZ effect, on the other hand, does not suffer from the observational biases of X-ray emission, or the restricted spatial coverage and the metallicity dependence of X-ray absorption. Also, the redshift independence of the SZ Effect allows one to study the CGM across cosmic time. Therefore, the SZ effect is a powerful technique to characterize the CGM and test self-similarity in the halo of galaxies, complementary to X-ray observations.  

Unlike spectroscopy, the tSZ effect cannot distinguish between the hot CGM and the partially ionized cool/warm ($10^{4-6}$\,K) CGM. However, the hot CGM has a higher temperature ($\rm T_e$) and higher dispersion measure (DM $= \int n_e dl$) compared to the cooler phases of the CGM \citep[e.g., the hot CGM of the Milky\,Way contributes to the DM of the Galactic halo $\approx$an order-of-magnitude more than all other phases of the CGM combined; see][]{Das2021a}. That makes the hot CGM the primary contributor to the tSZ effect of any galaxy halo. 

Using the \planck data, the detection of the tSZ effect has been reported in the CGM of massive (M$_\star > 10^{11.3}$ \msun) locally brightest galaxies (LBGs) and galaxy groups at $z\lesssim 0.1$ \citep{Planck2013,Greco2015,Lim2018} and the large-scale structure of the universe up to $z$=1 \citep{Chiang2020,Chiang2021}. Recently, the combined maps of \planck\!{+}\act~(ACT) and \planck\!{+}\spt (SPT) has been cross-correlated with galaxy catalogs at $0.5 \leqslant z \leqslant 1.5$, again focusing on massive galaxies and galaxy groups \citep{Schaan2021,Meinke2021,Vavagiakis2021,Pandey2022}.

Building upon these works, we extend the search for the tSZ effect in the CGM of lower-mass galaxies. In \S\ref{sec:analyses}, we discuss the data extraction and data analysis. The results and their physical interpretation are mentioned in \S\ref{sec:resdis}. We conclude and summarize our results in \S\ref{sec:conclusion}.

We have used the flat $\Lambda$CDM (cold dark-matter) cosmology of \cite{Planck2016}: local expansion rate $ H_0 =67.7\rm\, km\,s^{-1} Mpc^{-1}$, matter density $\Omega_m=0.307$, baryon density $\Omega_b=0.0486$, and dark energy density $\Omega_\Lambda = 1- \Omega_m = 0.693$ throughout the paper. The cosmological baryon fraction $f_{b,cosmo} = \Omega_b/\Omega_m$ is 0.158. Virial parameters are expressed in terms of overdensity $\Delta$, e.g., ${\rm M_\Delta = \frac{4}{3} \pi R_\Delta^3 }\Delta\rho_c(z)$ is the mass enclosed within a sphere of radius $\rm R_\Delta$, where the mean mass density is $\Delta$ times the critical density of the universe, $\rho_c(z) = 3H(z)^2/8\pi \rm G$. $H(z)$ is the Hubble parameter at redshift $z$, and G is the Newtonian constant of gravitation. The redshift evolution of the Hubble parameter is $E(z)^2 = H(z)^2/H_o^2 = \Omega_m\times(1+z)^3 + \Omega_\Lambda$.

\section{Data extraction and analyses}\label{sec:analyses}
\subsection{Compton-$y$ maps}\label{sec:ymap}
\begin{figure}
    \centering
    \includegraphics[trim = 0 55 0 55, clip, width=0.45\textwidth]{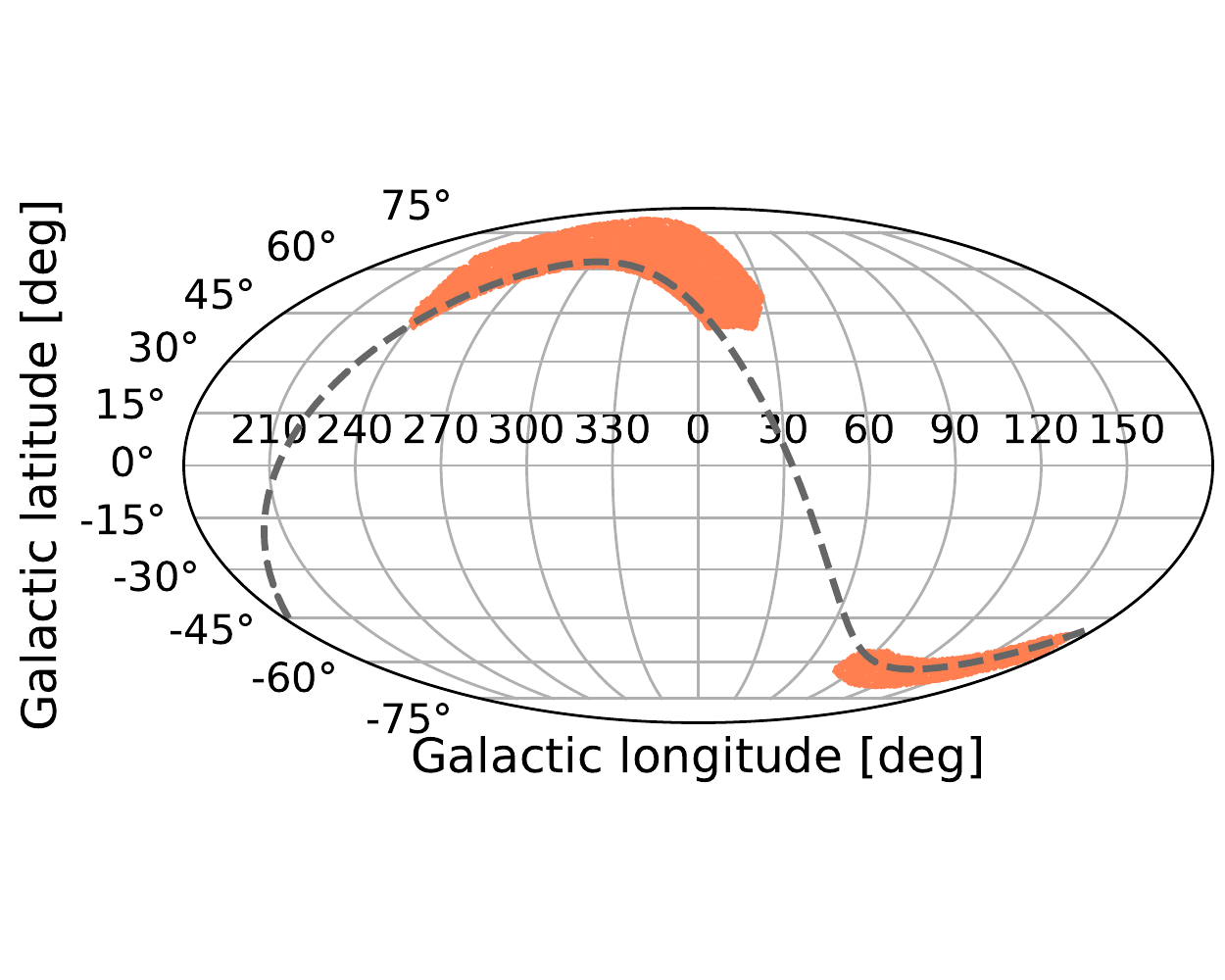}
    \caption{Positions of the Compton-$y$ maps (red patches) in Mollweide projection of galactic coordinates. \texttt{BN} and \texttt{D56} are in the northern and southern hemispheres, respectively. The equatorial plane is shown with the dashed gray curve.}
    \label{fig:skymap}
\end{figure}
We obtain the Compton-$y$ maps constructed by using the component separation method based on the internal linear combination (ILC) approach from \citet[hereafter \citetalias{Madhavacheril2020}]{Madhavacheril2020}. These maps have been extracted from the multi-frequency data of ACTPol\,DR4 (98 and 150 GHz), and of \planck (30, 44, 70, 100, 143, 217, 353, and 545 GHz). These maps are derived in two distinct, non-overlapping patches of the sky: \texttt{BN} ($-117^\circ < \rm RA<150^\circ$, $-2^\circ < \rm DEC < 19^\circ$; area = 1633 deg$^2$) and \texttt{D56} ($-9^\circ < \rm RA<40^\circ$, $-7^\circ < \rm DEC < 4^\circ$; area = 456 deg$^2$), shown in Figure \ref{fig:skymap}.  At small scales (multipole $l > 1000$) these $y$-maps are dominated by the ACT data. These maps cover only 5\% of the sky, but the signal-to-noise ratio (S/N) on small scales is orders of magnitude higher than previous \planck based $y$-maps. Also, the smaller beam size of ACT (full-width at half maxima, FWHM $\approx 1'$) compared to \planck (FWHM $\approx 10'$) allows us to resolve the galaxy halos at a smaller angular size than before.

To understand the effect of thermal dust on the Compton-$y$ measurements, we have used two types of $y$-maps from \citetalias{Madhavacheril2020} in our analyses - the maps before and after the deprojection of the cosmic infrared background (CIB). The spectral energy distribution (SED) of CIB is modeled as a modified blackbody with T$_{\rm CIB}$= 24\,K, consistent with the all-sky average of the SED fit to the CIB power spectra measurements from \planck\!. The FWHM of the Gaussian beams relevant for the $y$-maps before and after CIB deprojection are $1.\!'6$ and $2.\!'4$, respectively.  

\subsubsection{Galactic dust}\label{sec:galdust}
\begin{figure}
    \centering
    \includegraphics[width=0.475\textwidth]{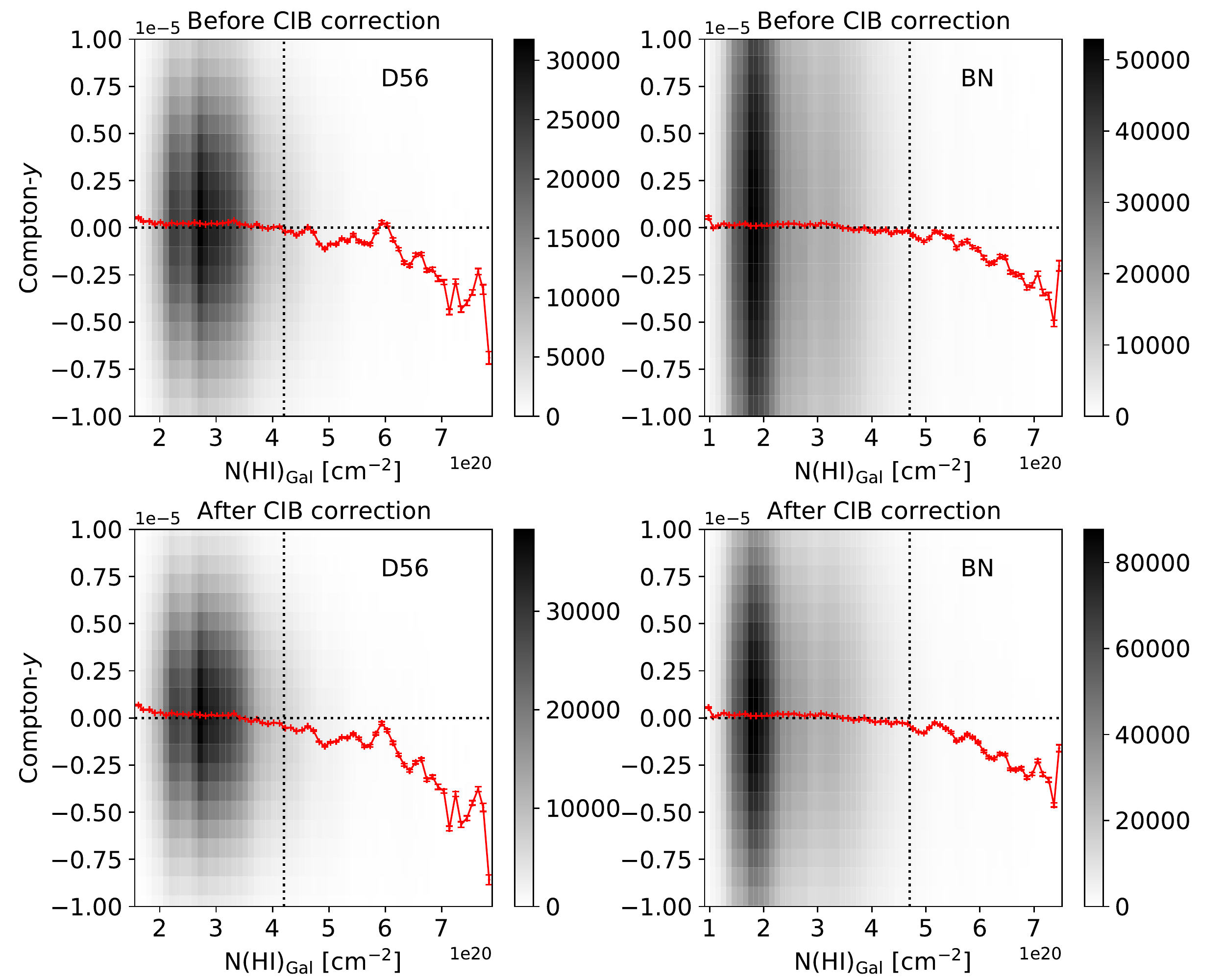}
    \caption{Compton-$y$ values before (top) and after (bottom) the CIB correction as a function of Galactic N(\hin) in the sky regions of \texttt{D56} (left) and \texttt{BN} (right). The colorbars denote the number of pixels in the $y$-maps at the corresponding values of Compton-$y$ and N(\hin)$\rm _{Gal}$ in the 2-D histograms. The mean Compton-$y$ as a function of the mean N(\hin)$\rm _{Gal}$ are calculated at a constant bin width of $\rm 10^{19} cm^{-2}$; these are shown with the red points and the red curves. The vertical dotted lines are drawn at N(\hin)$\rm_{cut}$ that we have implemented in our analyses. The horizontal dotted lines are drawn at y=0 to guide the eye.} 
    \label{fig:Galdust}
\end{figure}
We test if the $y$-maps are affected by the Galactic dust. Because the characteristic angular scale of the CIB and the Galactic dust is not necessarily the same, the $y$-maps after the CIB correction might have residual contamination by the Galactic dust emission. Therefore, we consider the $y$-maps both before and after the CIB correction. N(\hin)$\rm_{Gal}$, the neutral hydrogen column density of the Galaxy linearly correlates with E(B-V), the reddening due to Galactic dust \citep{Lenz2017}. Therefore, we use N(\hin)$\rm_{Gal}$ as a substitute for Galactic dust in our analyses. 
We obtain N(\hin)$\rm_{Gal}$ from the HI4PI survey \citep{Bekhti2016} and construct N(\hin)$\rm_{Gal}$ maps for the sky regions \texttt{BN} and \texttt{D56}. 

In the scenario of negligible contamination by Galactic dust, Compton-$y$ values would be uncorrelated with N(\hin)$\rm_{Gal}$. We find that the pixel-wise Compton-$y$ values are unlikely correlated with N(\hin)$\rm_{Gal}$ (Spearman correlation coefficients $\rm |C_{Sp}|<0.1$; 2-D histograms in Fig.\,\ref{fig:Galdust}). However, the mean Compton-$y$ shows a strong negative correlation with the mean N(\hin)$\rm_{Gal}$ ($\rm |C_{Sp}|> 0.9$) in both patches of the sky, both before and after the CIB correction (Fig.\,\ref{fig:Galdust}, red lines). The correlation becomes stronger at high N(\hin)$\rm_{Gal}$. The unphysical negative values of mean Compton-$y$ at high N(\hin)$\rm_{Gal}$ and the persisting correlation even after the CIB correction imply that the contribution of Galactic dust has not been completely accounted for in the process of CIB correction.  

To minimize the effect of Galactic dust in further analyses, we exclude the pixels in the $y$-maps corresponding to N(\hin)$_{\rm Gal}$ above a certain N(\hin)$_{\rm cut}$ based on the following conditions. The remaining mean $y$-values are in the order of $10^{-7}$ within error, and the correlation between the mean Compton-$y$ and the mean N(\hin)$\rm_{Gal}$ becomes weaker, while we can retain most ($>95\%$) of the data. We use N(\hin)$_{\rm cut}$ of  $4.7\times 10^{20} \rm cm^{-2}$ and  $4.2\times 10^{20} \rm cm^{-2}$ for the sky regions \texttt{BN} and \texttt{D56}, respectively. 



\subsection{Galaxy sample}\label{sec:galsample}
\begin{figure}
    \centering
    \includegraphics[trim=0 5 40 40, clip, width=0.45\textwidth]{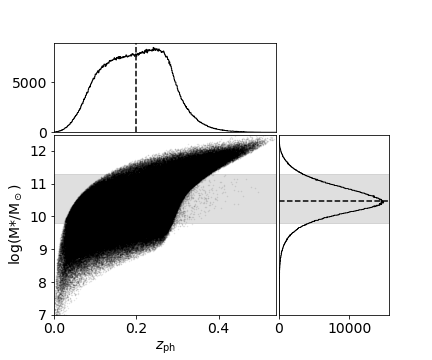}
    \caption{The redshift distribution of stellar masses of our galaxy sample. The horizontal and vertical dashed lines denote the median of the stellar mass and the redshift, respectively. The range of the stellar mass considered in this paper, $9.8 \leqslant \rm log(M_\star/M_\odot) \leqslant 11.3$ \msun, is highlighted with the horizontal gray band.}
    \label{fig:Mz}
\end{figure}
In order to cross-correlate with the Compton-$y$ maps, we consider the WISE$\times$SCOSPZ galaxy catalog that is constructed by cross-matching the all-sky samples of \wise and SuperCOSmos, and the photometric redshift (PZ) is estimated using artificial neural network algorithm \citep{Bilicki2016}. We extract the sky location, photometric redshift $z_{\rm ph}$, Galactic dust extinction-corrected \wise 3.4$\mu$m (W1) and 4.6$\mu$m (W2) magnitude of the galaxies in the \texttt{BN} and \texttt{D56} regions from the galaxy catalog. The catalog excludes targets with $\rm W1-W2>$0.9, thus making sure that the galaxies in the catalog are unlikely to host active nuclei. 

We calculate the stellar masses from the W1 and W2 magnitude using the scaling relation from \cite{Cluver2014}. The redshift distribution of the stellar masses is shown in Figure \ref{fig:Mz}. Because the detection of the tSZ effect has been previously reported in the CGM of M$_\star > 10^{11.3}$ \msun~galaxies, we focus on galaxies of lower stellar mass: $9.8 \leqslant \rm log(M_\star/M_\odot) \leqslant 11.3$. The median redshift and the median stellar mass of our sample are 0.2 and 10$^{10.48}$\msun, respectively. 

\begin{figure}
    \centering
    \includegraphics[width=0.475\textwidth]{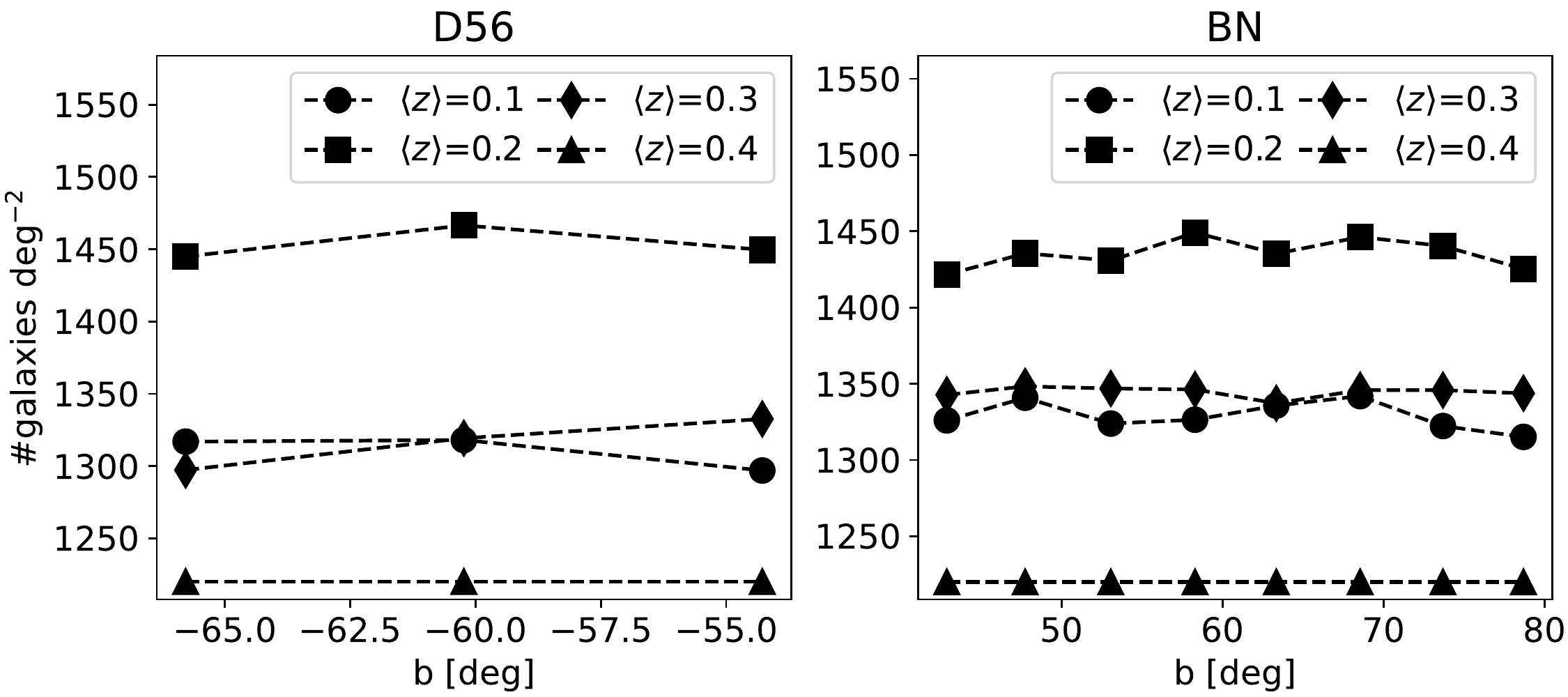}
    \caption{The surface density distribution of the $9.8 \leqslant \rm log(M_\star/M_\odot) \leqslant 11.3$ \msun~galaxies as a function of the galactic latitude and the redshift of the galaxies.}
    \label{fig:surfdens}
\end{figure}
The sky regions \texttt{BN} and \texttt{D56} are unlikely to be contaminated by the stars in the Galactic disk due to their high galactic latitude ($|b|>30^\circ$; see Fig.\,\ref{fig:skymap}). To test if it is otherwise, we calculate the surface density, i.e., the number of galaxies per unit area of the sky as a function of the galactic latitude and the redshift of the galaxies (Fig.\,\ref{fig:surfdens}). If the galaxy sample is dominantly contaminated by stars, the surface density would be anti-correlated with the galactic latitude. But the surface density is practically uniform, with $<10\%$ fluctuation (Fig.\,\ref{fig:surfdens}). At a given galactic latitude the surface density follows the redshift distribution shown in Figure\,\ref{fig:Mz}, but it is not correlated with the galactic latitude at a given redshift. It confirms that our galaxy sample is not contaminated by the stars in the Galactic disk.


We estimate the virial (or halo) masses of the galaxies, M$_{\rm vir}$ ($\equiv \rm M_{200}$) from the stellar masses using the stellar-to-halo mass relation (SHMR) from \citet[hereafter \citetalias{Zu2015}]{Zu2015}. The behavior of the empirical model of \citetalias{Zu2015} is governed by the results of \cite{Behroozi2010}, but \citetalias{Zu2015} has refitted the parameters to the $z=0$ data of SDSS\,DR7, and has allowed the scatter around the best-fit to vary with halo mass. 
We calculate the virial radii, R$_{\rm vir}$ ($\equiv \rm R_{200}$) using the Navarro-Frank-White (NFW) profile of the dark-matter \citep{Navarro1997} . We also determine the corresponding angular size, $\rm \theta_{\rm vir}$ ($\equiv \theta_{200}$ = R$_{\rm 200}/\rm D_A$), where D$\rm _A$ is the angular diameter distance. Using the concentration factor, $c_{\rm 200}$ as a function of the virial mass \citep{Neto2007}, we convert M$_{\rm 200}$ to M$_{\rm 500}$ and obtain the corresponding R$_{\rm 500}$ as well. 
 


\subsection{Masking}\label{sec:masking}
The tSZ effect of the intracluster medium (ICM) and the emission from compact radio sources can contaminate/bias our Compton-$y$ measurements. Therefore, we mask the $y$-maps at the positions of galaxy clusters and radio sources. We discuss the removal process in the following sections.  

\subsubsection{Galaxy clusters}\label{sec:clustermask}
We extract the position, redshift, and virial mass, M$_{\rm 200,cl}$ of the galaxy clusters from the tSZ cluster catalog \citep{Hilton2021}. These clusters have been individually detected at $>\!4\sigma$ in the 98 and 150\,GHz channels of ACT, and have been optically confirmed. We calculate the virial radii, R$_{\rm 200,cl}$, using the NFW profile of dark matter. We determine the angular size, $\theta_{\rm 200,cl}$ of the cluster halos from their R$_{\rm 200,cl}$ and redshift. We exclude ($\equiv \texttt{NAN}$) the $\theta_{\rm 200,cl}$ regions around the clusters in the $y$-maps. The cluster catalog does not include the nearby Virgo cluster and the Abell\,119 cluster that are prominently visible in the $y$-maps. We obtain the R$_{\rm 200,cl}$ and the redshift of these clusters from \cite{Kashibadze2020} and \cite{Piffaretti2011}, respectively, and remove the corresponding $\theta_{\rm 200,cl}$ regions around them. The galaxies in our sample that are residing within these clusters are naturally removed in this process. Because the field galaxies and galaxies within clusters evolve in different environments, their tSZ effect might be different from each other. The exclusion of the galaxies within clusters will make the interpretation of our results more straightforward.   

\subsubsection{Radio sources}\label{sec:radiomask}
For the sky regions \texttt{D56} and \texttt{BN}, the 1$\sigma$ noise level at 98\,GHz is 1.4\,mJy and 2.1\,mJy and at 150\,GHz it is 0.9\,mJy and 1.8\,mJy, respectively. The $y$-maps were constructed after subtracting the unresolved sources that are individually detected at $\geqslant 5\sigma$ \citepalias{Madhavacheril2020}. To account for the remaining fainter sources, we obtain the position, flux, and angular size of the radio sources from the latest catalog of the FIRST \citep[Faint Images of the Radio Sky at Twenty-cm;][]{FIRST1997} survey. We choose the sources that are unlikely to be spurious (\texttt{P$\rm _S<$0.05}; \texttt{P$\rm _S\equiv$} the  probability of falling in the sidelobe of a nearby brighter radio source). To make sure that each position in the sky is probed down to equal sensitivity, we consider the sources above a constant peak flux of 1.1\,mJy and an integrated flux of 0.7\,mJy at $\approx$1.4\,GHz. 

Our goal is to remove the radio sources from the $y$-maps down to the 1$\sigma$ noise level at 98 and 150\,GHz. We translate the corresponding flux at these frequencies to the flux at $\approx$1.4\,GHz, $\rm F_{1.4}$, using the median SED of $F_\nu \propto \nu^{-0.75}$ for radio-loud quasars \citep{Shang2011}\footnote{A fraction of the radio sources are stellar or unclassified and may have a different SED than the radio galaxies. Our goal is to exclude the brightest sources, and galaxies are brighter than other sources, validating our assumption of the SED.}. 
We exclude ($\equiv \texttt{NAN}$) the regions in the $y$-maps affected by the radio sources brighter than $\rm F_{1.4}$, i.e., the angular size of the sources convolved with the Gaussian beams relevant for the $y$-maps (see \S\ref{sec:ymap}).

We cross-match the WISE$\times$SCOSPZ catalog with the FIRST catalog, and find that $\approx$0.4\% of WISE$\times$SCOSPZ galaxies of our interest are radio-loud; we exclude them from our galaxy sample.

\subsection{Stacking}\label{sec:stacking}
We split the galaxy sample into stellar mass bins of different widths, $\Delta \rm M_\star$ = 0.2\,dex, 0.3\,dex, and 0.4\,dex, and stack the Compton-$y$ map in each mass bin separately. With wide mass bins, we can constrain the value of Compton-$y$ at higher significance. On the other hand, we get better insight into the mass dependence of Compton-$y$ from the narrow mass bins. We provide the stellar mass, mean redshift, the total number of galaxies, mean virial mass, and virial radius in each mass bin in Table\,\ref{tab:result}.
We perform the same procedure for both patches of the sky and then combine them. 

We stack the $y$-maps in three ways, at I) fixed angular size, II) fixed projected radius, and III) fixed fraction of virial radius; these are discussed in detail below. 

I) Fixed angular size: Following most of the previous studies, we stack the stamps\footnote{We call the cutout of $y$-map around a galaxy as ``stamp". It is extracted using the \texttt{thumbnail} subroutine of the \texttt{reprojection} module of \texttt{pixell} \citep{Naess2021}. Each ``stamp" is projected onto a local tangent plane in order to remove the effect of size and shape distortions in the input $y$-map.} of a fixed angular size:
\begin{equation}\label{eq:stackmethod1}
    y_{\rm stacked} = \langle y(\theta_i)\rangle;\hspace*{0.3 cm}  \theta_i = \theta_\perp = r_i/{\rm D}_{{\rm A},i}
\end{equation}
Here, $r_i$ and D$_{\rm A,i}$ are the projected radius and the angular diameter distance of the $i$-th galaxy in the subsample, respectively. $\langle y \rangle$ is the average Compton-$y$ value of all galaxies in that subsample at a fixed galactocentric angular distance perpendicular to the line-of-sight, $\theta_\perp$. We extract each stamp out to the median 6$\times \theta_{\rm 200}$ of the relevant subsample so that we can study $y$ out to a sufficiently large angular separation beyond the CGM, and estimate the tSZ background from the same stamp.

For galaxies at different redshift, this method stacks different projected radii for different galaxies, which is also a different fraction of R$_{\rm 200}$ for galaxies of similar halo masses. Thus, it is not straightforward to extract any physically meaningful information for the galaxies from the stacking at a given angular size. For this reason, we introduce the other two stacking  procedures. 

II)  Fixed projected radius: Here, the galaxy halos are stacked at the same projected radius:
\begin{equation}\label{eq:stackmethod2}
    y_{\rm stacked} = \langle y(\theta_i)\rangle; \hspace*{0.2 cm} \theta_i = r_{\perp}/{\rm D}_{{\rm A},i}
\end{equation}
Here $\langle y \rangle$ is the average Compton-$y$ value of all galaxies in the subsample at a fixed galactocentric physical distance perpendicular to the line-of-sight, r$_{\perp}$. This method works under the assumption that the radial pressure profiles of the CGM, P($r$) have the same shape in all galaxies in a subsample.

We extract each stamp out to the median 6$\times$R$_{\rm 200}$ in the relevant mass bin. 
Because the corresponding angular size of the stamp is not the same for each galaxy, the stamps have different numbers of pixels, $n_{\rm pixel}$, on each side, which cannot be directly stacked. Therefore, we reconstruct each stamp using 2-D piece-wise linear interpolation so that all stamps have the same $n_{\rm pixel}$. Thus, the stamps are under or over-sampled when the initial $n_{\rm pixel}$ is larger or smaller than the final $n_{\rm pixel}$, respectively. The final $n_{\rm pixel}$ is set to be the median of the initial $n_{\rm pixel}$ distribution of all stamps, corresponding to the angular size at the median D$_{\rm A}$ of a subsample. Thus, half of the stamps are under-sampled and the other half is over-sampled, minimizing the effect of the stamp reconstruction. The distance of a given pixel from the central pixel in a reconstructed stamp is equivalent to the projected physical distance between the corresponding sky point and the center of the relevant galaxy. 

III) Fixed fraction of virial radius: Here, we stack the galaxy halos at the same fraction of R$_{\rm 200}$ of each galaxy:
\begin{equation}\label{eq:stackmethod3}
    y_{\rm stacked} = \langle y(\theta_i)\rangle; \hspace*{0.1 cm} \theta_i =  r_i/{\rm D_{A,i}};\hspace*{0.1 cm} r_i = f_{200}\times {\rm R}_{200,i}
\end{equation}
Here $\langle y \rangle$ is the average Compton-$y$ value at galactocentric physical distances perpendicular to the line-of-sight, $r_i$, which are a given fraction, $f_{200}$, of R$_{\rm 200,i}$, the virial radius of the $i$-th galaxy in the relevant mass bin. This method works under the assumption that the radial pressure profiles of the CGM are self-similar, i.e., the pressure at a given fraction of R$_{\rm 200}$, P($f_{200}$) has the same shape in all galaxies that are being stacked. 

We extract the stamp out to 6$\times$R$_{\rm 200}$ of each galaxy. Because the corresponding angular size, 6$\times \theta_{\rm 200}$ is different for each galaxy, we follow the same steps of reconstructing $y$-maps as the previous method of stacking. The final $n_{\rm pixel}$, same as the median of the initial $n_{\rm pixel}$ distribution, corresponds to the median 6$\times \theta_{\rm 200}$ of a subsample. The distance of a given pixel from the central pixel in a reconstructed stamp is equivalent to the projected distance of the corresponding sky point from the relevant galaxy center in the unit of R$_{\rm 200}$. 

For galaxies at similar redshift, the first and the second stacking method would produce the same result. For galaxy samples with similar R$_{\rm 200}$, the second and the third stacking methods would be equivalent. For galaxies at similar redshift and with similar R$_{\rm 200}$, the third stacking method would converge to the first one.  

\subsection{Aperture photometry}\label{sec:apphot}
We have extracted the radial profile of Compton-$y$ from the stacked $y$-maps in two different ways: cumulative (the conventional method), and differential. As such, these provide us with complementary information.

The differential profile is a direct way to detect the signal at any projected distance. In the absence of the signal, the differential profile would be consistent with zero.
We define the differential profile in equation\,\ref{eq:ydf}

\begin{subequations}
\begin{equation}\label{eq:ydf}
     y(\theta_{\rm df}) = \bar y_{\rm stacked}(\theta_{\rm df}-\Delta \theta\leqslant \theta_\perp < \theta_{\rm df}+\Delta \theta) 
\end{equation}
Here, $\theta_{\rm df}$ and $2\Delta \theta$ are the angular size and width of the circular annulus, respectively. $2\Delta \theta$ is the FWHM of the Gaussian beam so that each annulus is just resolved. $\bar y$ is the average $y$ within the annulus. For the stacking method II,
\begin{equation}
\theta_{\rm df} = r_{\rm \perp,df}/\rm D_{\rm A,med}; \Delta \theta = \Delta r/\rm D_{\rm A,med}
\end{equation}
Here, $r_{\rm \perp,df}$ and 2$\Delta r$ are the physical radius and width of the circular annulus, and D$_{\rm A,med}$ is the median angular diameter distance of the galaxy subsample. For the third stacking method, 
\begin{equation}
\theta_{\rm df} = f_{\rm df} \times \theta_{\rm 200,med}; \Delta \theta = \Delta f \times  \theta_{\rm 200,med}
\end{equation} 
Here, $f_{\rm df}$ and 2$\Delta f$ are the physical radius and width of the annulus in the unit of R$_{\rm 200}$, and $\theta_{\rm 200,med}$ is the median angular size of the galaxy halos in the subsample.
\end{subequations}

The cumulative profile is an indirect way to detect the signal, especially when the data is not deep enough to directly detect the signal. We define the cumulative profile in equation\,\ref{eq:ycm}
\begin{subequations}
\begin{equation}
\begin{split}\label{eq:ycm}
y(\theta_{\rm cm}) = \bar y_{\rm stacked}(\theta_\perp \leqslant \theta_{\rm cm}) \\ 
{\rm Y}(\theta_{\rm cm}) = \int^{\theta_{\rm cm}}_0 y {\rm d\Omega} = y(\theta_{\rm cm})\times \pi \theta_{\rm cm}^2
\end{split}
\end{equation}
Here $\theta_{\rm cm}$ is the angular radius of the circular aperture, and $\bar y$ is the average $y$ within the aperture. For the stacking methods II and III, 
\begin{equation}
    \theta_{\rm cm} = r_{\rm \perp,cm}/\rm D_{\rm A,med}
\end{equation}
and 
\begin{equation}
    \theta_{\rm cm} = f_{\rm cm}\times\rm \theta_{\rm 200,med}
\end{equation}
\end{subequations}
Here $r_{\rm \perp,cm}$ is the physical radius of the aperture, and $f_{\rm cm}$ is the radius of the aperture in the unit of R$_{\rm 200}$. 

The minimum value of $\theta_{\rm cm}$ and ($\theta_{\rm df}+\Delta \theta$) by definition is half of the FWHM of the relevant Gaussian beam because any region smaller than this is not resolvable. For each mass bin it is mentioned in Table\,\ref{tab:result} in the unit of virial radius. Maximum $\theta_{\rm cm}$ and ($\theta_{\rm df}+\Delta \theta$) depends on the stamp size, which, as described in \S\ref{sec:stacking}, are median 6$\sqrt{2}\times \theta_{\rm 200}$, median 6$\sqrt{2}\times$R$_{\rm 200}$/$\rm D_{A,med}$, and median 6$\sqrt{2}\times \theta_{\rm 200}$ of the relevant galaxies in stacking methods I, II and III, respectively. 

Equation\,\ref{eq:ycm} is different from the cumulative estimate that was used in previous studies of the tSZ effect \citep[e.g.,][]{Schaan2021}. For an aperture of angular size $\theta$, they extract the background from the region between $\theta$ and $\sqrt 2\theta$ and subtract it from the signal. 
In our case, the \textit{shape} of the cumulative profile is more informative than the absolute \textit{values}. In the presence of any signal, the cumulative profile should monotonically increase with radius, and the rate of increase would indicate the strength of the signal in the outer part of the aperture. The profile would saturate when there is no signal. 

\subsection{Error estimation}\label{sec:err}
We estimate the combined statistical and systematic uncertainties in the Compton-$y$ using the bootstrap method. For a sample of N galaxies, we create a replica of the sample by replacing one galaxy with a randomly chosen galaxy from the rest (N-1) galaxies in that sample. For every stacked sample, we make a set of 1000 such replicas and obtain the mean, covariance matrix, 68.27\%, and 99.73\% confidence intervals of the distribution of those replicas. 

\subsection{tSZ background}\label{sec:2halo}
For the following calculation, we consider the radial profiles of Compton-$y$, $y_{\rm tot}$, derived from the CIB and Galactic dust-corrected, galaxy clusters and radio sources-masked stacked $y$-maps. $y_{\rm tot}$ has two components, the ``one-halo" radial profile of Compton-$y$, $y_{\rm 1h}$, and the tSZ background. This background consists of the ``two-halo" term, $y_{\rm 2h}$ and any zero-point offset, $y_{zp}$. The ``two-halo" term is the correlated tSZ signal of other halos around the galaxy of interest. The $y_{zp}$ comprises the global Compton-y signal arising from reionization, the intergalactic medium, the CGM of the Milky\,Way, and calibration uncertainties. We consider the ``two-halo" term calculated for an FWHM = 1.$\!'4$ beam from \cite{Vikram2017}, and add it to $y_{\rm 1h}$ in each radial bin with a variable amplitude, $\rm A_{2h}$ (equation\,\ref{eq:2halo}). 
\begin{equation}\label{eq:2halo}
y_{\rm tot}(r_\perp) = y_{\rm 1h}(r_\perp) + {\rm A_{2h}}y_{\rm 2h}(r_\perp)+ y_{zp} 
\end{equation}

\subsection{Modeling pressure and tSZ}\label{sec:P(r)}
In the observed tSZ effect, the thermal pressure of the free electrons in the CGM, ${\rm P_e}(r)$ is convolved with the effective beam of the $y$-maps that we use in the stacking.
\begin{equation}\label{eq:fit}
\begin{split}
    y_{\rm 1h}(r_\perp) 
    = (\sigma_{\rm T}/m_e c^2) \int_{-\infty}^{\infty} {\rm P_e}(r) dl \circledast Beam(r_\perp) \\
    {\rm where}\;\;     r^2 = r_\perp^2 + l^2,\\
     Beam(\theta) = \frac{1}{\sqrt{2\pi}\sigma}exp(-\theta^2/2\sigma^2),
    \\ \rm and,\;\; \sigma = FWHM/2\sqrt{2ln(2)}
\end{split}
\end{equation}
Here, $r_\perp$ and $l$ are physical distances perpendicular to and along the line-of-sight, respectively. 

We model ${\rm P_e}(r)$ with a generalized NFW (GNFW) profile (equation\,\ref{eq:pressure}) that was first defined by \cite{Nagai2007} to describe the intracluster medium. 
\begin{equation}\label{eq:pressure}
\begin{split}
{\rm P_e}(x) = {\rm P_{500}}\Big(\frac{\rm M_{500}}{3\times 10^{14} h_{70}^{-1}\rm M_\odot}\Big)^{\alpha_p+\alpha_p'(x)}\mathbb{P}(x)
\\ {\rm where}\;\; x = r/\rm R_{500},
\\ {\rm P_{500}} = 500\rho_c(z)f_{b,cosmo}\Big(\frac{\mu}{\mu_e}\Big)\rm \Big(\frac{GM_{500}}{2R_{500}}\Big),
\\ h_{70} =  H_o/70\,\rm km\,s^{-1}\,Mpc^{-1},
\\ {\rm and}\;\; \mathbb{P}(x) = {\rm P_o} (c_{500}x)^{-\gamma}(1 + (c_{500}x)^\alpha)^{-(\beta - \gamma)/\alpha}
\end{split}
\end{equation} 
Here, $\mu$ is the mean molecular weight, and $\mu_e$ is the mean molecular weight per free electron. $\alpha_p$ corresponds to a modification of the
standard self-similarity and $\alpha_p'(x)$ introduces a break in the self-similarity. $c_{500}$ is the concentration factor, $\alpha, \beta$ and $\gamma$ are the slopes of the pressure profile at $x\approx r_s$, $x\gg r_s$, and $x\ll r_s$, respectively, where, $r_s = {\rm R_{500}}/c_{500}$.

For local ($z<0.2$) massive ($\rm M_{500} > 10^{14} M_\odot$) galaxy clusters, \citeauthor{Arnaud2010} (\citeyear{Arnaud2010}; hereafter \citetalias{Arnaud2010}) considered a hybrid average profile of pressure, combining the profiles from X-ray observations within 0.03--1\,R$_{500}$ and simulations in the 1--4\,R$_{500}$ region. For their best-fit estimate of $c_{500}$, $r_s$ of our sample galaxies are smaller than the angular resolution of the $y$-maps that we consider in our analysis. Therefore, by construction, we cannot constrain the slopes of the pressure profile independently. 
Thus, we freeze $c_{500}, \alpha, \beta, \gamma$ and $\alpha_p$ at their empirically derived best-fitted values in \citetalias{Arnaud2010}: 1.177, 1.0510, 5.4905, 0.3081, and 0.12, respectively, and neglect the weak radial dependence $\alpha_p'(x)$ as it has a second order effect \citep[also adopted in][for tSZ profile of $\rm M_\star > 10^{11.3} M_\odot$ galaxies]{Greco2015}.  We freeze $\mu$ at 0.59 and $\mu_e$ at 1.14, the values adopted by \cite{Nagai2007}. We set the initial $\rm P_o$ as 8.403$h_{70}^{-1.5}$ \citepalias[best-fitted value in][]{Arnaud2010} and allow it to vary. 

We use Markov Chain Monte Carlo (MCMC) calculations \citep{Metropolis1953,Hastings1970} to estimate the posterior probability distribution functions of $\rm P_o$, $\rm A_{2h}$ and $y_{zp}$, using the Affine-Invariant Ensemble Sampler algorithm implemented in \texttt{emcee} \citep{Foreman-Mackey2013}. We assume uniform priors on $0\leqslant\rm P_o\leqslant200$ and $0\leqslant\rm A_{2h}\leqslant 10$ and do not put any constraint on $y_{zp}$. We assume the likelihood to be Gaussian. We run multiple \texttt{emcee} ensembles and keep on adding independent sets of chains until I) the number of chains is larger than 50 times the integrated auto-correlation time, $\tau$, of each parameter, and II) the Gelman-Rubin convergence parameter reaches values within 0.9--1.1 \citep{Gelman1992}. We remove the first 2$\times\tau_{max}$ steps from the sampler to get rid of the burn-in phase. We consider the most likely values of the free parameters (i.e., values corresponding to the minimum $\chi^2$) and the covariance matrices of their posterior probability distributions for the following calculations. 

\begin{figure*}
    \centering
    \includegraphics[width=0.42\textwidth]{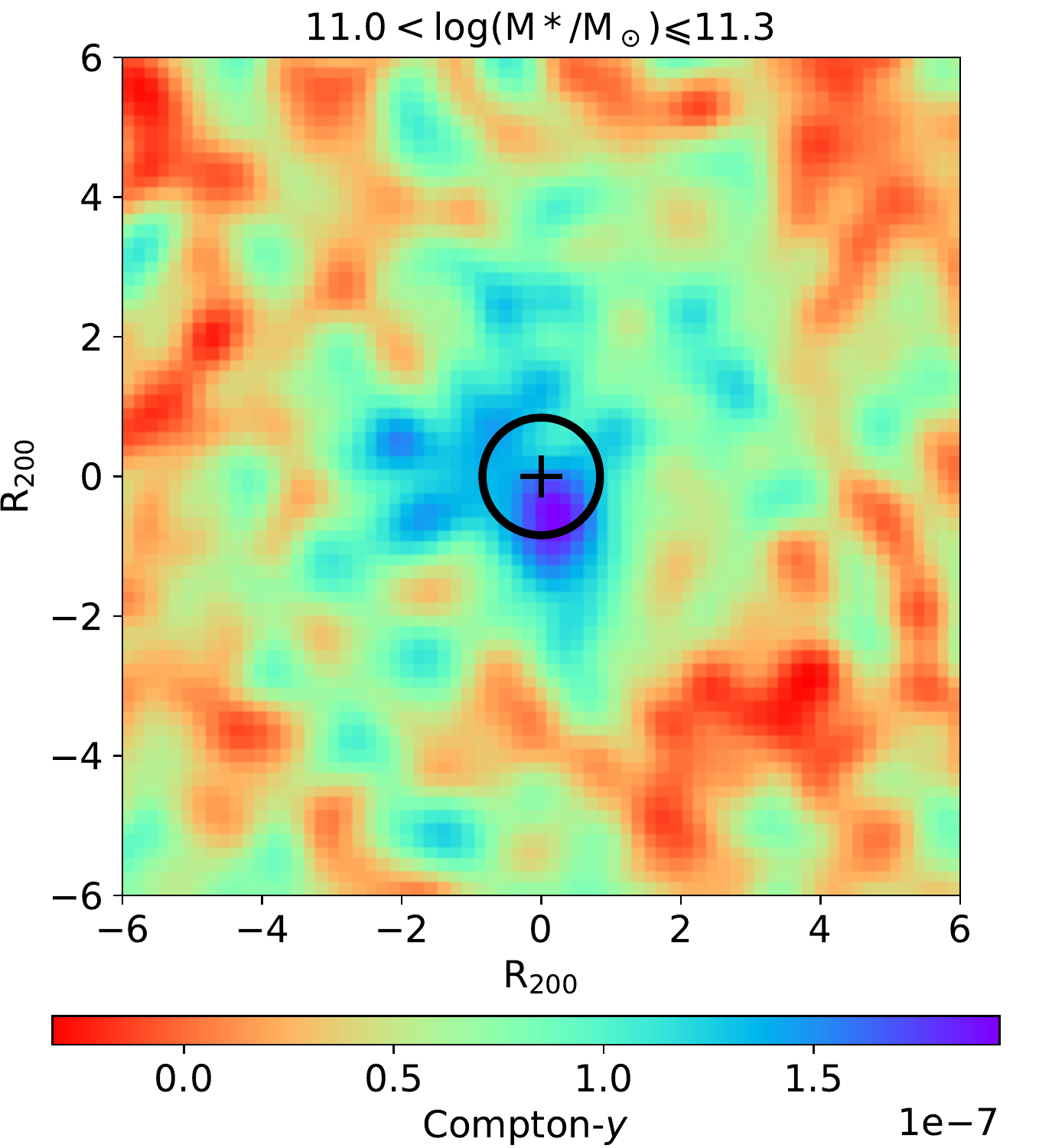}
    \includegraphics[width=0.49\textwidth]{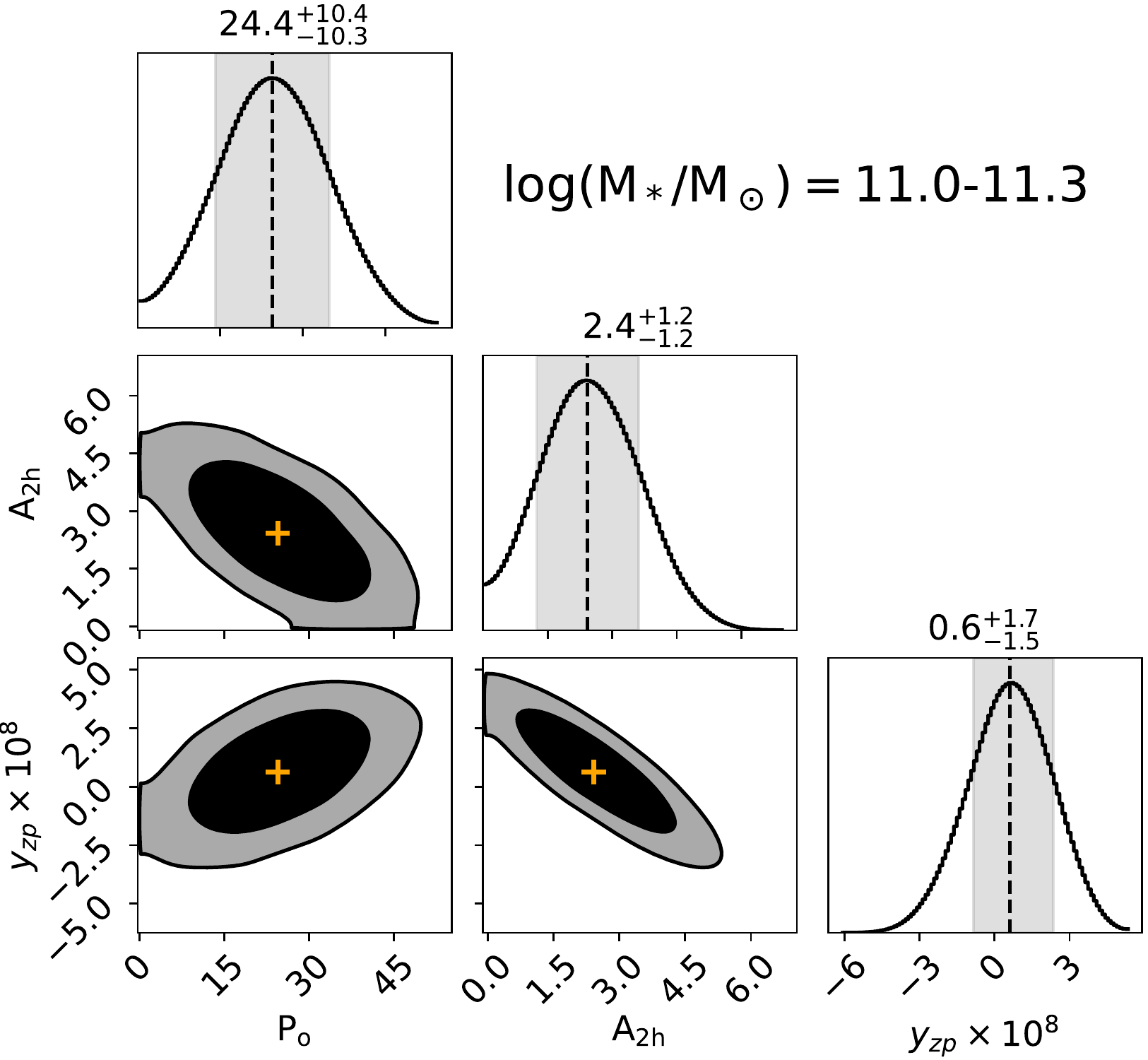}
    \includegraphics[width=0.475\textwidth]{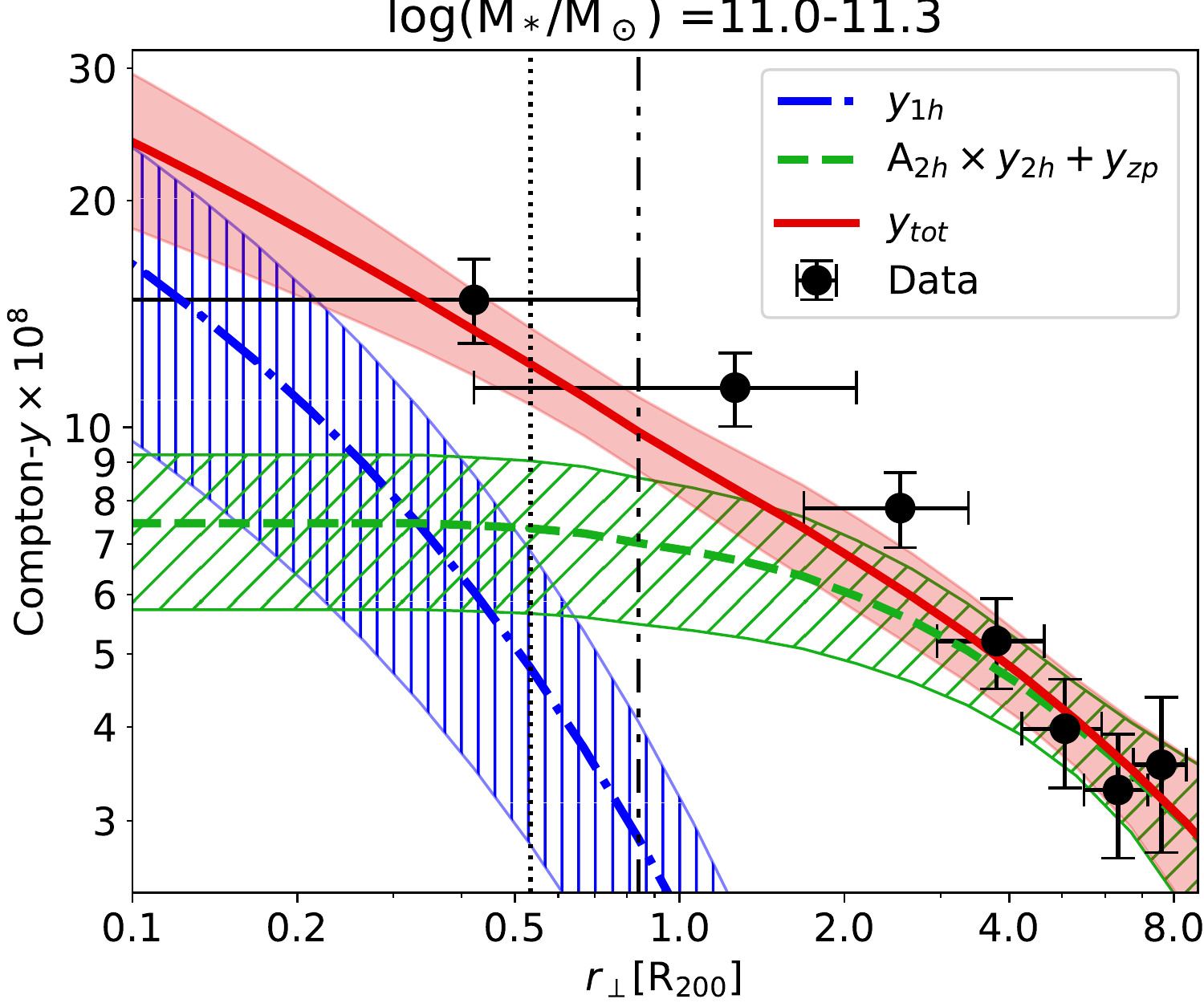}
    \includegraphics[width=0.475\textwidth]{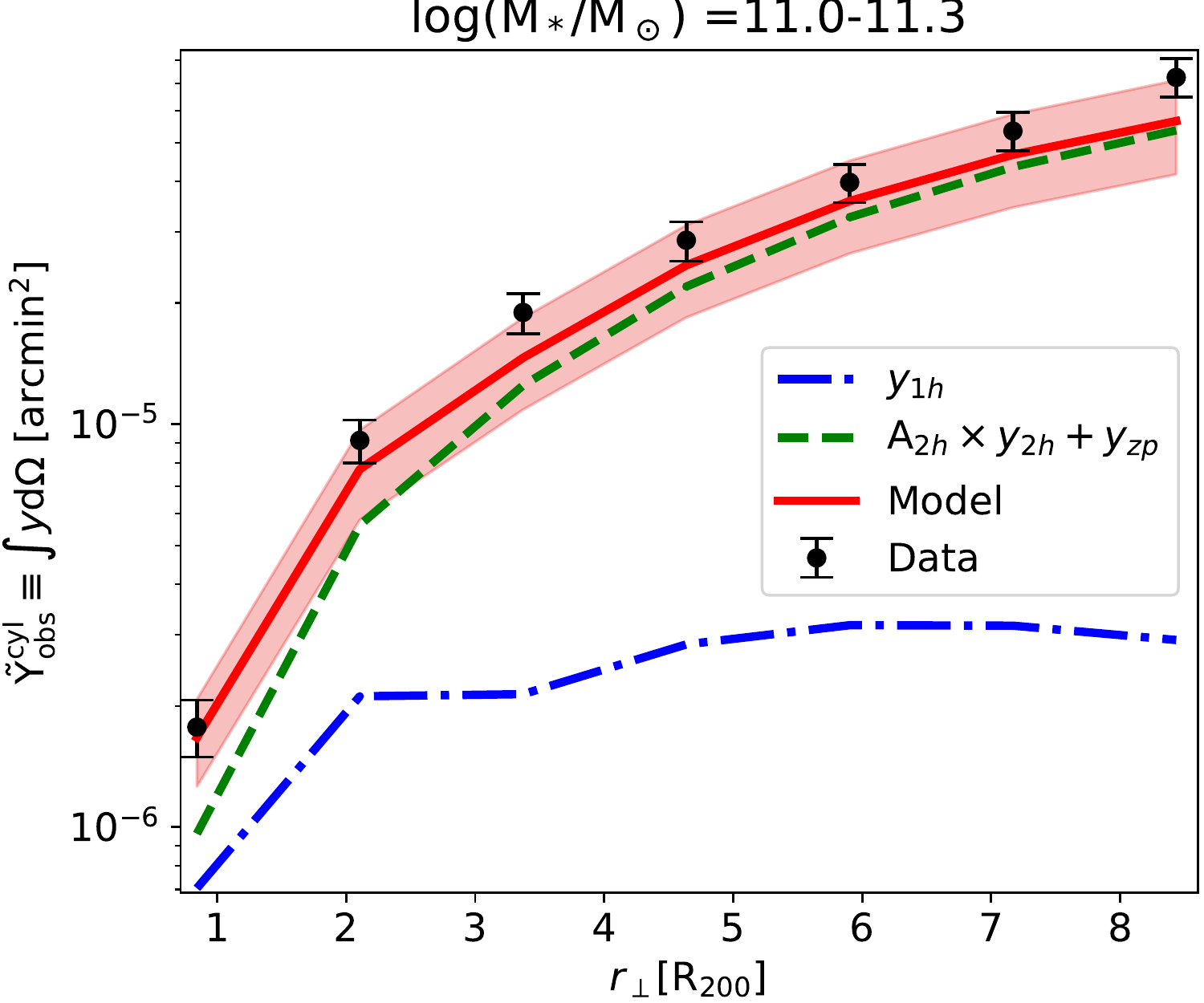}
    \caption{Results for the $\rm log(M_\star/M_\odot)=11.0-11.3$ mass bin. \textbf{Top left:} CIB and Galactic dust-corrected, galaxy clusters and radio sources-masked Compton-$y$ map stacked using method\,III (fixed $f_{200}$). The center of stacked galaxies (i.e., the central pixel) is marked with a plus (‘+’) sign. The Gaussian
    beam (FWHM of 2.$\!'$4) is shown with a circle at the center of each map. The colorbar is shown below the map. \textbf{Top right:} Posterior probability distributions of the amplitude of the thermal pressure profile: $\rm P_o$, the amplitude of $y_{2h}$: $\rm A_{2h}$, and zero-point offset: $y_{zp}$, obtained by fitting the GNFW pressure model to our tSZ measurements. The vertical dashed lines and the shaded regions in the diagonal plots correspond to the most likely value and 68\% confidence interval of the marginalized distribution of the free parameters; the values are mentioned in the respective titles. The most likely values are marked with `+' in the contour plots. The contours correspond to 68\% and 95\% confidence intervals.  \textbf{Bottom left:} Differential radial profile of Compton-$y$. The black circles with error bars are extracted from the map shown above (see \S\ref{sec:apphot}). The dash-dot-dotted vertical line denotes the FWHM of the beam (2.$\!'$4). By construction, the width of each annulus is the same as the beam size, shown with the error bars along the x-axis. The error bars along the y-axis denote 68\% confidence intervals. The dotted vertical line is drawn at R$_{500}$. The dash-dotted blue curve and the dashed green curve are the best-fit ``one-halo" term, $y_{1h}$ and tSZ background (${\rm A_{2h}}y_{2h} + y_{zp}$), respectively, with the hatched area around them corresponding to the 1$\sigma$ uncertainty. The solid red curve and the shaded area around it are the best-fit model with 1$\sigma$ uncertainties. \textbf{Bottom right:} Cumulative radial profile of Compton-$y$, i.e. $\int y \rm d\Omega$, normalized at an angular diameter distance of 500\,Mpc. The black circles with error bars are extracted from the map shown above (see \S\ref{sec:apphot}). Error bars correspond to 68\% confidence intervals. The solid red curve and the shaded areas around it are calculated from the model and its 1$\sigma$ uncertainties, best fitted to the differential radial profile of Compton-$y$ shown on the left.}
    \label{fig:bestfit}
    \vspace*{-0.1 in}
\end{figure*}

\subsection{Thermal energy}
From the best-fit pressure profiles in \S\ref{sec:P(r)} we calculate $\rm Y^{sph}_{R500}$, i.e., the thermal pressure of the free electrons\footnote{\textcolor{black}{For a fully ionized monoatomic ideal gas, the thermal energy density is $\Big(\frac{3\mu_e}{2\mu}\Big)\rm P_e \sim 2.9 P_e$. The self-similarity (or the lack thereof) of $\rm Y^{sph}_{R500}$ is not affected by the proportionality factor. Because the goal of this paper is to study $\rm Y^{sph}_{R500}$ as a function of mass, we do not explicitly discuss the thermal energy further.}} integrated within a spherical volume of radius R$_{500}$. Following conventions, it is multiplied by $\sigma_{\rm T}/m_e c^2$ and divided by ${\rm D_A}(z)^2$ in order to express it in the unit of solid angle subtended by the spherical volume of consideration in the sky. 
\begin{subequations}\label{eq:Ysph}
\begin{equation}
    {\rm Y^{sph}_{R500}} = (\sigma_{\rm T}/m_e c^2) \int_{0}^{\rm R_{500}} {\rm P_e}(r)\,4\pi r^2 dr /{\rm D_A}(z)^2
\end{equation}
\begin{equation}
    {\rm and,\, \tilde Y^{sph}_{R500} =  Y^{sph}_{R500}} E(z)^{-2/3} {\rm (D_A}(z)/500\,\rm Mpc)^2,
\end{equation}
\end{subequations}
$\rm \tilde Y^{sph}_{R500}$ takes into account different redshift of galaxies and scale $\rm Y^{sph}_{R500}$ to $z=0$ through $E(z)$, and normalize $\rm Y^{sph}_{R500}$ to a fixed angular diameter distance of 500\,Mpc.

\subsection{Baryonic mass and baryon fraction}\label{sec:masscal}
We assume that the hot CGM, i.e., the ionized gas within $\rm R_{200}$ of the stacked galaxies in each mass bin, is at the virial temperature, $\rm T_{vir}$ ($\equiv \rm T_{200}$). We estimate $\rm T_{200}$ from the virial mass and the virial radius (equation \ref{eq:Tvir}). We estimate the density profiles from the best-fitted pressure profiles in \S\ref{sec:P(r)} and calculate the mass of the hot circumgalactic gas using equation \ref{eq:mass}.
\begin{subequations}
\begin{equation}\label{eq:mass}
\begin{split}
   {\rm M_{gas,200}} = \int_{0}^{\rm R_{200}} \mu_e m_p n_e(r) 4\pi r^2 dr \\
    {\rm where},\;\; n_e(r)  = {\rm P_e}(r) \rm k_B^{-1} T_{200}^{-1} \\
\end{split}
\end{equation}
\begin{equation}\label{eq:Tvir}
   {\rm  and, T_{200}} = \frac{\mu m_p \rm G M_{200}}{\rm 2k_B R_{200}} 
\end{equation}
\end{subequations}
Here, $m_p$, $n_e$, and $\rm k_B$ are the rest mass of the proton, the number density of electrons, and Boltzmann's constant, respectively. Finally, we calculate the baryon fraction, $f_b = {\rm (M_\star + M_{gas,200})/ M_{200}}$.

\subsection{Null test}
For the null test, we create a mock catalog of galaxies in each patch of the sky, i.e., \texttt{BN} and \texttt{D56}. We construct a 2-D random distribution spanning the range of RA and DEC of our sample and replace the true location (RA, DEC) of the galaxies with values drawn from that distribution. Then, we repeat the stacking, aperture photometry, and error estimation of Compton-$y$ for this mock catalog of galaxies. 

\section{Results and discussion}\label{sec:resdis}
In the following sections, we discuss the results derived from CIB and Galactic dust-corrected, galaxy clusters and radio sources-masked $y$-maps stacked using the third stacking method. We have discussed the individual effect of CIB, Galactic dust, galaxy clusters, and radio sources on stacking and compared three stacking methods in Appendix\,\ref{sec:proofofconcept}. 
\subsection{Best-fit model}
In Figure\,\ref{fig:bestfit}, we show the results of stacking (top left panel) and pressure modeling for the $\rm M_\star = 10^{11.0-11.3} M_\odot$ mass bin. $\rm A_{2h}$ is greater than 1, and $y_{zp}$ is consistent with zero within 1$\sigma$ uncertainty (top right panel), implying a significant contribution of the ``two-halo" term and negligible contribution of the zero-point offset. $\rm A_{2h}$ and $y_{zp}$ are anti-correlated as expected for a given tSZ background. In the bottom left panel, we show the differential radial profile of Compton-$y$ extracted from the stacked $y$-map (black circles with error bars) and the profile calculated from the best-fitted pressure model (red curve). The ``one-halo" term (blue dash-dotted curve) contributes at a small radius, while the tSZ background (dashed green curve) dominates at a large radius. The excess in the data compared to the tSZ background in the innermost radial bin indicates the detection of the tSZ effect in the CGM of $\rm M_\star = 10^{11.0-11.3} M_\odot$ galaxies. 
We show the results for other mass bins in Appendix\,\ref{sec:crosscheck}.

\subsection{Dependence on stellar mass}\label{sec:YMstar}
\begin{figure*}
    \centering
    \includegraphics[width=0.475\textwidth]{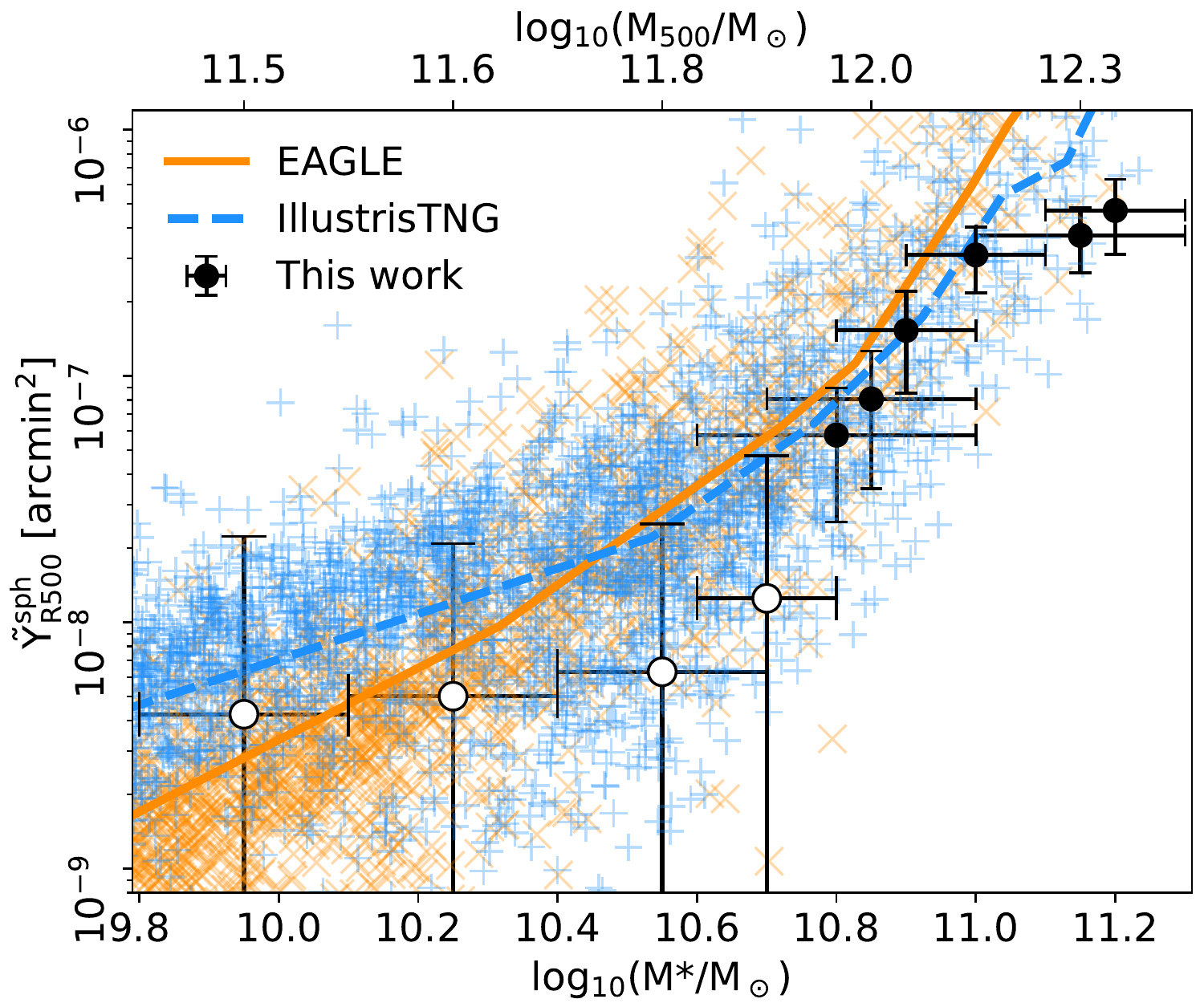}
    \includegraphics[width=0.475\textwidth]{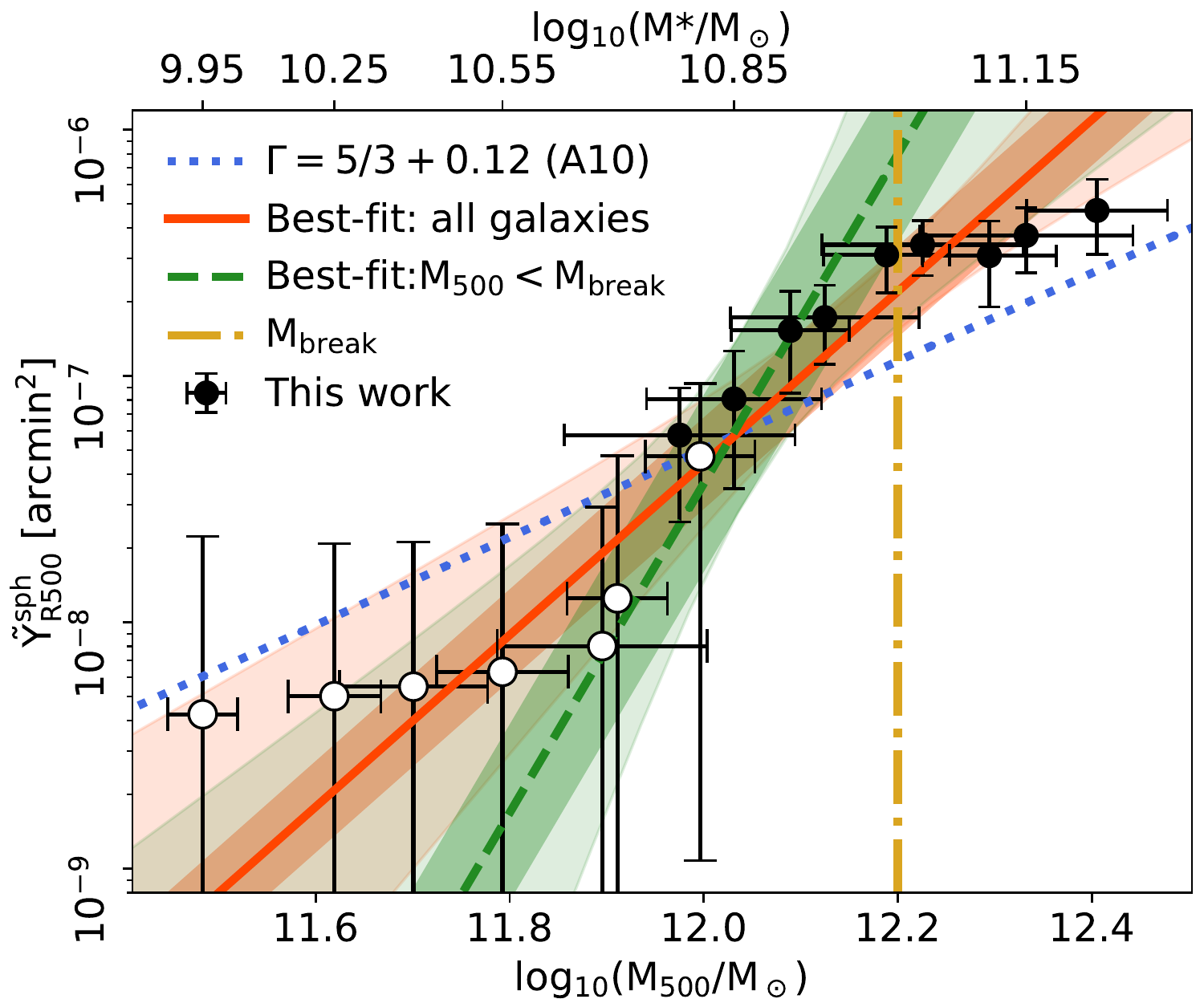}
    \caption{$\rm \tilde Y^{sph}_{R500}$, {a measure of} the thermal energy within R$_{500}$, as a function of the stellar mass, M$_\star$ (left) and halo mass, M$_{500}$ (right). {The upper x-axes have been labeled with corresponding M$_{500}$ (left) and M$_\star$ (right).} The error bars in the y-axes are 1$\sigma$ uncertainties. Our measurements of $\rm \tilde Y^{sph}_{R500}$ constrained at more/less than 90\% confidence are plotted in filled/unfilled black circles. Left: The error bars in the x-axes correspond to the range of M$_\star$ in the mass bin of consideration. We have plotted a subset of the results (Table\,\ref{tab:result}) to help visualize the pattern. The orange `$\times$' and blue `+' symbols are $\rm \tilde Y^{sph}_{R500}$ predicted from cosmological zoom-in simulations EAGLE and IllustrisTNG, respectively \citep{Kim2022}, with solid orange and dashed blue curves showing the median. Right: The error bars in the x-axes correspond to the 1$\sigma$ scatter in M$_{500}$. The dotted blue line is the best-fit result of $\rm \tilde Y^{sph}_{R500} \propto M_{500}^{5/3+\alpha_p}$ for $\rm M_{500} > 10^{14}$\msun~halos \citep[see \S\ref{sec:P(r)}]{Arnaud2010}, extrapolated to lower mass. The solid red and the dashed green curves are the best-fit results for all galaxies and $\rm M_{500} \leqslant M_{break}$ galaxies, respectively. {$\rm M_{break}$ is marked with the vertical dash-dotted yellow line. The \textcolor{black}{dark} and \textcolor{black}{light} shaded areas denote intrinsic and total scatter in the regression, \textcolor{black}{respectively}}. See \S\ref{sec:YMstar} and \S\ref{sec:YMhalo} for details.} 
    \label{fig:YM}
\end{figure*}

In Figure\,\ref{fig:YM} (left panel) we show the best-fit thermal pressure of electrons integrated within a spherical volume as a function of the stellar mass of galaxies. Qualitatively, the thermal energy increases with stellar mass and flattens at $\rm M_\star>10^{11}$\msun.

We compare our measurements with the predictions from cosmological zoom-in simulations EAGLE (4262 galaxies) and IllustrisTNG \citep[5768 galaxies;][hereafter \citetalias{Kim2022}]{Kim2022}. 
While EAGLE and IllustrisTNG have different feedback prescriptions, their predicted median $\rm \tilde Y^{sph}_{R500}$ are consistent with each other. Our measurements for $\rm M_\star<10^{11}$\msun~including the upper limits are consistent with the median of the simulations. Our measurements for $\rm M_\star >10^{11}$\msun~are smaller than the median, but are $\approx$consistent within the scatter (`$\times$' and `+' symbols). 

The simulations in \citetalias{Kim2022} were tuned to reproduce the results of \cite{Planck2013} within $5\times \rm R_{500}$ of $\rm M_\star>10^{11.3}$\msun~locally brightest galaxies (LBGs; galaxies without any brighter neighbor within 1\,Mpc). LBGs have a different luminosity function than the whole population of galaxies. But we do not have any such preference. Therefore, the discrepancy between \citetalias{Kim2022} and our results at $\rm M_\star = 10^{11.0-11.3}$\msun~could be due to sample selection. It should be noted that the galaxies in \citetalias{Kim2022} were at $z=0$, whereas our $\rm M_\star>10^{11}$\msun~galaxy sample is at $z \approx 0.28$ (Table\,\ref{tab:result}). Also, the star-forming galaxies have larger $\rm \tilde Y^{sph}_{R500}$ than quiescent galaxies in \citetalias{Kim2022}. Therefore, the discrepancy could be due to any redshift evolution, and/or our galaxy sample being dominated by quiescent galaxies. We will test these possibilities in future work. 

\begin{figure}
    \centering
    \includegraphics[width=0.475\textwidth]{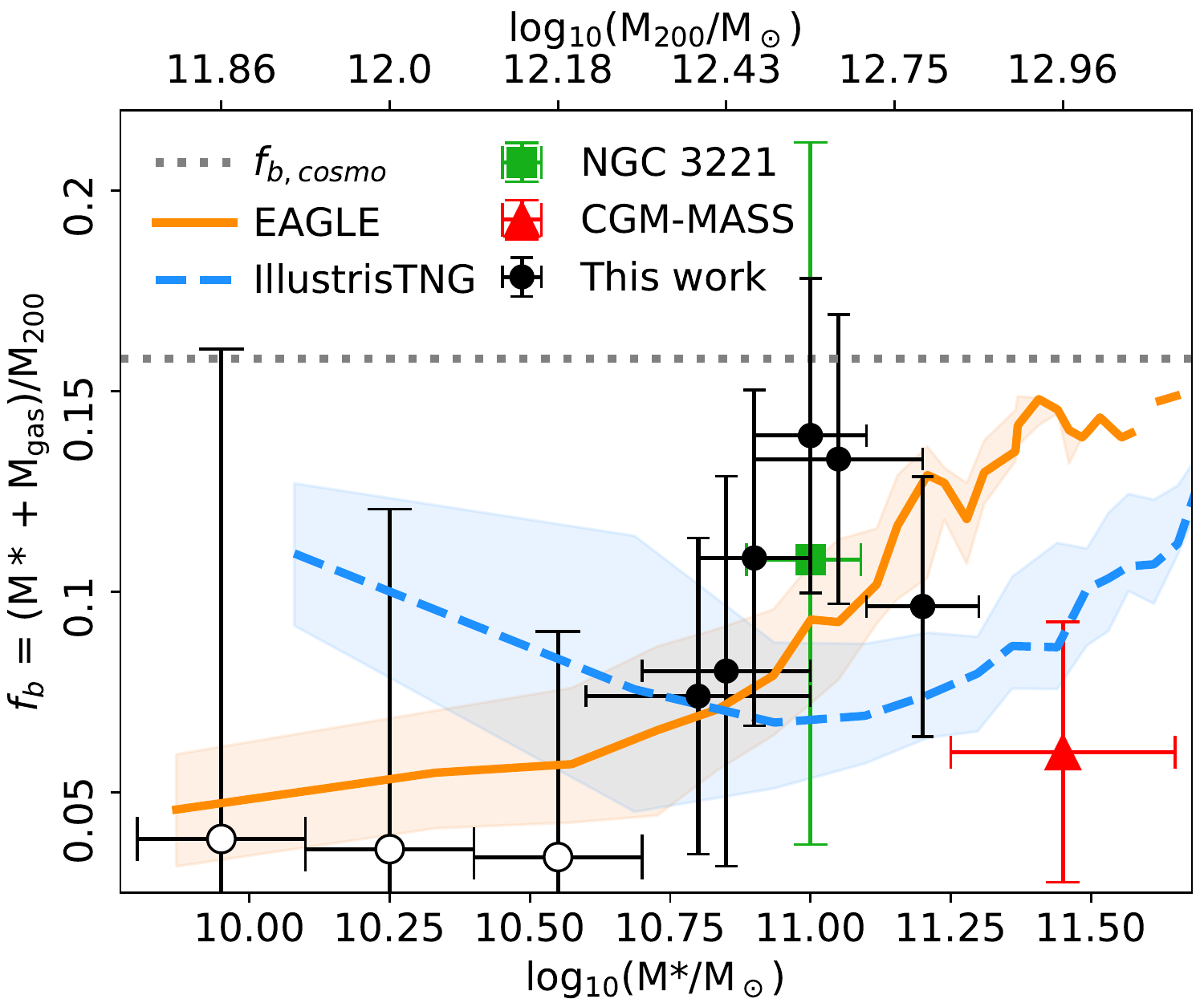}
    \caption{The baryon fraction, $f_b$, as a function of the stellar mass. {The upper x-axis has been labeled with corresponding M$_{200}$.} Our measurements constrained at more (less) than 90\% confidence are plotted in filled (unfilled) black circles. Note that we have plotted a subset of the results (Table\,\ref{tab:result}) to help visualize the pattern. The error bars in the x-axes correspond to the range of M$_\star$ in the mass bin of consideration. The green square and the red triangle are based upon the X-ray emission measurement of NGC\,3221 \citep{Das2019b,Das2020a} and of six galaxies in the CGM-MASS survey \citep{Li2018}, respectively. The solid orange and dashed blue curves are median predictions from EAGLE and IllustrisTNG, with the shaded regions showing 1$\sigma$ scatter \citep{Davies2020}. The horizontal dotted gray line corresponds to the cosmological baryon fraction \citep{Planck2016}.}
    \label{fig:fb}
\end{figure}
\subsection{Self-similarity}\label{sec:YMhalo}
In the right panel of Figure\,\ref{fig:YM} we show the volume integrated thermal pressure of electrons as a function of the virial mass (M$_{500}$) of galaxies. \citetalias{Arnaud2010} introduced a modification to the self-similarity to fit the X-ray observations of $\rm M_{500} > 10^{14}$\msun~halos: $\rm \tilde Y^{sph}_{R500} \propto M^{5/3+\alpha_p}_{500}$ with $\alpha_p=0.12$ (see \S\ref{sec:P(r)}). We extrapolate this relation to lower mass and compare it (dotted blue line) with our results (black circles). 

{With the expectation that $\rm M_{500}$ and $\rm \tilde Y^{sph}_{R500}$ follows a power-law relation: $\rm \tilde Y^{sph}_{R500} = Y_{norm}\times M_{500}^\Gamma$, we fit a straight line to ln($\rm M_{500}$) and ln($\rm \tilde Y^{sph}_{R500}$) using the python package \href{https://github.com/jmeyers314/linmix}{linmix} based on the hierarchical Bayesian model of \cite{Kelly2007}. $\rm M_{500}$ and $\rm \tilde Y^{sph}_{R500}$ are assumed to be drawn from a 2-d log-normal distribution $\mathcal{N}_2(\mu,\Sigma)$ with the mean $\mu = (\xi,\eta)$ which are the unobserved \textit{true} values of ln($\rm M_{500}$) and ln($\rm \tilde Y^{sph}_{R500}$), and the covariance matrix $\Sigma$ containing the measured errors of ln($\rm M_{500}$) and ln($\rm \tilde Y^{sph}_{R500}$). $\xi$ and $\eta$ are connected through $\rm P(\eta|\xi) = \mathcal{N}(\alpha + \beta \xi, \sigma^2)$, where the regression parameters $\alpha, \beta$, and $\rm \sigma^2$ are the intercept, slope, and Gaussian intrinsic scatter of $\eta$ around the regression line, respectively. The priors on $\alpha, \beta$, and $\rm \sigma^2$ are uniform.} 

{The best-fit value of $\beta$, i.e., the \textit{true} value of $\Gamma$ is $3.27_{-0.69}^{+0.91} \equiv 5/3 + 1.60_{-0.69}^{+0.91}$ (68\% confidence interval; solid red line in Figure\,\ref{fig:YM}, right panel). It implies that our sample galaxies deviate from the self-similar relation by $\approx 2.3\sigma$.} The deviation from self-similarity is stronger than the $\alpha_p$ of \citetalias{Arnaud2010}, i.e., {$1.60_{-0.69}^{+0.91} > 0.12$}. The largest possible value of the second-order effect, $\alpha_p'(r/\rm R_{500})$ that we have ignored in our analysis (see \S\ref{sec:P(r)}), was 0.10 in \citetalias{Arnaud2010}. The equivalent $\alpha_p'$ in our case is {$1.48_{-0.69}^{+0.91}$}. Therefore, our samples not only deviate from self-similarity, but it also deviates more than their massive counterparts in \citetalias{Arnaud2010}. Our finding of steeper $\rm \tilde Y^{sph}_{R500}-M_{500}$ relation is consistent with the best-fit model for $\gtrsim$25,000 simulated halos in the mass range of $10^{12}\rm  M_\odot < M_{500} <5\times 10^{13} M_\odot$ in IllustrisTNG \citep{Pop2022}. 


It should be noted that $\rm \tilde Y^{sph}_{R500}$ is practically constant at $\rm M_{500} > 10^{12.2}$\msun~(see the values of $\rm \tilde Y^{sph}_{R500}$ in the last two rows of $\Delta \rm M_\star=$ 0.3\,dex and 0.2\,dex in Table\,\ref{tab:result}). 
To test if this flattening affects the best-fitted $\rm \tilde Y^{sph}_{R500}-M_{500}$ relation, we fit the $\rm M_{500}\leqslant 10^{12.2}$\msun~halos, and obtain {$\Gamma=5.49_{-1.95}^{+4.57}$} (dashed green line in Figure\,\ref{fig:YM}, bottom panel), {which is larger than the slope for self-similarity, \citetalias{Arnaud2010}, and the slope we obtained for all our galaxies: $3.27_{-0.69}^{+0.91}$}. 
That means the overall $\rm \tilde Y^{sph}_{R500}-M_{500}$ relation of our sample might follow a broken power-law with $\rm M_{break} \approx 10^{12.2}$\msun, where the relation is weaker than the self-similarity above $\rm M_{break}$ but is stronger than the self-similarity below $\rm M_{break}$. It has not been predicted or observed before.  

Our result {of the global fit and the fit} for M$_{500}\leqslant 10^{12.2}$\msun~galaxies {are qualitatively similar: a deficit of the tSZ effect at a given stellar (and halo) mass compared to the self-similarity.} It indicates two possibilities regarding the galactic outflows - 1) too strong feedback: the galactic outflow spewed the ionized gas in the CGM outside $\rm R_{500}$, or 2) too weak feedback: the feedback was not strong enough to prevent cooling and eventual precipitation of the halo gas to the galactic disk. Distinguishing between these options and identifying the actual physical scenario is beyond the scope of our paper. It would be part of our future endeavors. 

The interpretation of our result for $\rm M_{500}> M_{break}$ galaxies depends on which side of the mass is considered the point of reference. If we consider $\rm M_{break}$ as the reference, the flattening of $\rm \tilde Y^{sph}_{R500}$ implies a deficit of $\rm \tilde Y^{sph}_{R500}$ in the CGM of {the most massive galaxies in our sample} compared to the self-similar relation \citepalias[and][]{Arnaud2010}. In that case, the interpretation would be similar to that  discussed in the previous paragraph. If we consider the most massive galaxies in our sample as a reference, the flattening of $\rm \tilde Y^{sph}_{R500}$ implies an excess of $\rm \tilde Y^{sph}_{R500}$ in the CGM of galaxies above $\rm M_{break}$. It indicates a non-negligible contribution of galactic feedback within $\rm R_{500}$ that keeps the CGM ionized but does not throw the gas outside the halo. However, there are only four data points at $\rm M_{500}> M_{break}$, and they span only 0.2\,dex of virial mass. Also, only 5\% of our galaxy samples are more massive than M$_{\rm break}$.
Therefore, we do not comment further on $\rm M_{500}> M_{break}$ galaxies to avoid over-interpretation.  

\subsection{Baryon budget}
In Figure\,\ref{fig:fb}, we show the baryon fraction, $f_b$ as a function of the stellar mass. We compare our measurements with the predictions from cosmological zoom-in simulations EAGLE (3544 galaxies) and IllustrisTNG \citep[5344 galaxies;][]{Davies2020}, and with previous measurements based upon X-ray observations. Finally, we discuss the implications of the variation of the baryon fraction with stellar mass in our sample.  

The X-ray emission-based measurement of NGC\,3221, the only L$^\star$ galaxy other than Milky\,Way with a detected CGM \citep[green square;][]{Das2019b,Das2020a} is consistent with our results. The measurements for more massive super-L$^\star$ galaxies in X-ray emission \citep[red triangle;][]{Li2018} are also in line with our findings. Instead of assuming an SHMR, the virial masses of these individual galaxies have been directly measured from the velocity dispersion of the neutral gas in these galaxies. The measurement uncertainty in the virial mass is the largest source of uncertainty in $f_b$ of these individual galaxies. The median of the $f_b$ distribution predicted in simulations (solid orange and dashed blue curves, Fig.\,\ref{fig:fb}) are not consistent with our measurements, but we cannot rule these predictions out modulo the scatter around the median and the uncertainties in our measurements. 

In previous studies on galaxy groups and clusters based on the tSZ effect and X-ray emission, $f_b$ has been found to decrease with decreasing halo mass \citep[e.g.,][]{Lim2018}. However, we find that the trend does not continue in low-mass halos. Starting from the largest stellar mass in Fig.\,\ref{fig:fb}, $f_b$ increases with decreasing stellar mass, reaches its peak and becomes consistent with the cosmological $f_b$ at M$_\star = 10^{10.9-11.2}$\msun, and falls off at lower stellar mass. It is arguably the first time that the baryon fraction has been found to show such a non-monotonic behavior with stellar (and virial) mass in observational data. This ``inversion", i.e., increasing $f_b$ with decreasing stellar mass has been predicted in IllustrisTNG simulations \citep[blue dashed curve, Fig.\,\ref{fig:fb}]{Davies2020}, although the ``inversion" in the simulation occurs at a different stellar mass than our measurements. 

It should be noted that we have assumed the gas to be at the virial temperature. The X-ray emission measurements of $\rm M_\star>10^{10.9}$\msun~galaxies \citep{Das2020a,Li2018} are consistent with that, but it might not be true for lower mass galaxies. While the exact mass of sub-virial phases in the CGM probed in UV absorption lines are extremely sensitive to the conditions in ionization modeling, the prevalence of those lines indicates that the volume-average temperature of these halos might not be as high as their corresponding virial temperatures. A lower temperature would increase the mass of the CGM calculated from the tSZ effect, thereby increasing $f_b$ in $\rm M_\star<10^{10.9}$\msun~ halos.  

In galaxy groups and clusters, as AGN feedback is the dominant mechanism of outflows, the decreasing $f_b$ with decreasing halo mass \citep[e.g.,][]{Lim2018} indicates a higher strength of the AGN feedback relative to gravitation in low-mass groups. The halo mass of our sample is where the stellar feedback could become more influential than the AGN feedback \citep[e.g.,][]{Harrison2017}. Therefore, the ``inversion" in $f_b$ from $\rm M_\star = 10^{11.6}$\msun~to $10^{10.9}$\msun~galaxies in Fig.\,\ref{fig:fb} might be the scenario where the stellar feedback enriches the CGM and prevents overcooling but unlike AGN feedback, it is not strong enough to throw the gas outside $\rm R_{200}$. In $\rm M_\star<10^{10.9}$\msun~galaxies, given the ubiquitous presence of cool CGM, we cannot conclusively comment on the baryon sufficiency and the interplay between feedback and cooling. It would require simulations focused on the large-scale impact of stellar feedback, and observations in X-ray and kinetic SZ (kSZ) effect\footnote{A measure of $n_e$ from the CMB power spectrum, without any temperature dependence} {\citep[e.g., as done in][]{Schaan2021,Amodeo2021}} to constrain these speculations.

\begin{deluxetable*}{cccccccc}
\tablecaption{Properties of our galaxy sample and the corresponding tSZ effect. The number of galaxies, the mean redshift, the mean virial mass, and the mean virial radius with 1$\sigma$ scatters in each stellar mass bin are quoted in the second--fifth column. R$_{\rm min}$ denotes the smallest aperture in the unit of R$_{200}$. The error bars in $\rm \tilde Y$ and $\rm M_{gas}$ are 1$\sigma$ and the upper limits (for values constrained at $<$90\% confidence) are 3$\sigma$. {$\rm M_{gas}$ is calculated assuming the CGM to be at virial temperature.}  \label{tab:result}}
\tablewidth{0pt}
\tablehead{
\colhead{log(M$_\star$)} & \colhead{$\#$galaxy} & \colhead{$z$} & \colhead{log(M$_{200}$)} & \colhead{R$_{200}$} & \colhead{R$_{\rm min}$} & \colhead{$\rm \tilde Y^{sph}_{R500}$} & \colhead{$\rm M_{gas,200}$}   
\\
$\rm [M_\odot]$ & & & [\msun] & [kpc] & [R$_{200}$] &  [$\rm \times 10^{-8} arcmin^{2}$] & [$\rm \times 10^{11} M_\odot$] }
\decimalcolnumbers
\startdata
\multicolumn{8}{c}{$\rm \Delta M_\star = 0.3$\,dex}\\
\hline
9.8-10.1 & 93907 & 0.158$\pm$0.063 & 11.86$\pm$0.04 & 185$\pm$5 & 1.09 & $<$5.86 & $<$2.85 \\
10.1-10.4 & 175255 & 0.172$\pm$0.061 & 12.00$\pm$0.05 & 206$\pm$7 & 1.05 & $<$5.26 & $<$2.74 \\
10.4-10.7 & 187761 & 0.192$\pm$0.060 & 12.18$\pm$0.07 & 237$\pm$12 & 0.99 & $<$6.28 & $<$2.74 \\
10.7-11.0 & 124892 & 0.233$\pm$0.056 & 12.43$\pm$0.09 & 287$\pm$19 & 0.94 & 8.1$\pm$4.6 & 1.51$\pm$1.31\\
10.8-11.1 & 101155 & 0.249$\pm$0.053 & 12.53$\pm$0.10 & 310$\pm$21 & 0.92 & 17.3$\pm$6.1 & 2.79$\pm$1.34\\
10.9-11.2 & 79367 & 0.265$\pm$0.050 & 12.64$\pm$0.10 & 336$\pm$25 & 0.88 & 34.2$\pm$8.6 & 4.73$\pm$1.55\\
11.0-11.3 & 60108 & 0.281$\pm$0.047 & 12.75$\pm$0.11 & 366$\pm$29 & 0.84 & 37.3$\pm$11.0 & 4.37$\pm$1.71\\
\hline
\multicolumn{8}{c}{$\rm \Delta M_\star = 0.2$\,dex}\\
\hline
10.6-10.8 & 106715 & 0.211$\pm$0.059 & 12.31$\pm$0.05 & 261$\pm$10 & 0.97 & $<$11.75 & $<$3.61 \\
10.7-10.9 & 91250 & 0.226$\pm$0.057 & 12.40$\pm$0.06 & 279$\pm$11 & 0.95 & $<$18.58 & $<$4.77 \\
10.8-11.0 & 75326 & 0.242$\pm$0.054 & 12.50$\pm$0.06 & 301$\pm$13 & 0.93 & 15.3$\pm$6.8 & 2.61$\pm$1.30 \\
10.9-11.1 & 59469 & 0.258$\pm$0.050 & 12.60$\pm$0.07 & 326$\pm$15 & 0.90 & 31.1$\pm$9.3 & 4.55$\pm$1.55\\
11.0-11.2 & 45726 & 0.274$\pm$0.047 & 12.71$\pm$0.07 & 355$\pm$18 & 0.86 & 30.8$\pm$11.7 & 3.84$\pm$1.96\\
11.1-11.3 & 34280 & 0.290$\pm$0.044 & 12.83$\pm$0.07 & 389$\pm$21 & 0.82 & 47.1$\pm$15.9 & 4.93$\pm$2.16 \\
\hline
\multicolumn{8}{c}{$\rm \Delta M_\star = 0.4$\,dex}\\
\hline
10.5-10.9 & 212095 & 0.210$\pm$0.060 & 12.29$\pm$0.11 & 258$\pm$20 & 0.97 & $<$7.21 & $<$2.58 \\
10.6-11.0 & 182042 & 0.225$\pm$0.058 & 12.38$\pm$0.12 & 275$\pm$23 & 0.95 & 5.8$\pm$3.2 & 1.17$\pm$0.92\\
\enddata 
\end{deluxetable*}

\section{Conclusion, summary and future directions}\label{sec:conclusion}
In this paper, we have looked for the tSZ effect in the CGM of WISE$\times$SuperCosmos galaxies with $\rm M_\star = 10^{9.8-11.3}M_\odot$ using Compton-$y$ maps from \textit{ACT}${+}$\planck data. We have studied the effect of dust, galaxy clusters, and radio sources on the stacked $y$-maps and corrected them. We have considered three methods of stacking, and two methods of aperture photometry, and zeroed in on the most complete and informative one. We have taken into account the beam smearing, the ``two-halo" term, and any zero-point offset. Below, we summarize our science results: 
\begin{itemize}
    \item We have been able to constrain the tSZ effect in the CGM of $\rm M_\star = 10^{10.6-11.3}M_\odot$ galaxies by modeling the thermal pressure with a generalized NFW profile. There was also indirect evidence of the tSZ effect in the $\rm M_\star < 10^{10.6}M_\odot$ galaxies before correcting for CIB (Fig.\,\ref{fig:CIBeffect}, top row), but we cannot quantify the signal after correcting for CIB. 
    \item  The dependence of $\rm \tilde Y^{sph}_{R500}$ ({a measure of the} thermal energy within $\rm R_{500}$) on the virial mass, $\rm M_{500}$ is {$>2\sigma$} stronger than that in self-similarity and the deviation from the same observed in higher mass halos in literature. 
    \item Galaxies in the 10$^{10.9-11.2}$\msun~range of stellar mass could be baryon sufficient. The baryon fraction of the galaxies exhibits a non-monotonic trend with stellar mass, implying a complex interplay of the gravitational potential and galactic outflow. The non-monotonic trend has not been predicted before, indicating an insufficient treatment of the large-scale effect of galactic feedback in low-mass galaxies in simulations.
\end{itemize}

Upcoming telescopes, e.g., Simons Observatory and CMB-S4 will improve the quality of SZ signal with increased sky coverage, frequency coverage, sensitivity, and angular resolution. Ongoing and future galaxy surveys by, e.g., Dark energy spectroscopic instrument (DESI) and Vera C. Rubin Observatory will increase the sample size of galaxies. This will allow us to measure the tSZ effect as well as the kSZ effect in low-mass galaxies and constrain the density and pressure profiles independently, leading to a better understanding of the effect of galactic feedback on the CGM.  

\section*{acknowledgments}
{We thank the anonymous referee for constructive feedback.} We thank Dr.\,Vinu Vikram, Dr.\,Junhan Kim, and Dr. Jonathan Davies for sharing the data of their \cite{Vikram2017}, \cite{Kim2022} and \cite{Davies2020} papers, respectively. S.D. acknowledges support from the Presidential Graduate Fellowship of The Ohio State University and the KIPAC Fellowship of Kavli Institute for Particle Astrophysics and Cosmology, Stanford University. Y.C. acknowledges the support of the National Science and Technology Council of Taiwan through grant NSTC 111-2112-M-001-090-MY3. S.M. is grateful for support from NASA through the ADAP grant 80NSSC22K1121. This research has made use of NASA's Astrophysics Data System Bibliographic Services. 

\facilities{WISE, SuperCOSMOS, \planck, \act, \jvla}
\software{\texttt{AstroPy} \citep{Astropy2022}, \texttt{corner} \citep{Foreman-Mackey2016}, \texttt{emcee} \citep{Foreman-Mackey2013},  \texttt{Halotools} \citep{Hearin2016}, \texttt{Jupyter} \citep{jupyter2016}, \texttt{Matplotlib} \citep{Hunter2007}, \texttt{NumPy} \citep{numpy2020}, \texttt{pandas} \citep{pandas2020},  \texttt{Pixell} \citep{Naess2021}, \texttt{Python math} \citep{mathpy2020}, \texttt{SciPy} \citep{scipy2022}}

\appendix 
\counterwithin{figure}{section}

\section{Proof of concept and techniques}\label{sec:proofofconcept}

\begin{figure*}
    \centering
    \includegraphics[trim=0 50 0 0, clip,width=\textwidth]{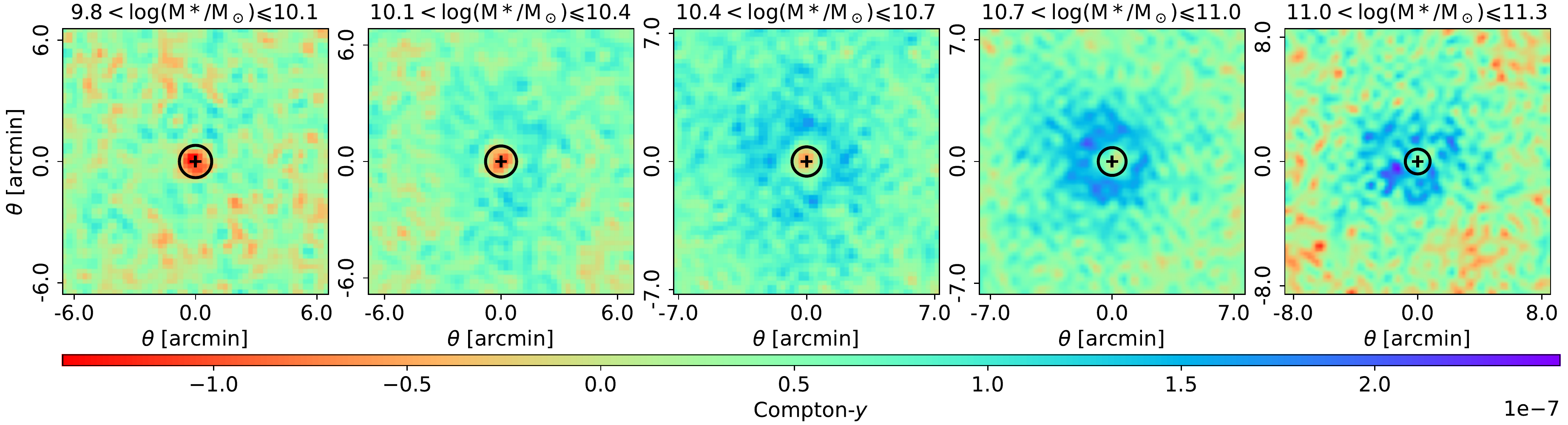}
    \includegraphics[width=\textwidth]{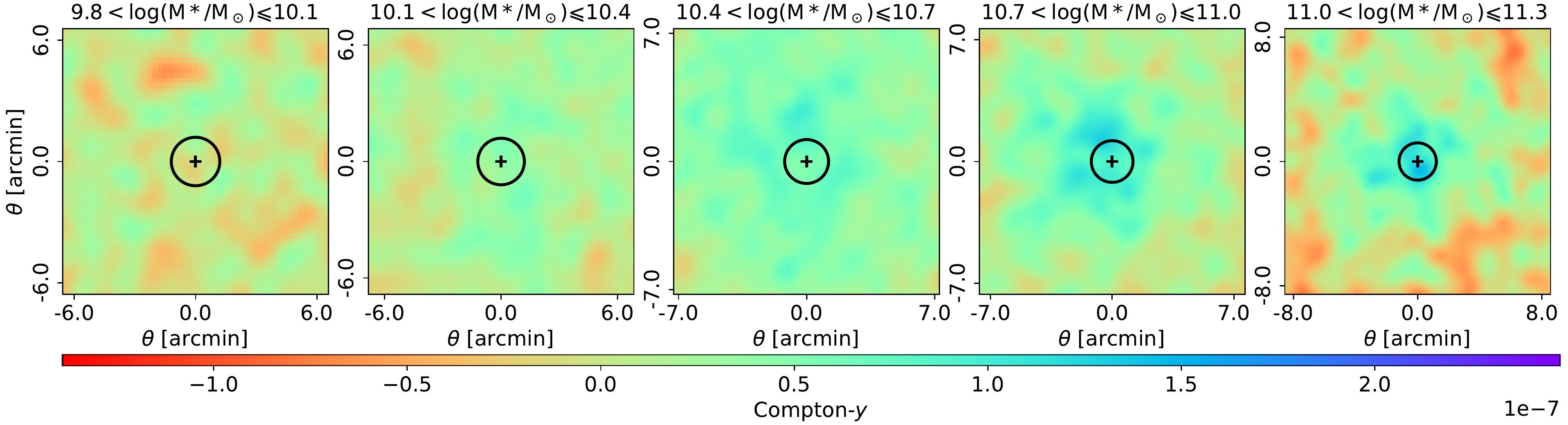}
    \includegraphics[width=\textwidth]{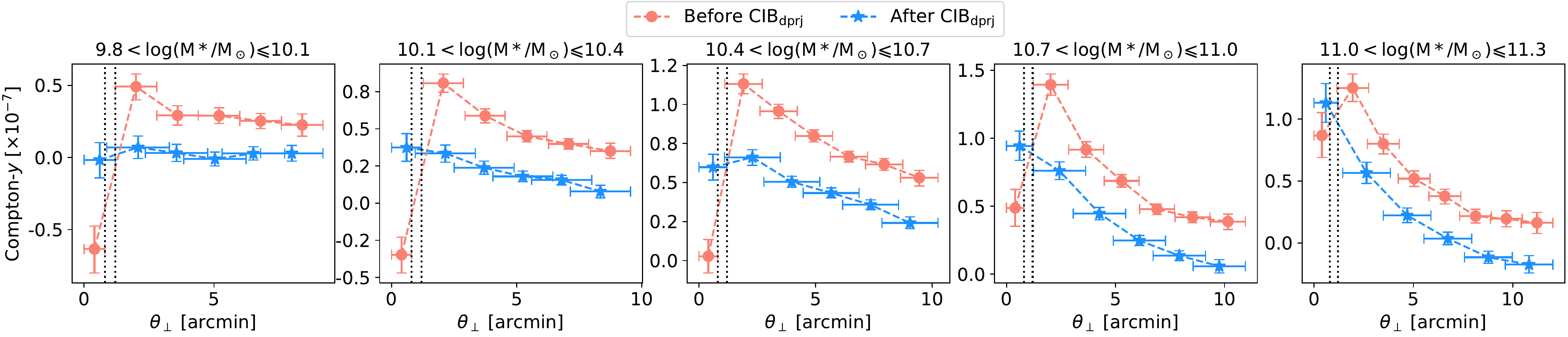}
    \caption{\textit{Top and middle row:} Stacked Compton-$y$ maps before and after the CIB deprojection (see \S\ref{sec:ymap}). The range of M$_\star$ in each mass bin is mentioned in the title of each map. The center of stacked galaxies (i.e., the central pixel) is marked with a plus (`+') sign. The Gaussian beam with an FWHM of $1.\!'6$ and $2.\!'4$ before and after the CIB deprojection, respectively, is shown with a circle at the center of each map. The colorbar is shown at the bottom. \textit{Bottom row:} Differential radial profiles of the stacked Compton-$y$ maps with 1$\sigma$ error before and after the CIB deprojection. The width of each annulus is the same as the FWHM of the Gaussian beam, shown with the error bars along the x-axis.}
    \label{fig:CIBeffect}
\end{figure*}

\subsection{Effect of thermal dust}

In this section, we discuss how the cosmic infrared background (CIB) and the Galactic dust affect the stacking results. 

\begin{figure*}
    \centering
    \includegraphics[trim=0 50 0 0, clip,width=\textwidth]{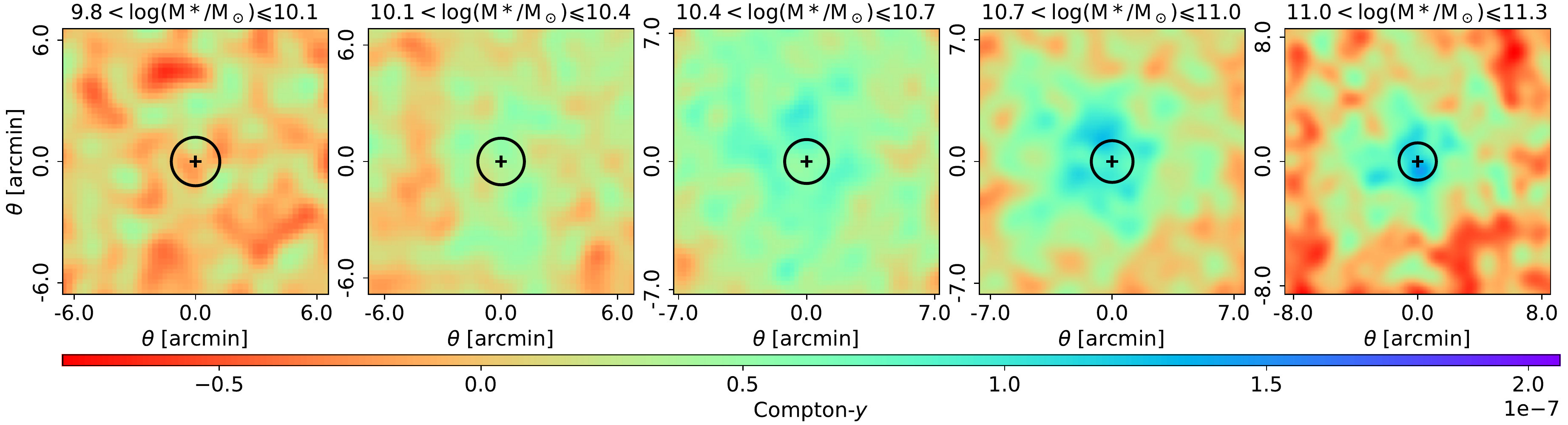}
    \includegraphics[width=\textwidth]{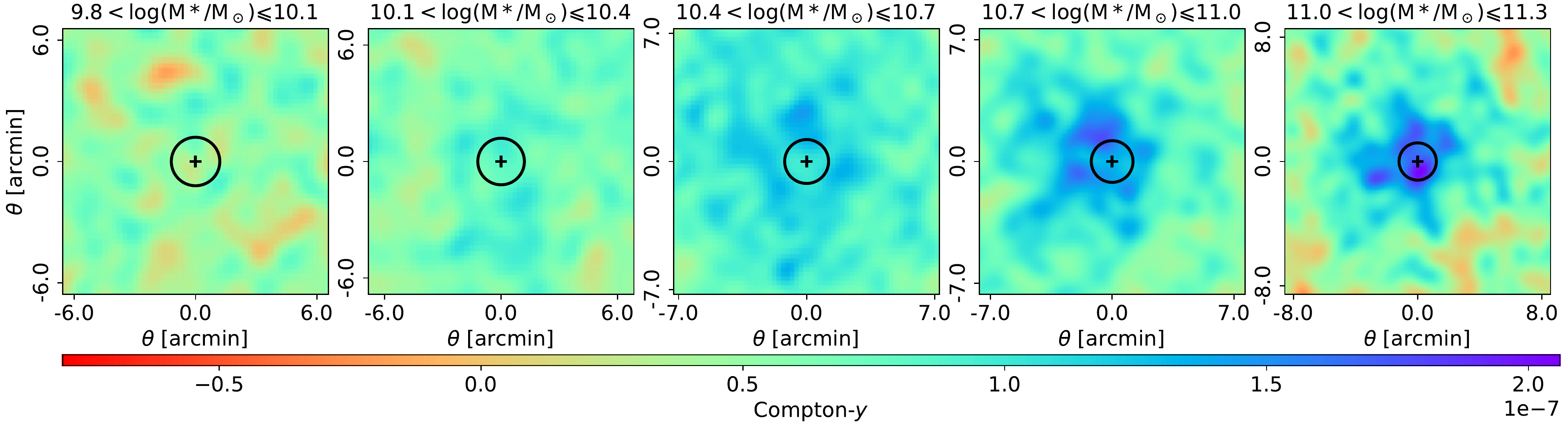}
    \includegraphics[width=\textwidth]{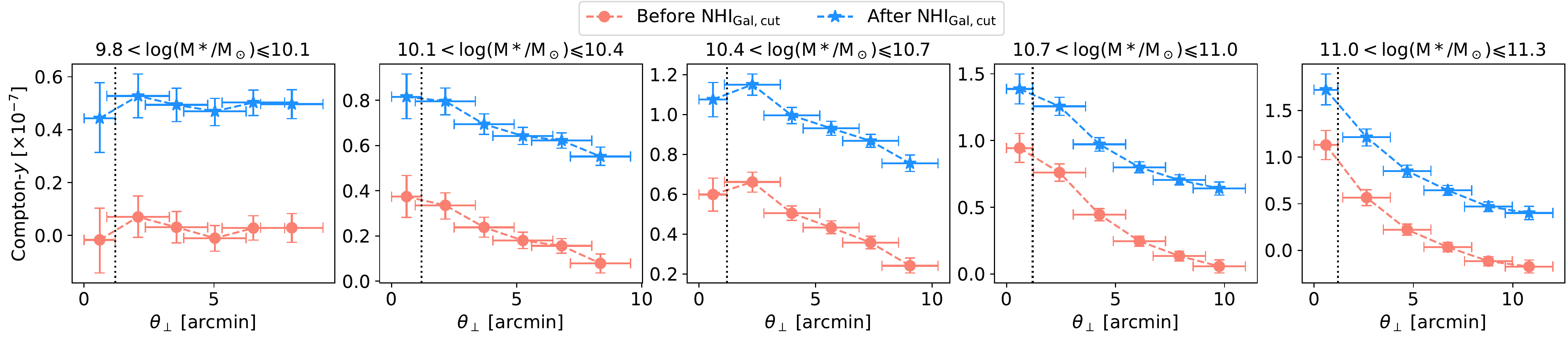}
    \caption{The effect of Galactic dust correction by applying the cut in N(\hin)$\rm_{Gal}$ (see \S\ref{sec:galdust}) on stacking. \textit{Top and middle row:} Stacked CIB corrected Compton-$y$ maps before and after the cut. The range of M$_\star$ in each mass bin is mentioned in the title of each map. The center of stacked galaxies (i.e., the central pixel) is marked with a plus (`+') sign. The Gaussian beam (FWHM of $2.\!'4$) is shown with a circle at the center of each map. The colorbar is shown at the bottom. \textit{Bottom row:} Differential radial profiles of the stacked Compton-$y$ maps with 1$\sigma$ error before and after the cut. The width of each annulus is the same as the FWHM of the Gaussian beam, shown with the error bars along the x-axis.}
    \label{fig:NHIeffect}
\end{figure*}

\subsubsection{CIB}
We show the stacked $y$-maps and the corresponding differential radial profiles before and after the CIB deprojection in Figure\,\ref{fig:CIBeffect}. The $y$-value in the central region of the maps before CIB deprojection  is suppressed (top row). It is also reflected by the radial profiles (red circles; bottom row), where the $y$-value in the innermost radial bin is smaller than the $y$-values in most (if not all) of the outer radial bins. Because the thermal dust of the galaxies has a different spectral energy distribution (SED) than that of the tSZ effect, the dust emission could be manifested as a dearth of Compton-$y$. This effect becomes stronger as the stellar mass decreases, to the extent of the innermost radial bin falling to negative Compton-$y$, which is unphysical. It attests to the fact that less massive galaxies are generally dustier \citep[e.g.,][]{Cortese2012}. Also, the $y$-values in the outer radial bins are enhanced due to the thermal dust emission by the CIB that mimics the tSZ signal. Both of these effects are corrected in the CIB deprojected maps (middle row) and the corresponding radial profiles (blue stars; bottom row). In the following sections, we consider the CIB-corrected results only.

\subsubsection{Galactic dust}
We show the stacked, CIB deprojected $y$-maps and the corresponding differential radial profiles before and after correcting for Galactic dust in Figure\,\ref{fig:NHIeffect}. The Compton-$y$ maps after applying the cut in N(\hin)$_{\rm Gal}$ are systematically brighter than those before applying the cut (middle vs. top row). The correction factor is $\approx$constant at small angular separation ($\theta <5'$), and it increases at larger angular separation. It leads to a flatter radial profile of Compton-$y$, effectively visible in the $\rm 10.7 < log (M_\star/M_\odot) \leqslant 11.3$ bins (red circles vs. blue stars; bottom row). It also shows that the negative values of Compton-$y$ at large angular separation ($\theta >8'$) in the most massive ($\rm 11.0 < log (M_\star/M_\odot) \leqslant 11.3$) bin is due to the residual effect of Galactic dust even after the CIB correction.  In the following sections, we consider the results corrected for CIB as well as the Galactic dust.

\subsection{Effect of masking}
In this section, we discuss how the galaxy clusters and the compact radio sources affect the stacking results. 

\subsubsection{Galaxy clusters}
\begin{figure*}
    \centering
    \includegraphics[trim=0 50 0 0, clip,width=\textwidth]{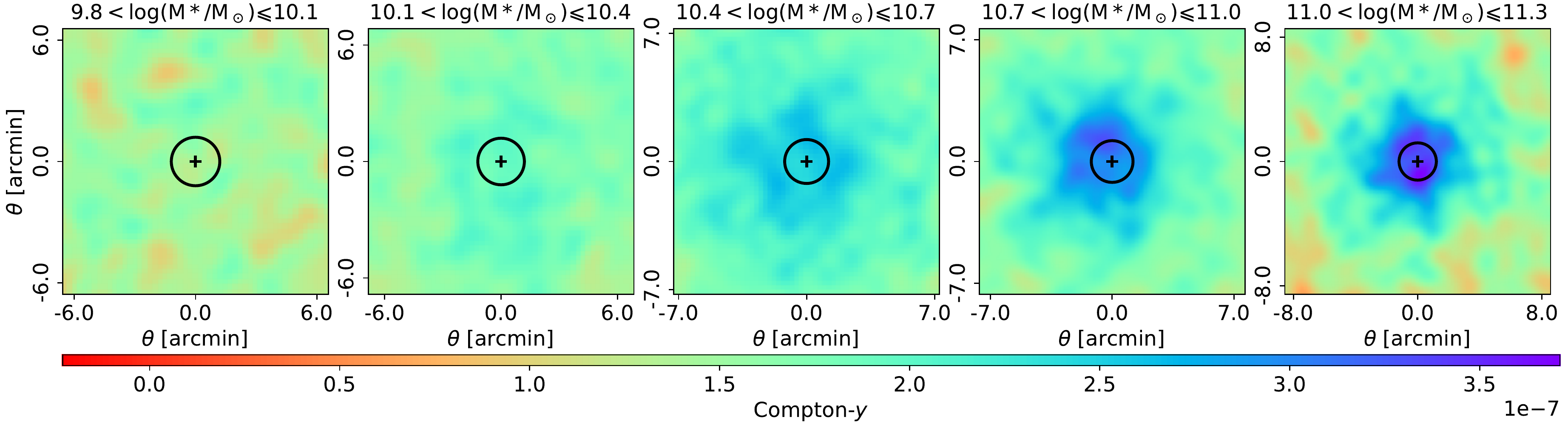}
    \includegraphics[width=\textwidth]{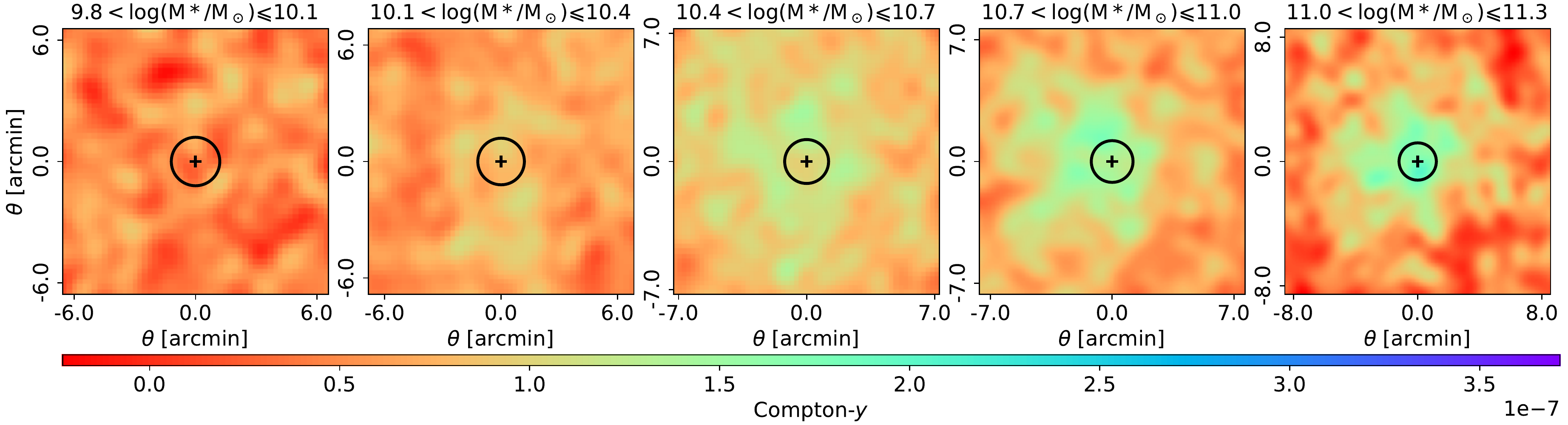}
    \includegraphics[width=\textwidth]{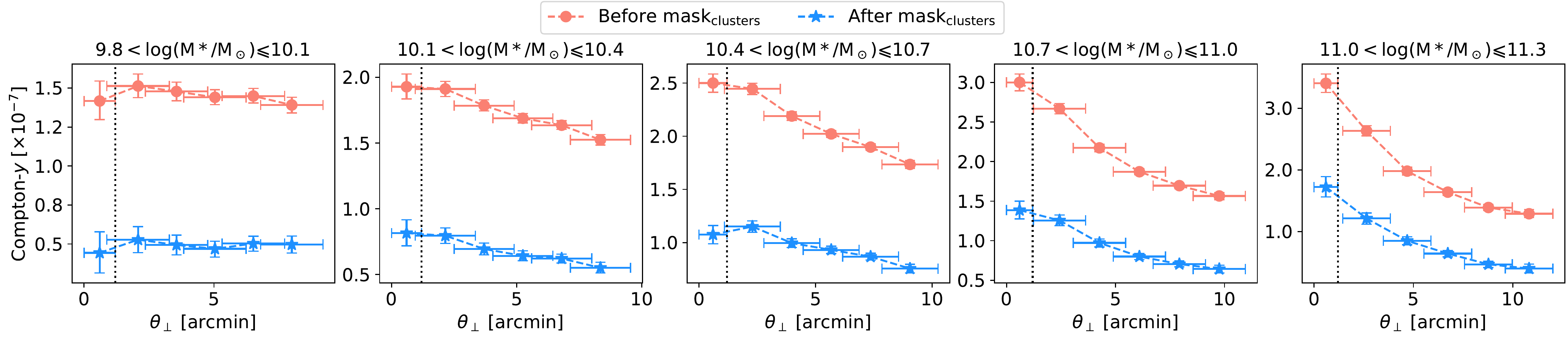}
    \caption{The effect of masking the galaxy clusters (see \S\ref{sec:clustermask}) on stacking. \textit{Top and middle row:} Stacked CIB and Galactic dust- corrected Compton-$y$ maps before and after masking. The range of M$_\star$ in each mass bin is mentioned in the title of each map. The center of stacked galaxies (i.e., the central pixel) is marked with a plus (`+') sign. The Gaussian beam (FWHM of $2.\!'4$) is shown with a circle at the center of each map. The colorbar is shown at the bottom. \textit{Bottom row:} Differential radial profiles of the stacked Compton-$y$ maps with 1$\sigma$ error before and after masking. The width of each annulus is the same as the FWHM of the Gaussian beam, shown with the error bars along the x-axis.}
    \label{fig:clustereffect}
\end{figure*}
In Figure\,\ref{fig:clustereffect}, we show the stacked $y$-maps and the corresponding differential radial profiles before and after excluding the galaxy clusters. The Compton-$y$ values are significantly larger in the presence of galaxy clusters, reflected by the stark differences between the stacked $y$-maps (top and middle row). This difference could be due to the tSZ emission from the ICM of the galaxy clusters. However, the clusters not only enhance the overall $y$-values, but they also steepen the radial profiles (red circles; bottom row), which would effectively lead to an overestimation of the tSZ signal. This contamination is corrected for in the masked profiles (blue stars; bottom row). In the following section, we consider the results masked for the galaxy clusters. 



\subsubsection{Radio sources}
\begin{figure*}
    \centering
    \includegraphics[trim=0 50 0 0, clip,width=\textwidth]{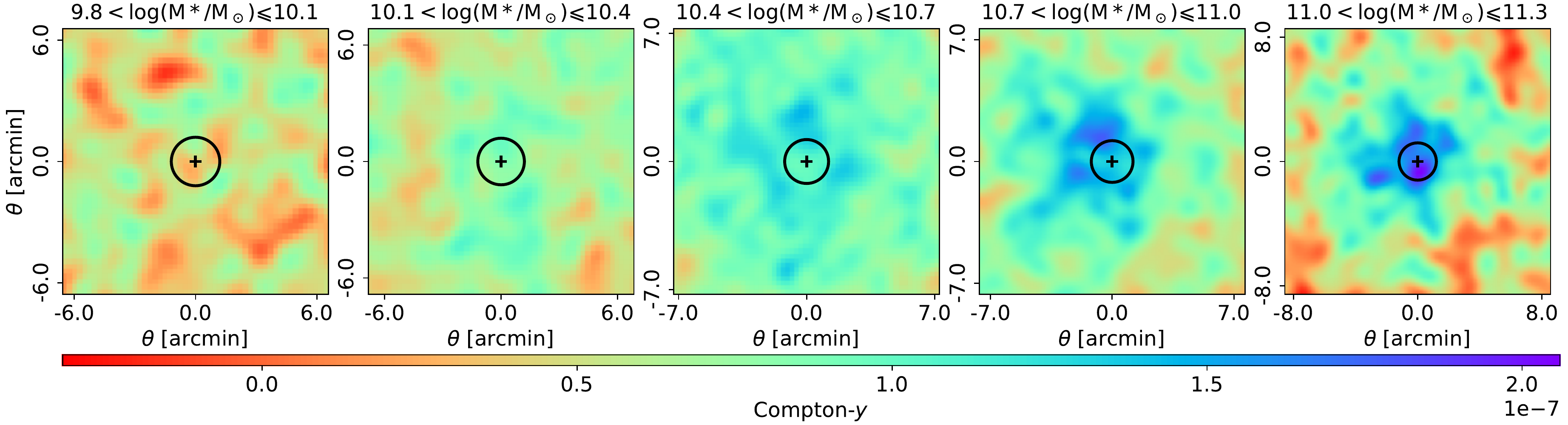}
    \includegraphics[width=\textwidth]{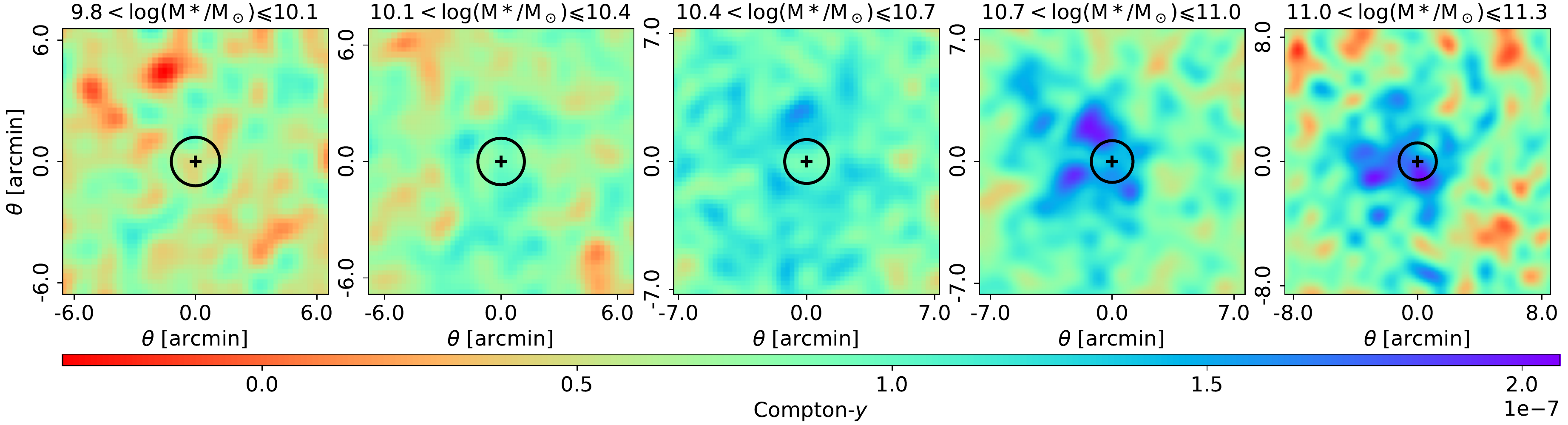}
    \includegraphics[width=\textwidth]{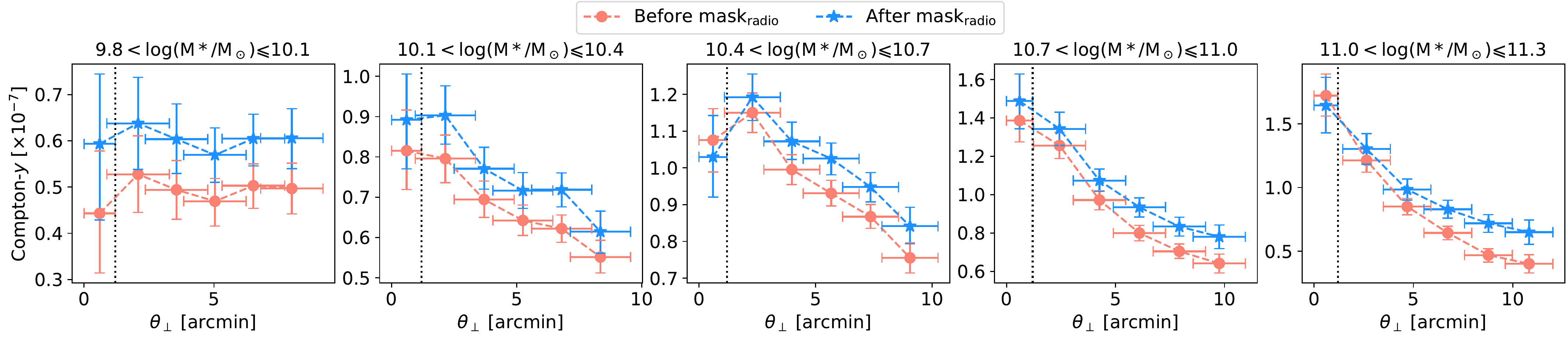}
    \caption{The effect of masking the radio sources (see \S\ref{sec:radiomask}) on stacking. \textit{Top and middle row:} Stacked CIB and Galactic dust-corrected, galaxy cluster-masked Compton-$y$ maps before and after masking. The range of M$_\star$ in each mass bin is mentioned in the title of each map. The center of stacked galaxies (i.e., the central pixel) is marked with a plus (`+') sign. The Gaussian beam (FWHM of $2.\!'4$) is shown with a circle at the center of each map. The colorbar is shown at the bottom. \textit{Bottom row:} Differential radial profiles of the stacked Compton-$y$ maps with 1$\sigma$ error before and after masking. The width of each annulus is the same as the FWHM of the Gaussian beam, shown with the error bars along the x-axis.}
    \label{fig:radioeffect}
\end{figure*}
In Figure\,\ref{fig:radioeffect}, we show the stacked $y$-maps and the corresponding differential radial profiles before and after removing the radio sources from the WISE$\times$SCOSPZ galaxy sample as well as from the $y$-map. Overall, the stacked $y$-maps are brighter after removing the radio sources (top and middle row). Same as the thermal dust, the compact radio sources, which include both stars and galaxies, have different SED than that of the tSZ effect. Thus, the emission from these sources could be manifested as a deficiency of Compton-$y$.
However, this effect is not uniform in all stellar mass bins and at all angular separations. In $\rm log(M_\star/M_\odot) \leqslant {10.4}$ galaxies, the $y$-values are systematically higher after masking the radio sources (red circles vs. blue stars; bottom row). Thus, it increases the zero-point offset but does not affect the tSZ signal as such. 
In $\rm log(M_\star/M_\odot) > {10.4}$ galaxies, $y$-values at the innermost radial bins are practically unchanged, while the $y$-values in the outer bins are higher after masking the radio sources, thus flattening the radial profiles. Therefore, if the radio sources are not removed, the background would be underestimated and hence the tSZ signal would be overestimated. This is corrected in the masked profiles. In the main text, we discuss the results masked for galaxy clusters as well as the radio sources.

\begin{figure*}
\centering
\includegraphics[trim= 0 50 0 0, clip,width=\textwidth]{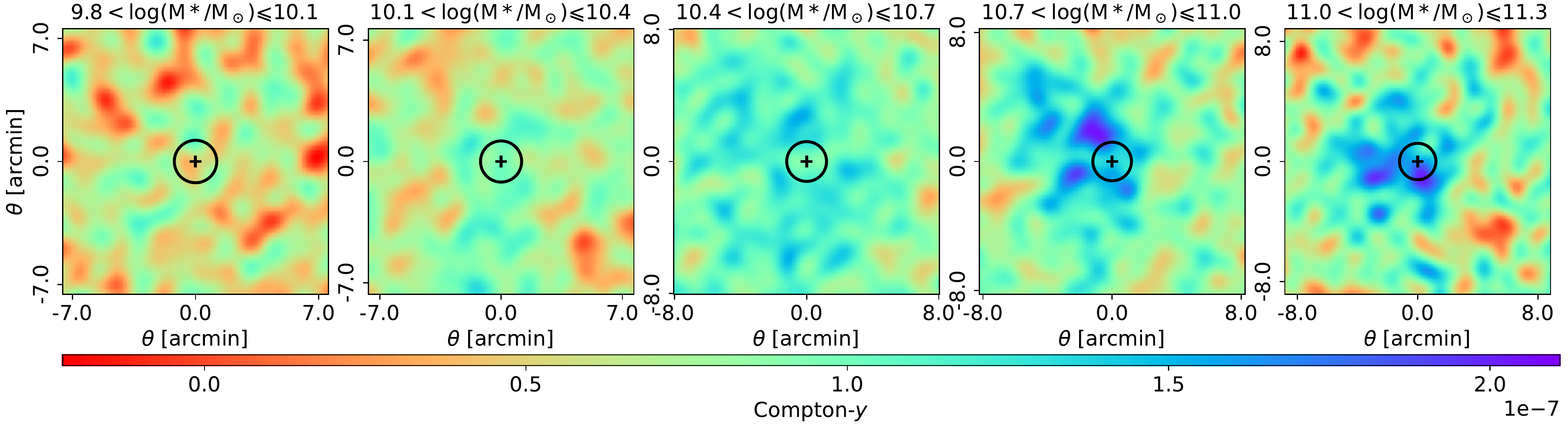}
\includegraphics[trim= 0 50 0 0, clip,width=\textwidth]{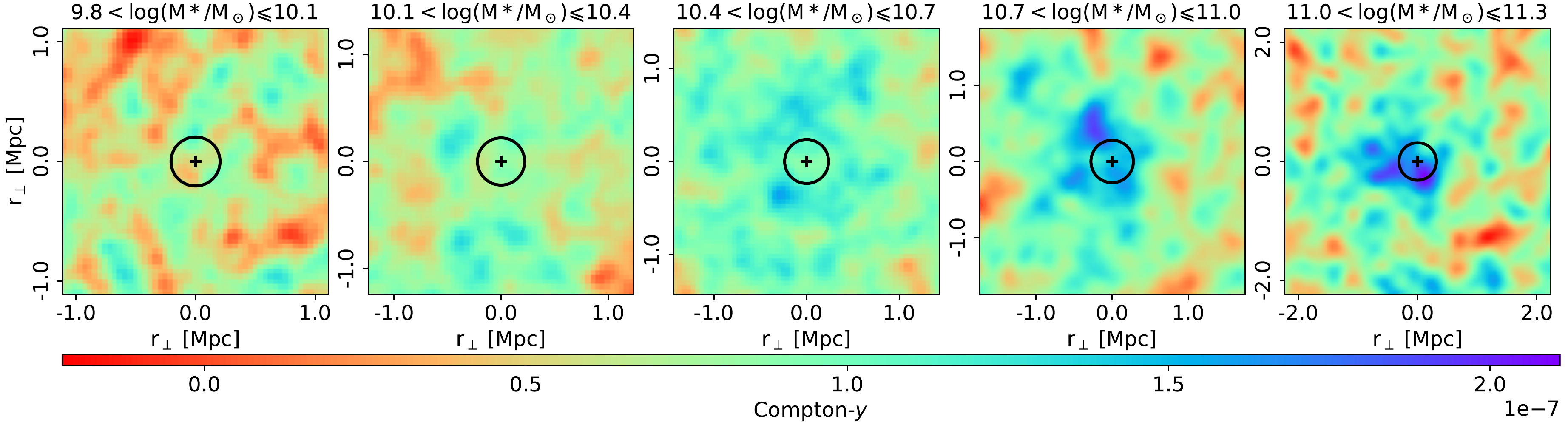}
\includegraphics[width=\textwidth]{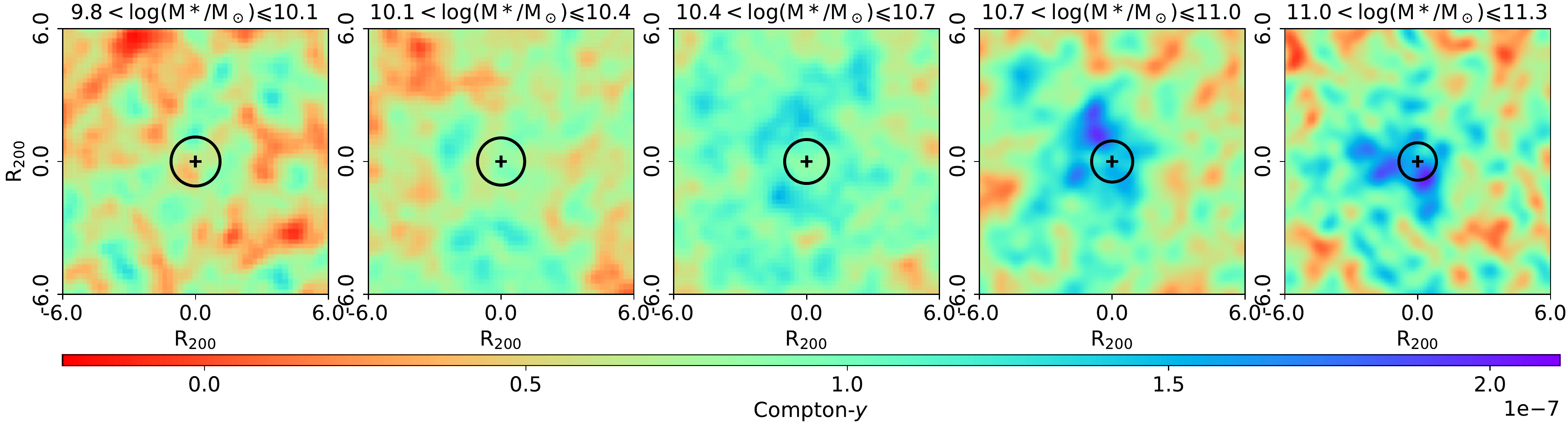}
\includegraphics[width=\textwidth]{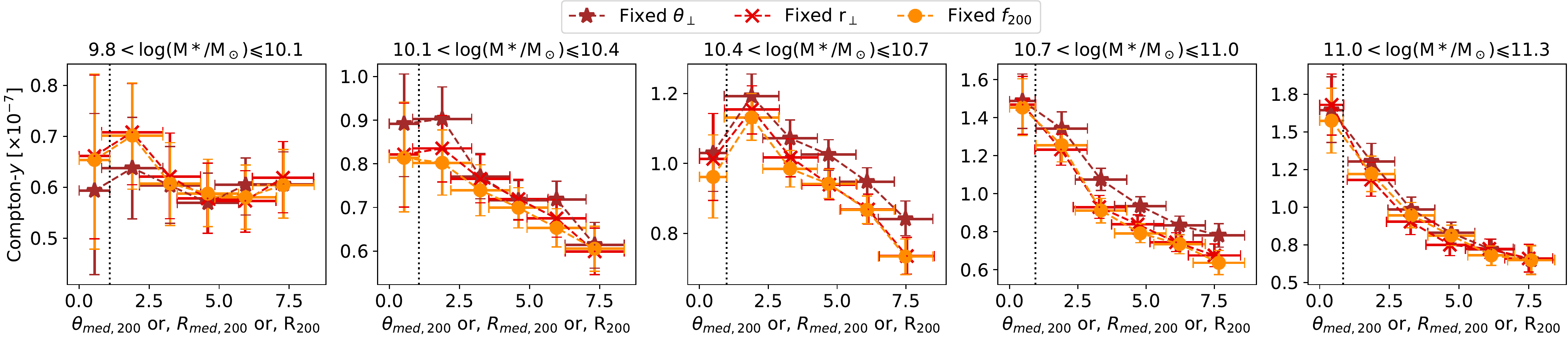}
\caption{\textbf{Top three rows:} Compton-$y$ maps stacked using method\,I (fixed $\theta_\perp$; top), II (fixed $r_\perp$; middle) and III (fixed $f_{200}$; bottom). The size of the maps are $6\times \theta_{\rm med,200}$, $6\times \rm R_{med,200}$ and $6\times \rm R_{200}$. The range of M$_\star$ in each mass bin is mentioned in the title of each map. The center of stacked galaxies (i.e., the central pixel) is marked with a plus (`+') sign. The Gaussian beam (FWHM of 2.$\!'$4) is shown with a circle at the center of each map. The colorbar is shown below the third row. \textbf{Bottom row:} Comparison of the differential radial profiles of Compton-$y$ among the three stacking methods. The range of M$_\star$ in each mass bin is mentioned in the title of each panel. The units of galactocentric distance are $\rm\theta_{\rm med,200}$, $\rm R_{med,200}$, and R$_{200}$ for the stacking methods I (brown stars), II (red crosses), and III (orange circles), respectively. The vertical dotted lines in each panel correspond to the angular resolution of the $y$-maps. The width of each annulus is the same as the beam size, shown with the error bars along the x-axis. The error bars in the y-axes denote statistical+systematic 68\% confidence intervals. See \S\ref{sec:comparestack} for details.}
\label{fig:compare_stack_method}
\end{figure*}

\subsection{Comparison among stacking methods}\label{sec:comparestack}
We show the CIB and Galactic dust-corrected, galaxy clusters and radio sources-masked $y$-maps stacked using the three stacking methods in Figure\,\ref{fig:compare_stack_method} (top three rows). The difference among the three stacking methods in the 2-D spatial distribution of Compton-$y$ in the maps is visible. We also show the corresponding differential and cumulative radial profiles in the bottom two rows of Figure\,\ref{fig:compare_stack_method}. The units of galactocentric distance in these radial profiles are the average $\theta_{200}$, average R$_{200}$, and R$_{200}$ for the stacking methods\,I (fixed $\theta_\perp$), II (fixed $r_\perp$) and III (fixed $f_{200}$), respectively. The radial profiles for the second and the third stacking methods are consistent with each other (red crosses and blue stars), but the profiles for the first stacking method are different (brown stars). In the differential profiles of Compton-$y$, the difference among the stacking methods is not uniform in all mass bins and at all radii. In the two lowest mass bins, the $y$-values are noise dominated, and the differences among the stacking methods in individual radial bins are consistent within error. In the intermediate ($\rm 10.4 < log(M_\star/M_\odot) \leqslant 11.0$) mass bins, the outer radial bins are more affected by the stacking methods than the innermost radial bin. In the most massive bin, the differences among the stacking methods, although present, are not visually prominent. 
Thus, it is crucial to incorporate the redshift and the virial radius into the stacking, otherwise, the signal might be incorrectly estimated. This is one of the major advancements in this work, compared to previous studies. 

Because the second stacking method involves the information about the galaxy redshift while the first stacking method does not, the $\approx 16-45$\% scatter in $\rm D_A^{-1}$ in individual mass bins causes the difference between the first and the second stacking methods (see equations \ref{eq:stackmethod1} and \ref{eq:stackmethod2}). The similar scatter in $\rm D_A^{-1}$ and in $\rm R_{200}/D_A$ (see equations \ref{eq:stackmethod2} and \ref{eq:stackmethod3}) in all mass bins result in similar radial profiles for the second and the third stacking methods. In further discussion, we consider the third stacking method only, because it takes into account the redshift and the virial radius, and the results using this method are the most straightforward to interpret in terms of virial properties of galaxies. 

\section{Radial profiles of Compton-$y$}\label{sec:crosscheck}

\begin{figure*}
    \centering
    \includegraphics[width=\textwidth]{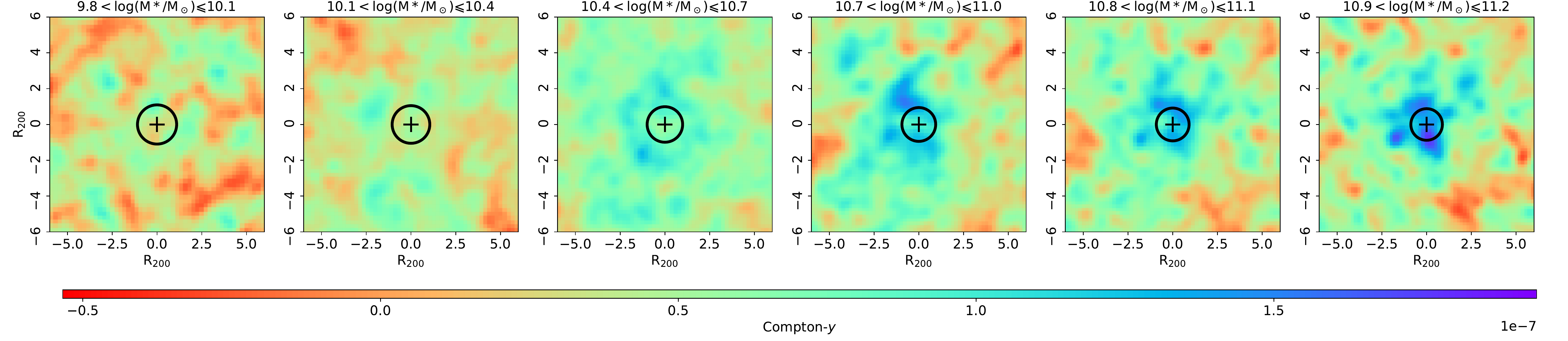}
    \includegraphics[width=0.15\textwidth]{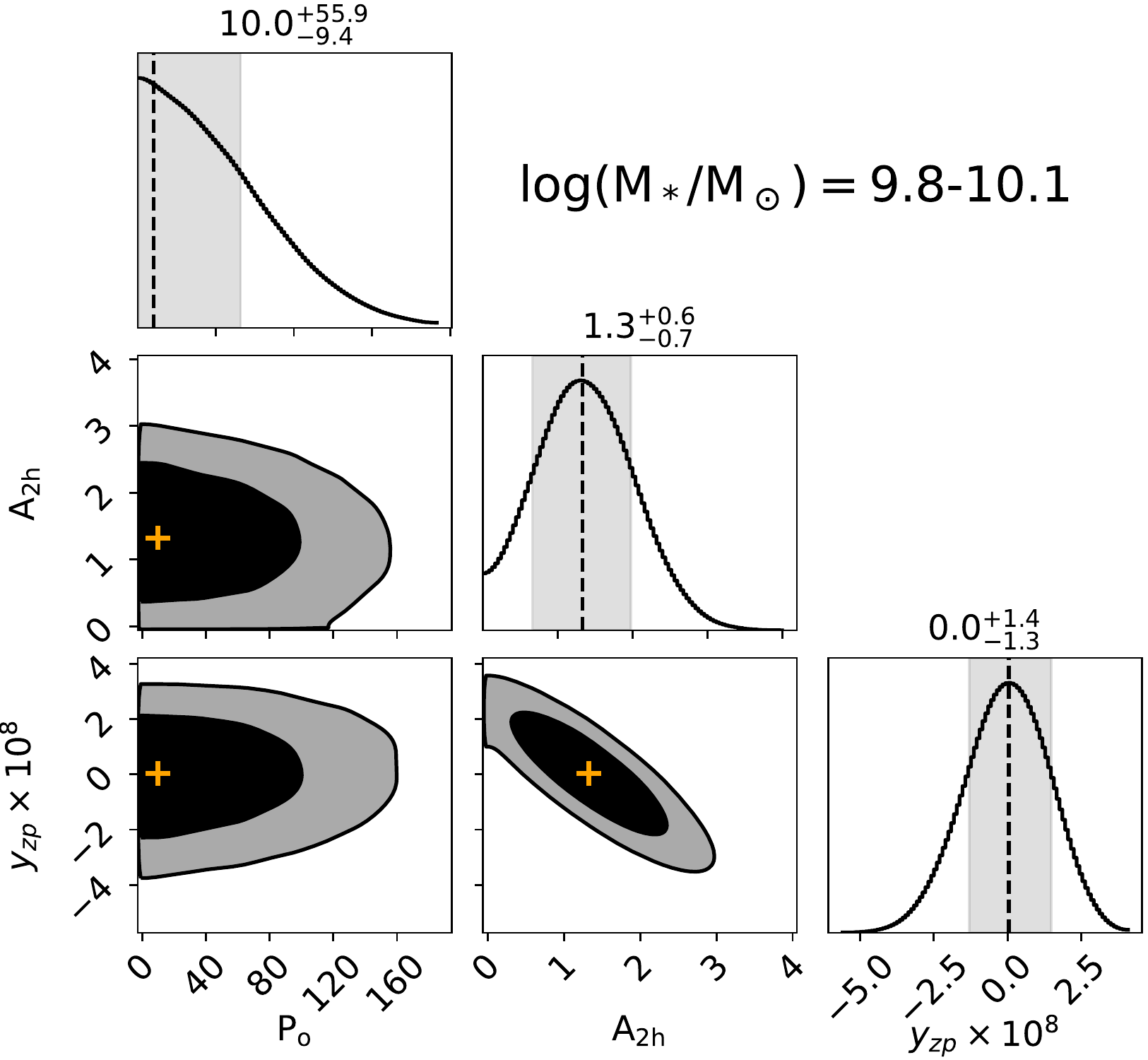}
    \includegraphics[width=0.15\textwidth]{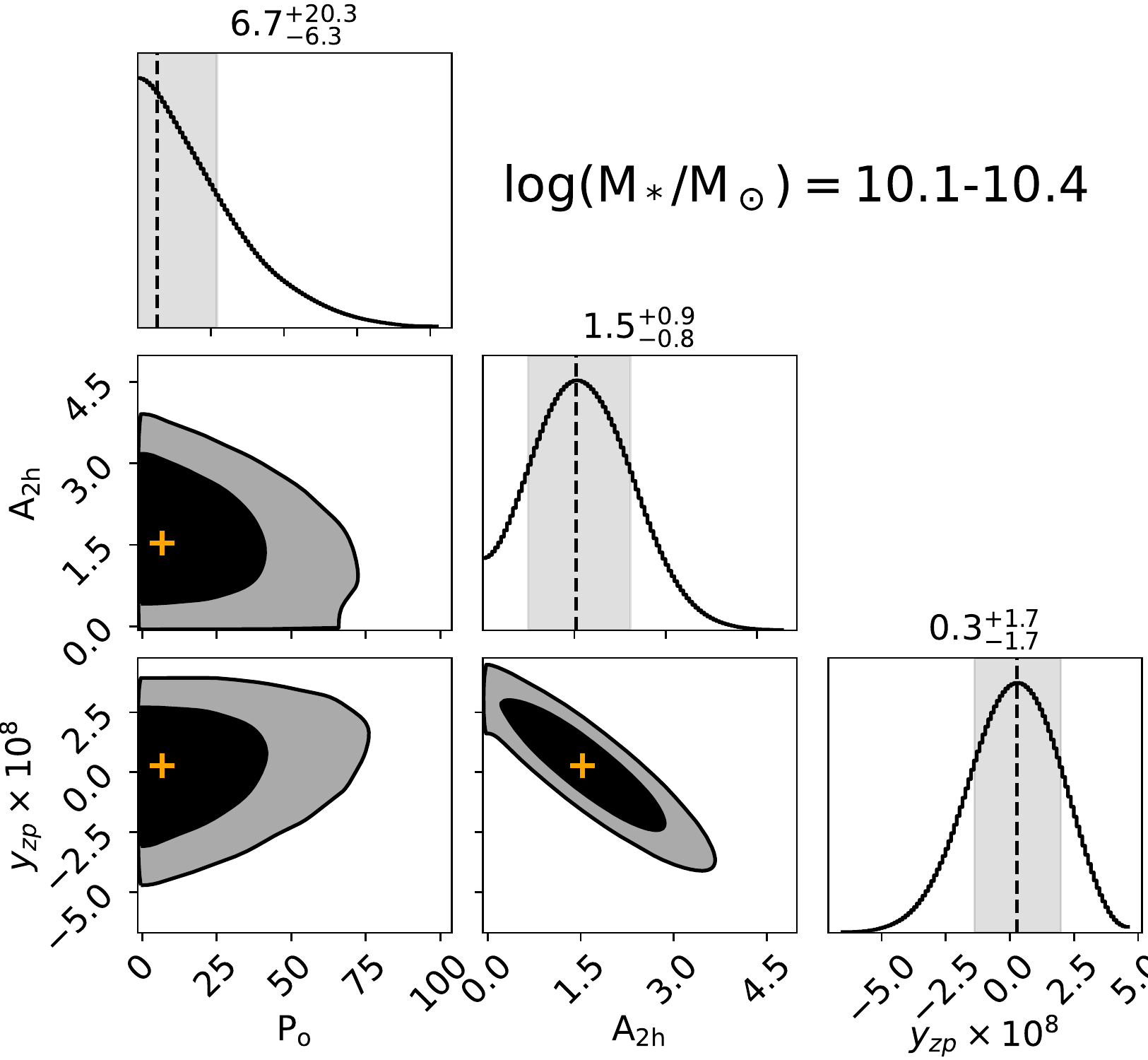}
    \includegraphics[width=0.15\textwidth]{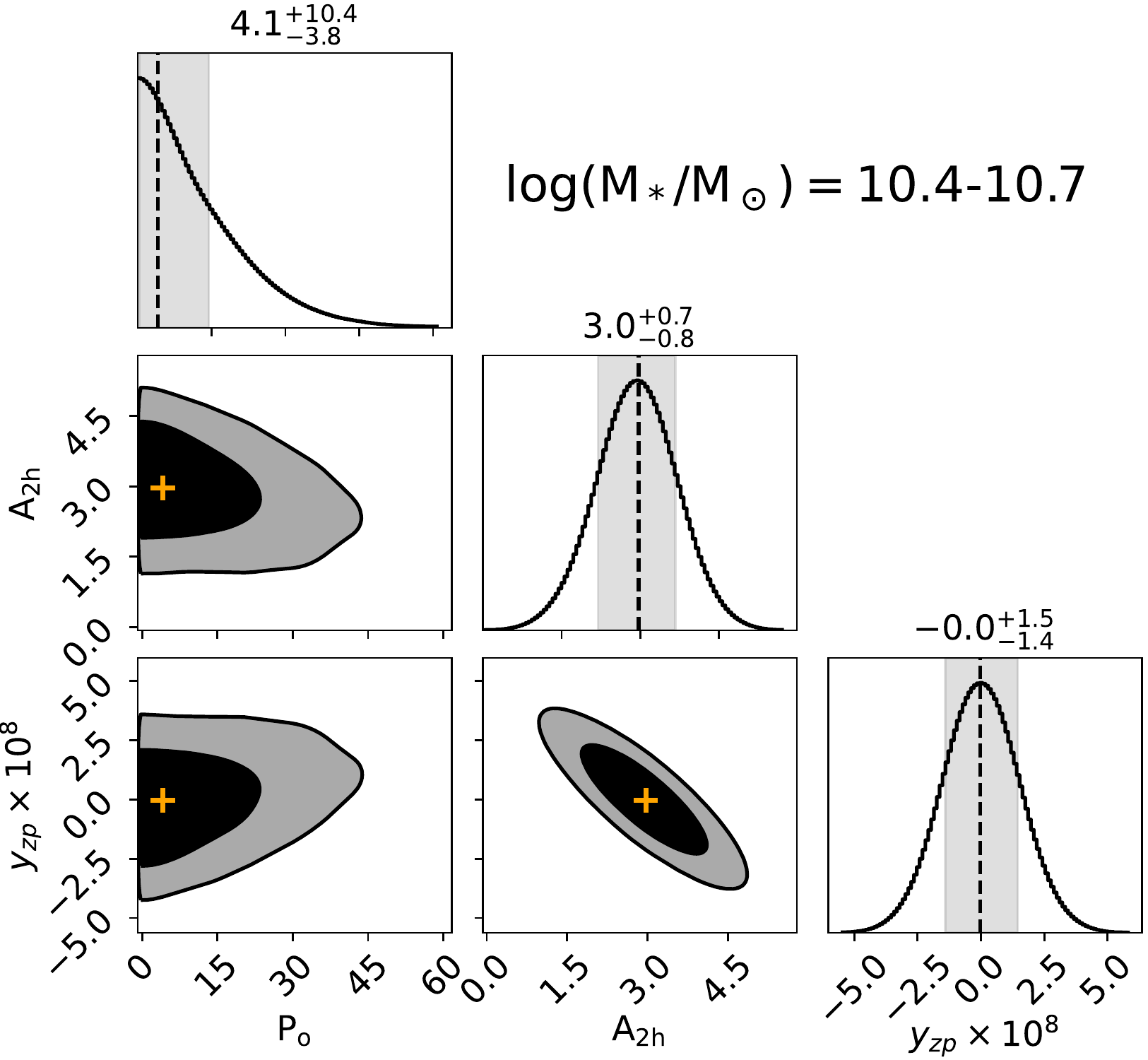}
    \includegraphics[width=0.15\textwidth]{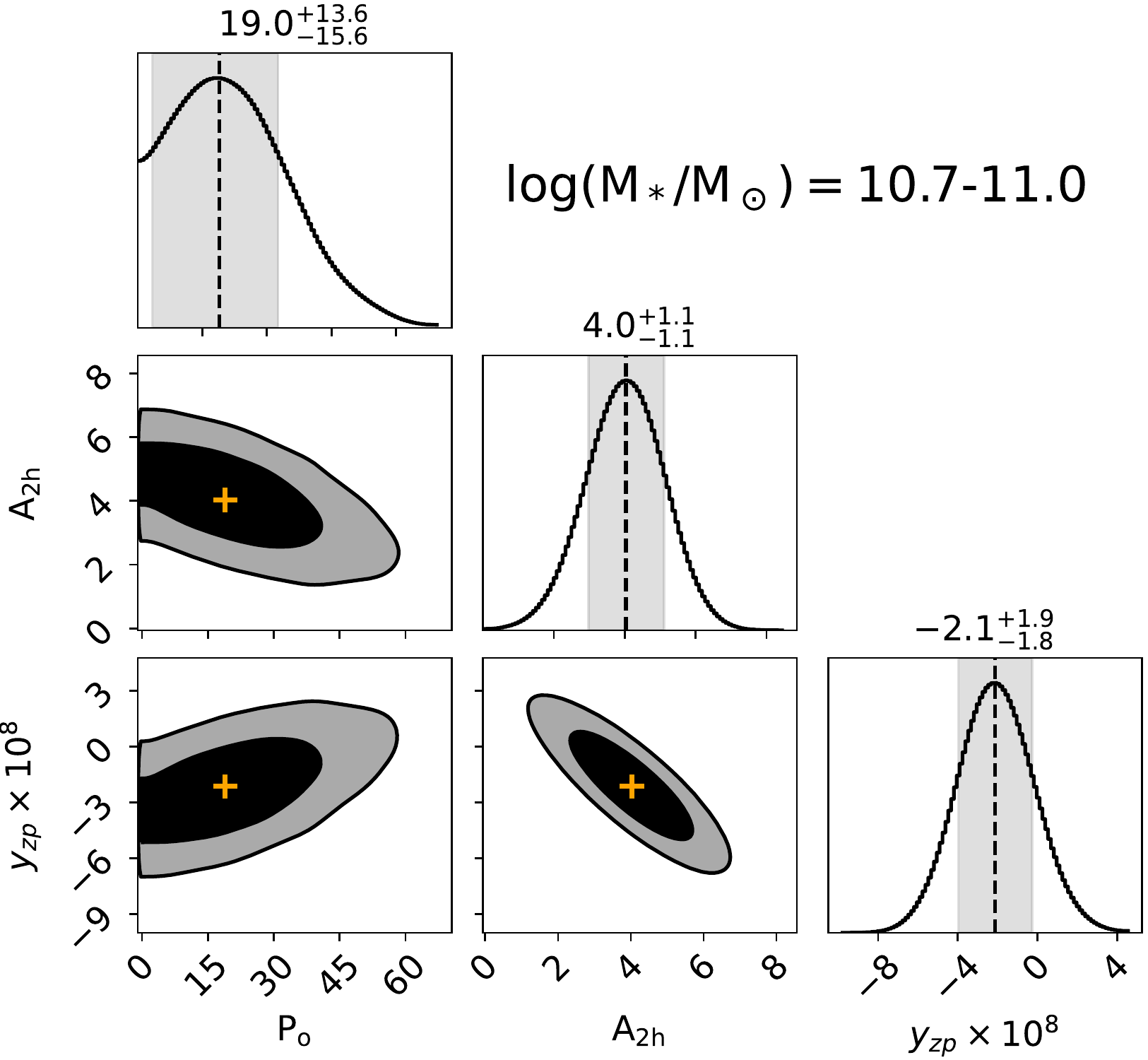}
    \includegraphics[width=0.15\textwidth]{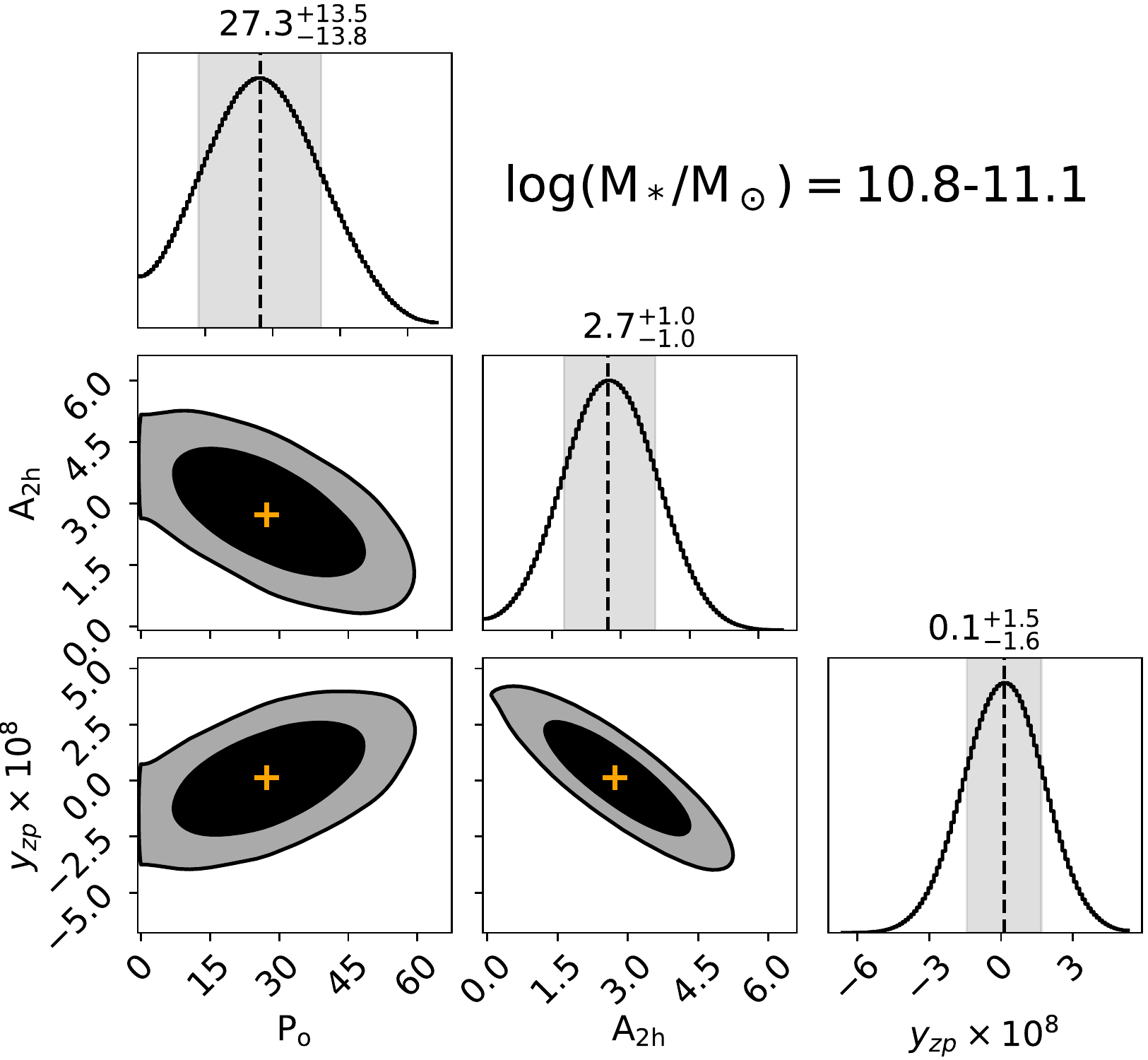}
    \includegraphics[width=0.15\textwidth]{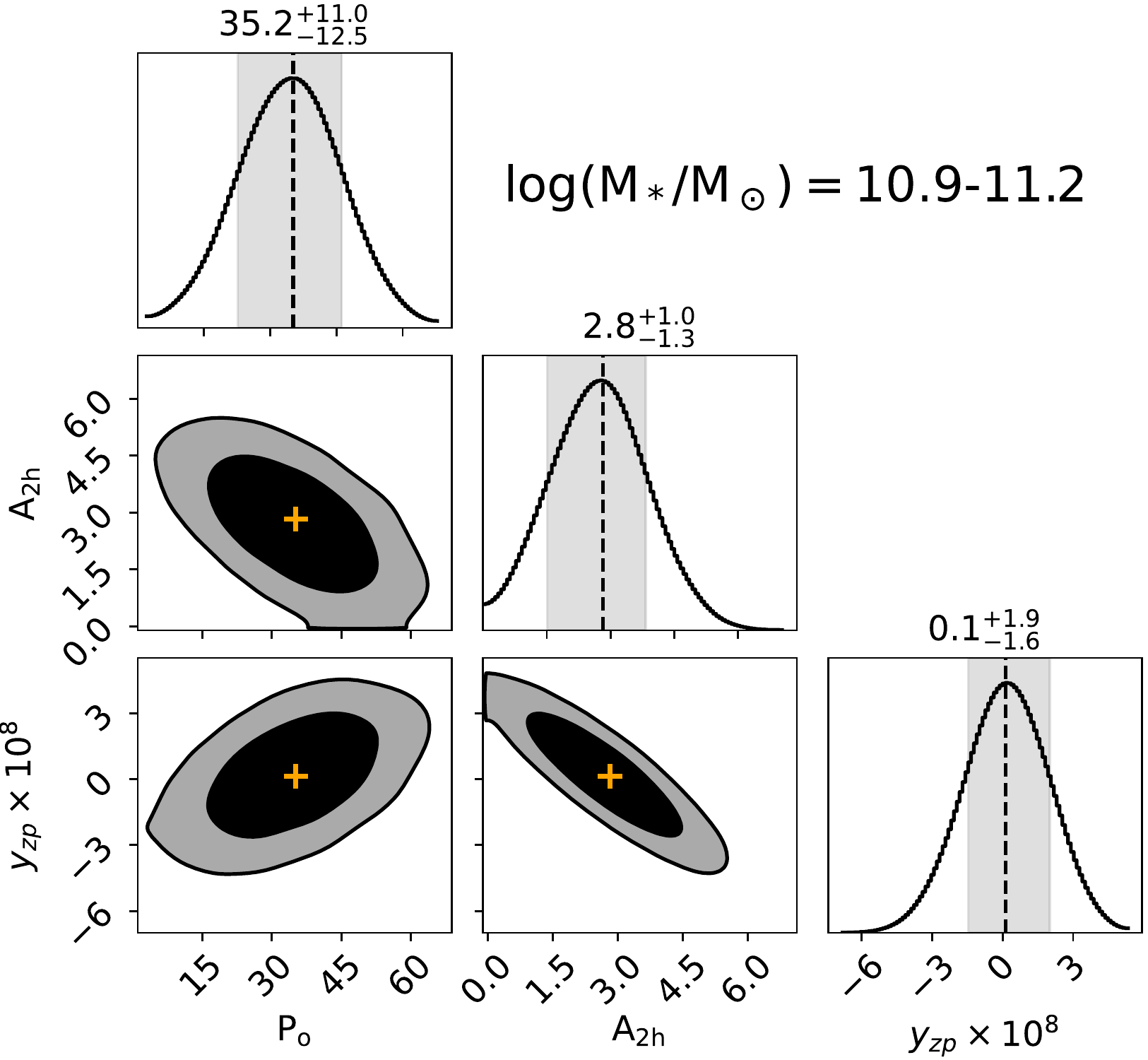}

    \includegraphics[width=0.15\textwidth]{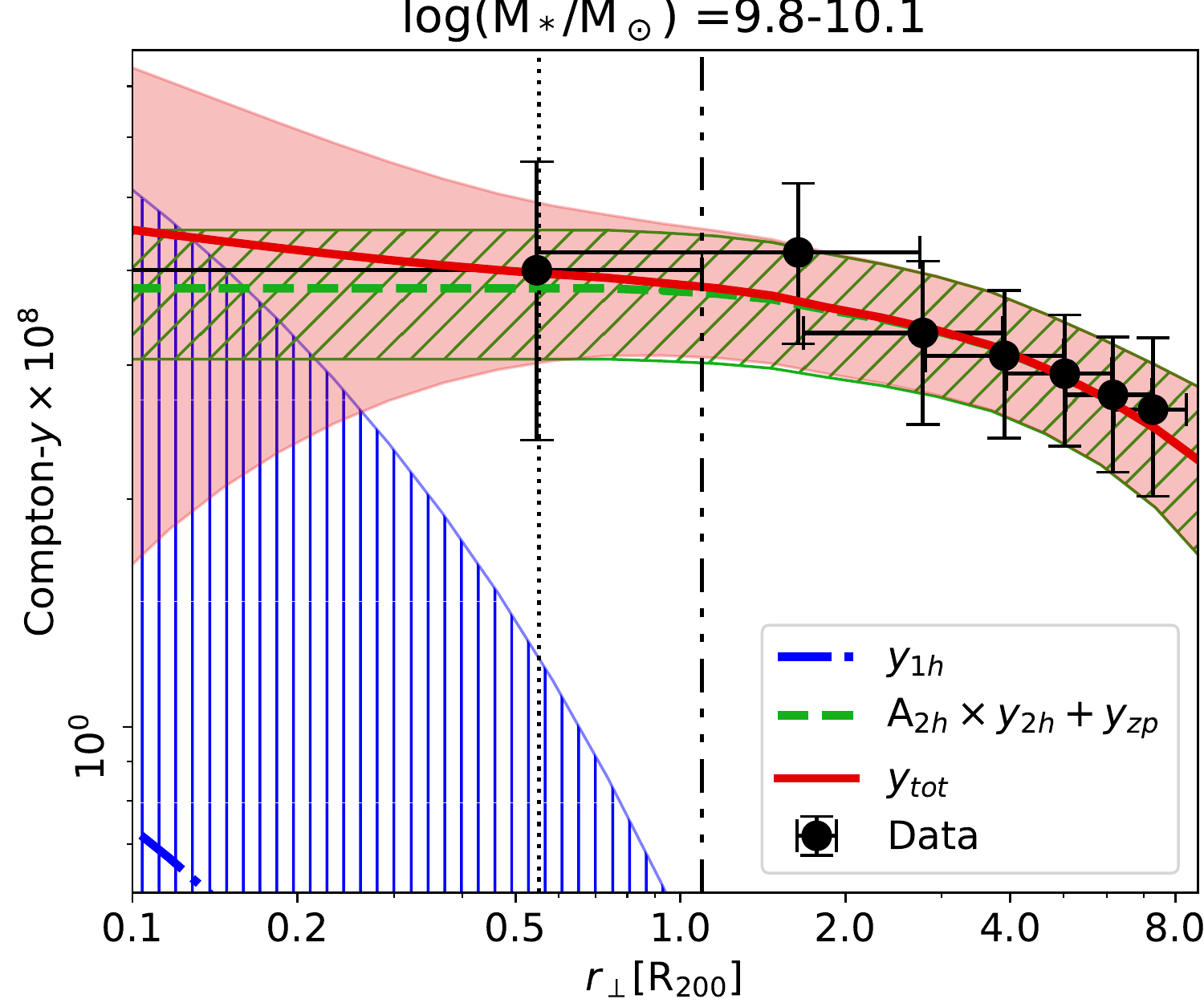}
    \includegraphics[width=0.15\textwidth]{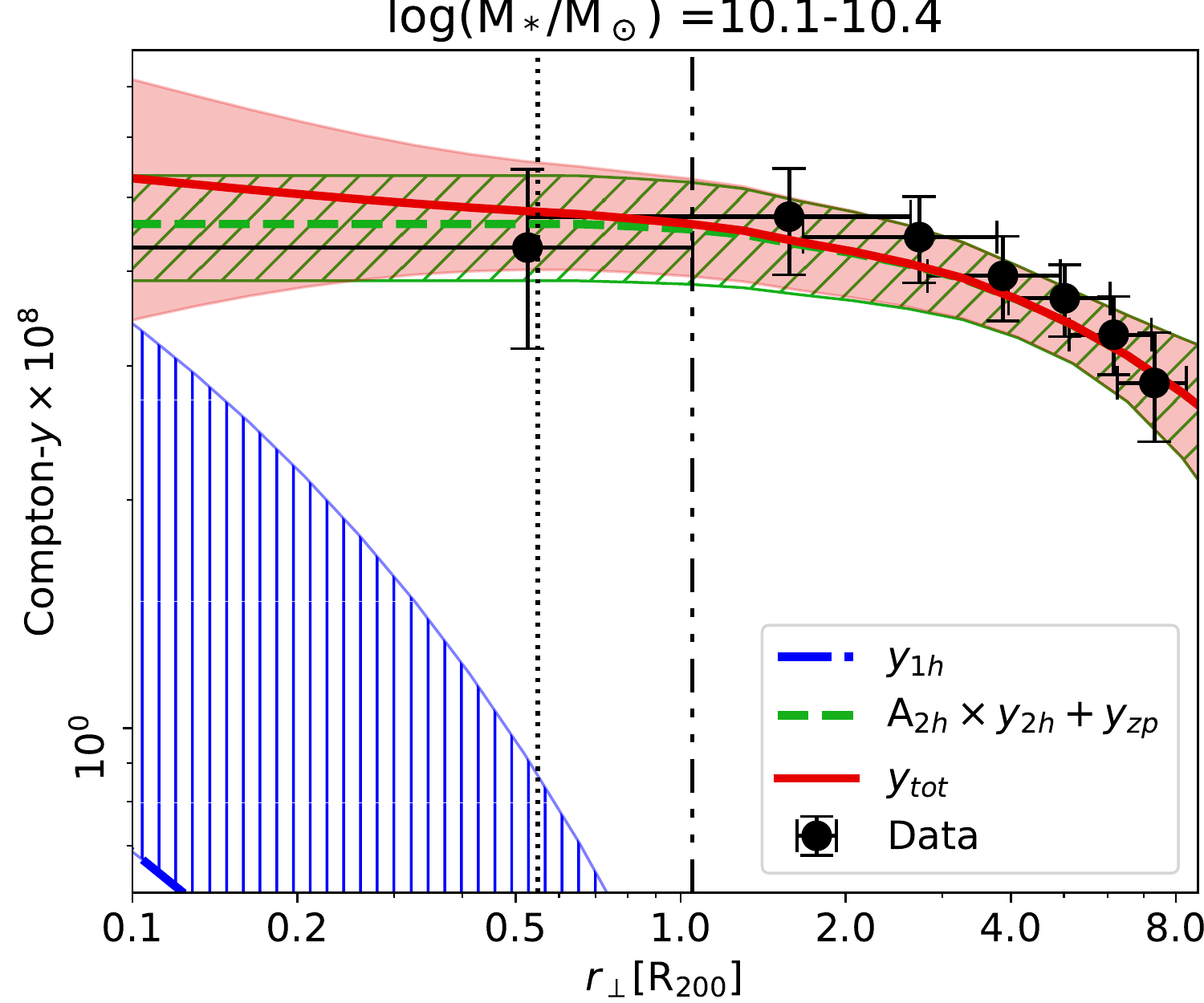}
    \includegraphics[width=0.15\textwidth]{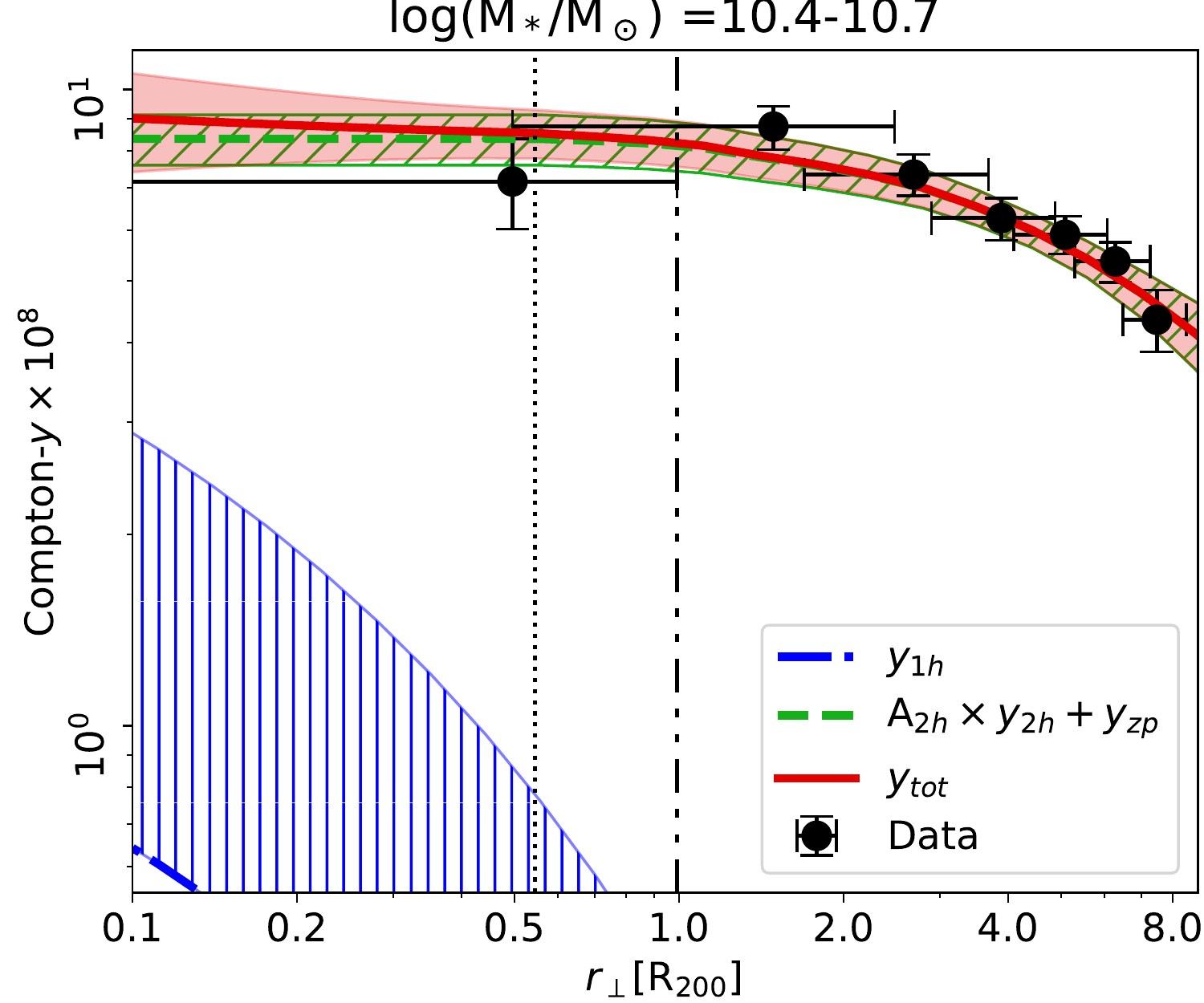}
    \includegraphics[width=0.15\textwidth]{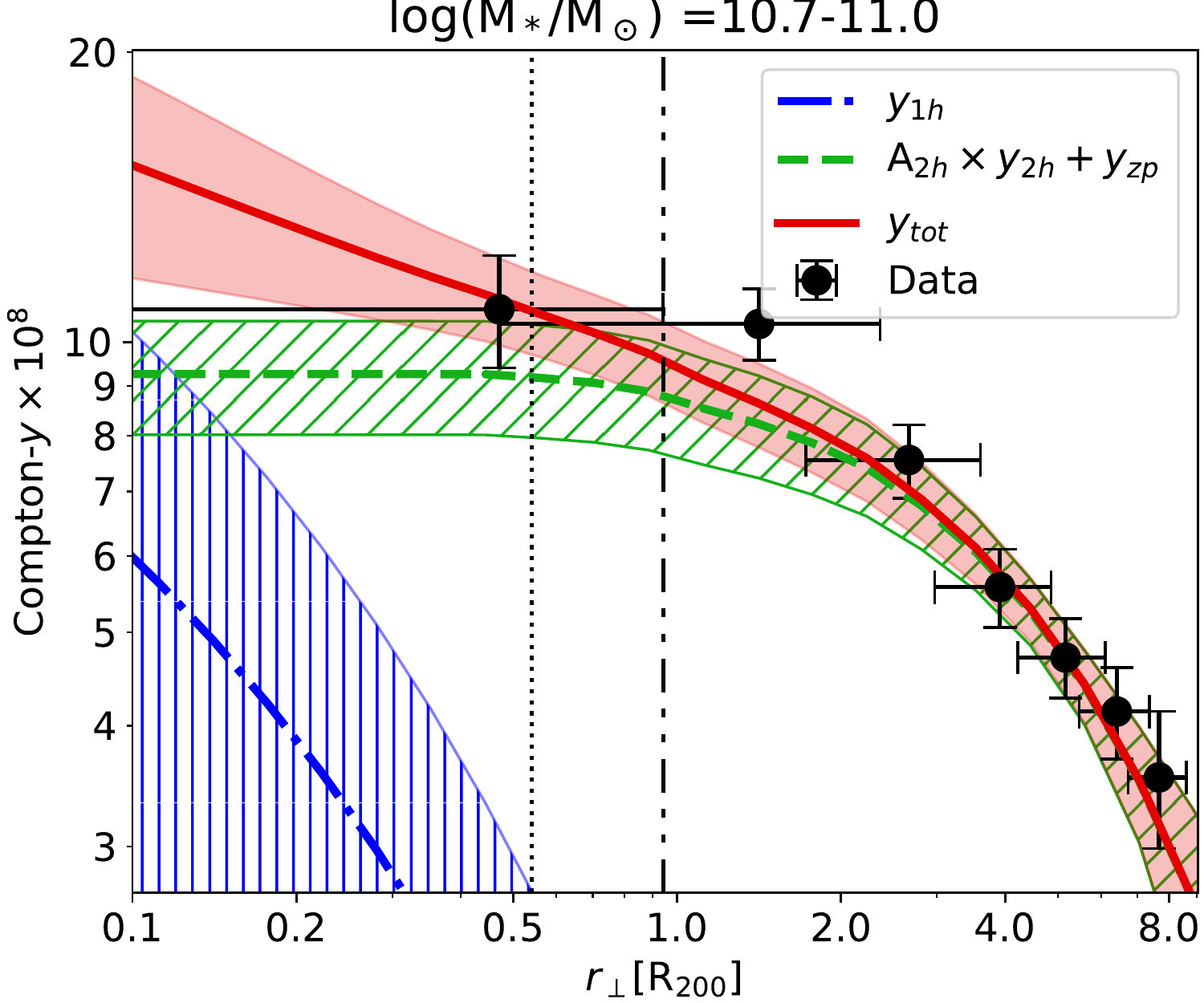}
    \includegraphics[width=0.15\textwidth]{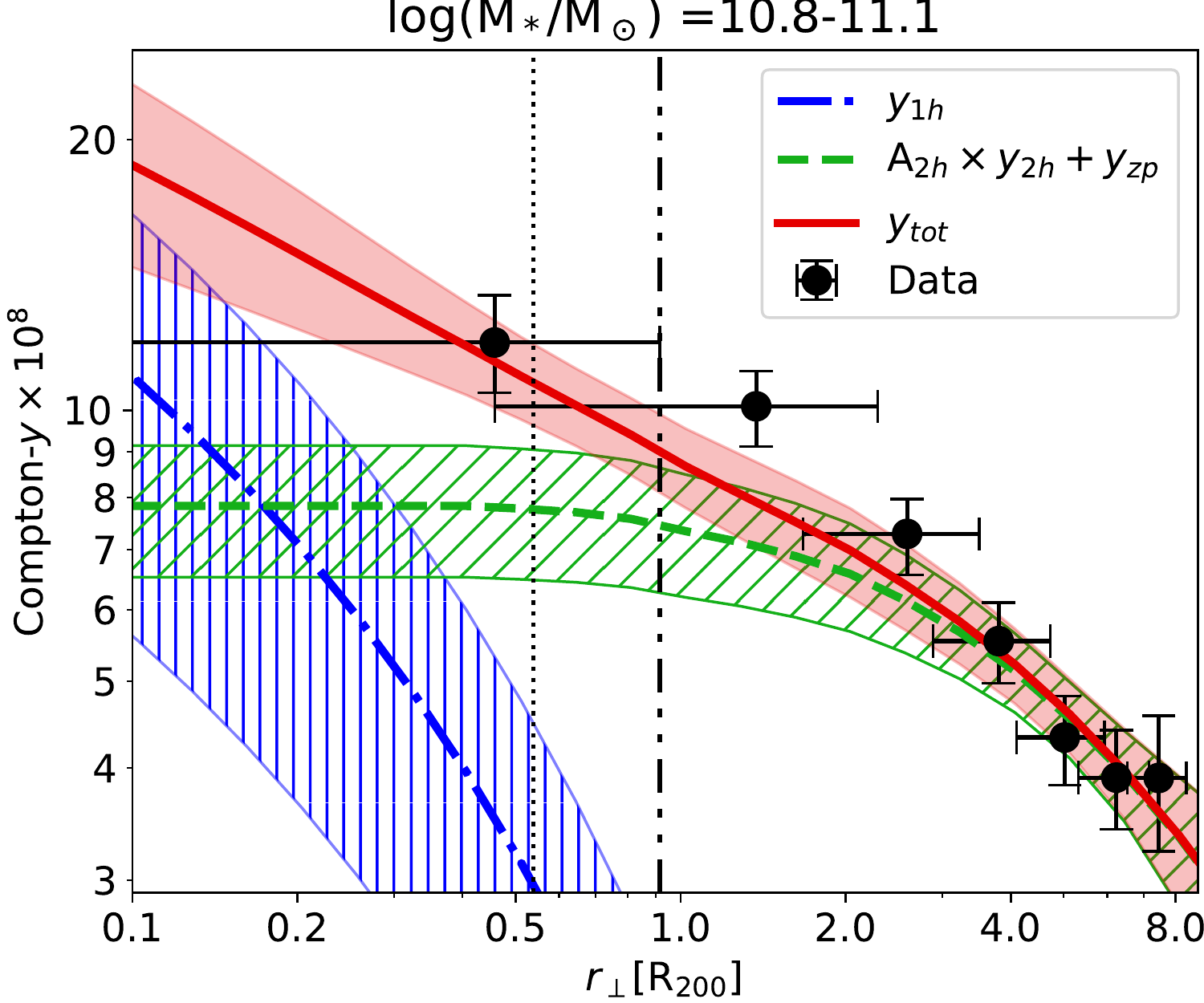}
    \includegraphics[width=0.15\textwidth]{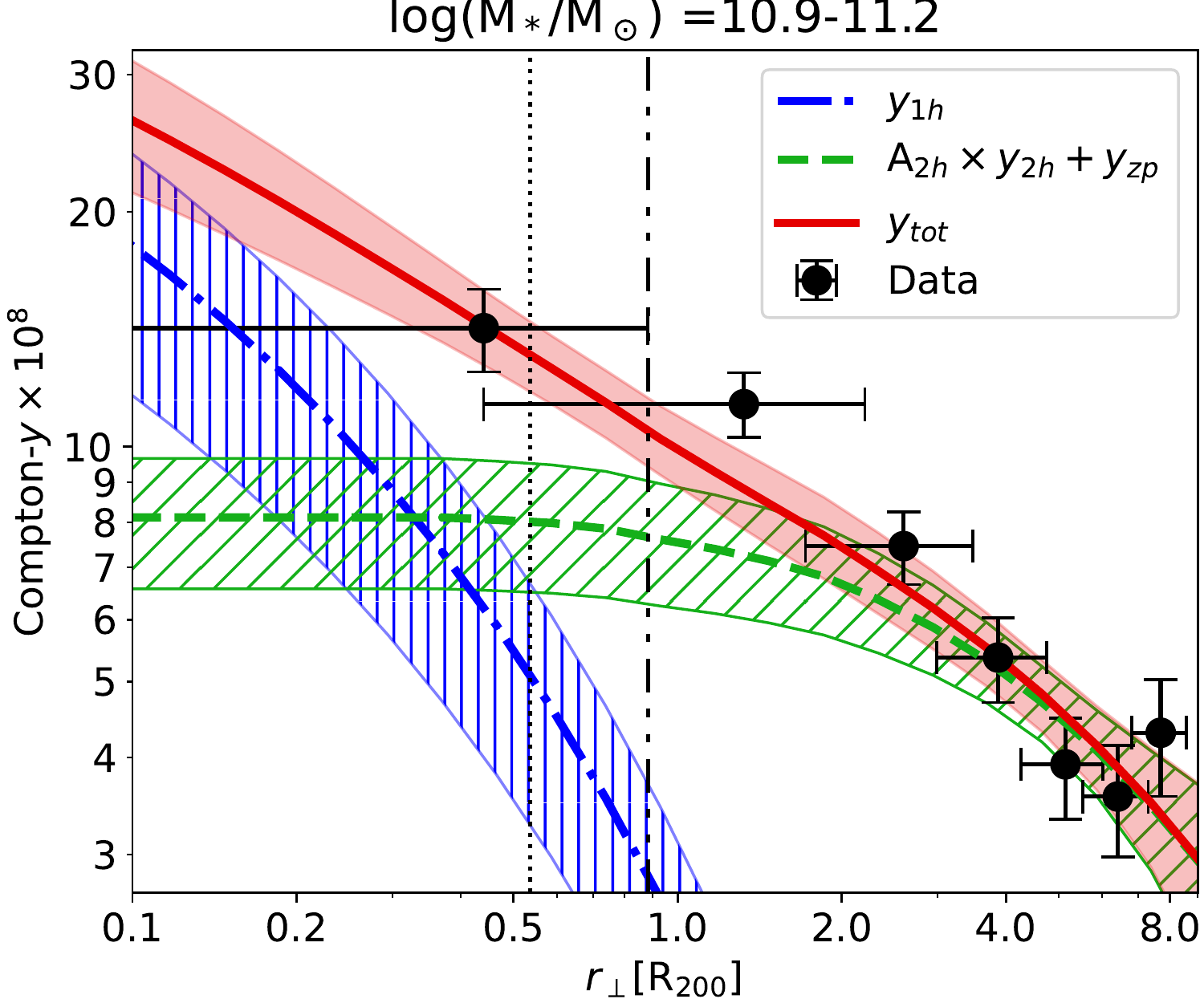}

    \includegraphics[width=\textwidth]{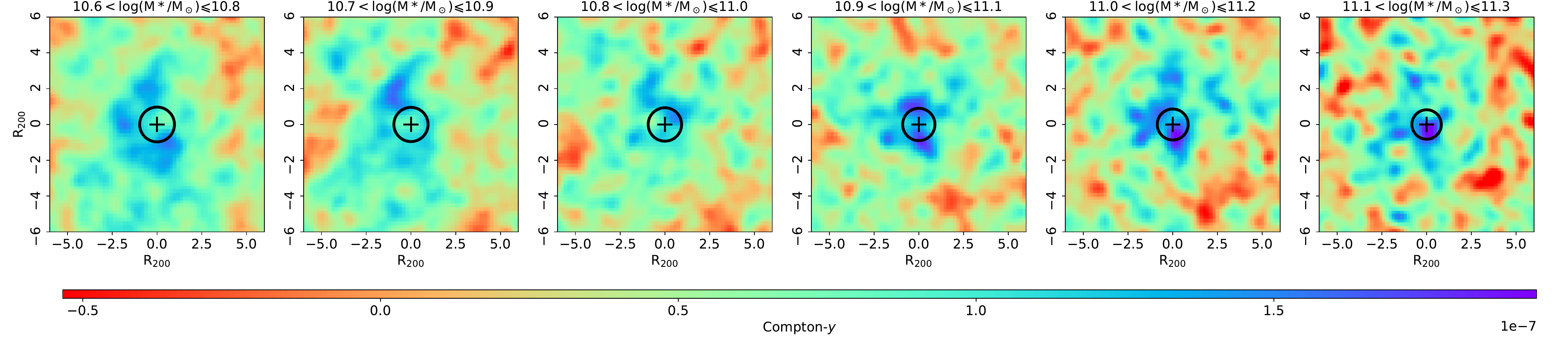}
    \includegraphics[width=0.15\textwidth]{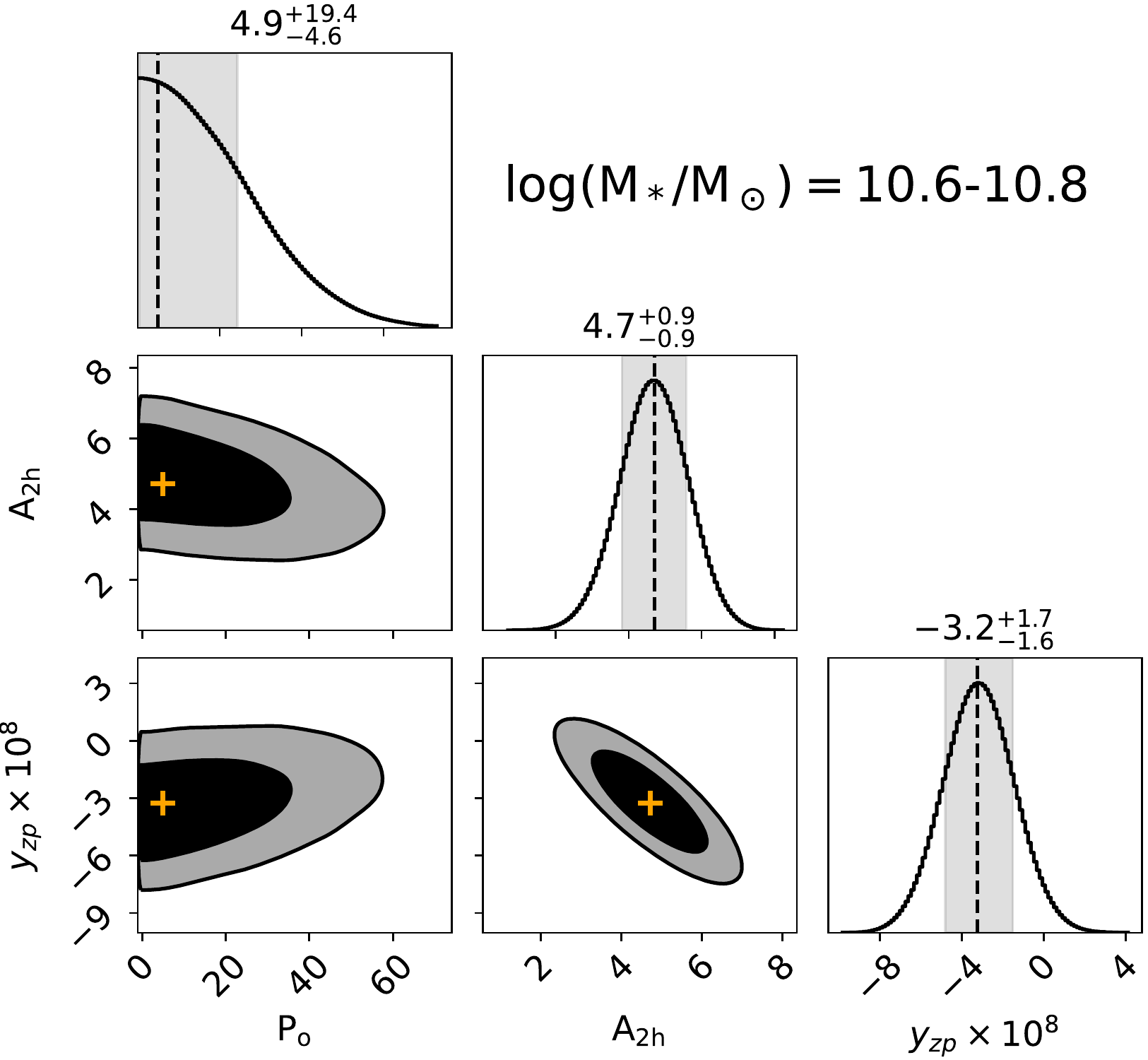}
    \includegraphics[width=0.15\textwidth]{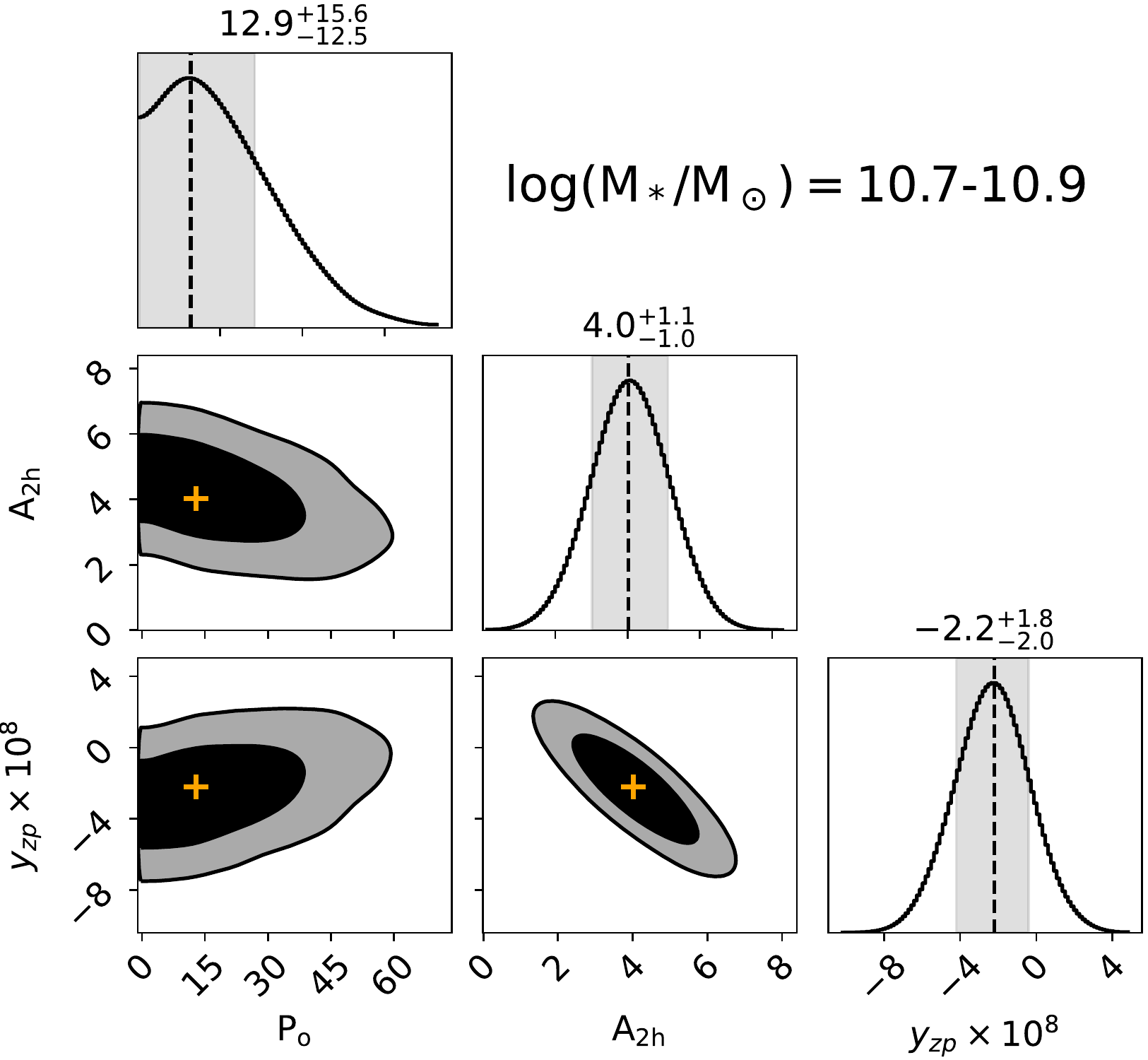}
    \includegraphics[width=0.15\textwidth]{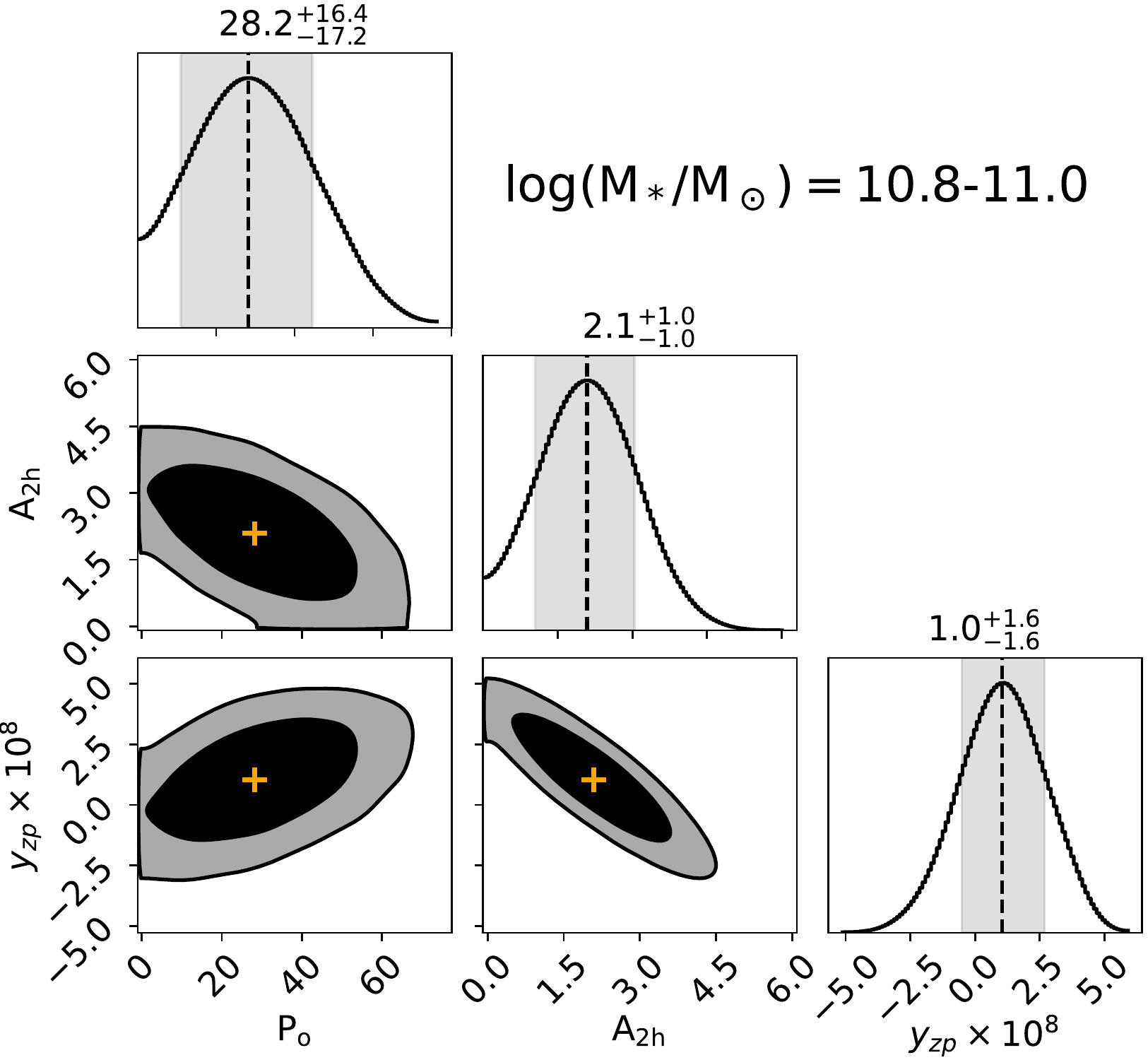}
    \includegraphics[width=0.15\textwidth]{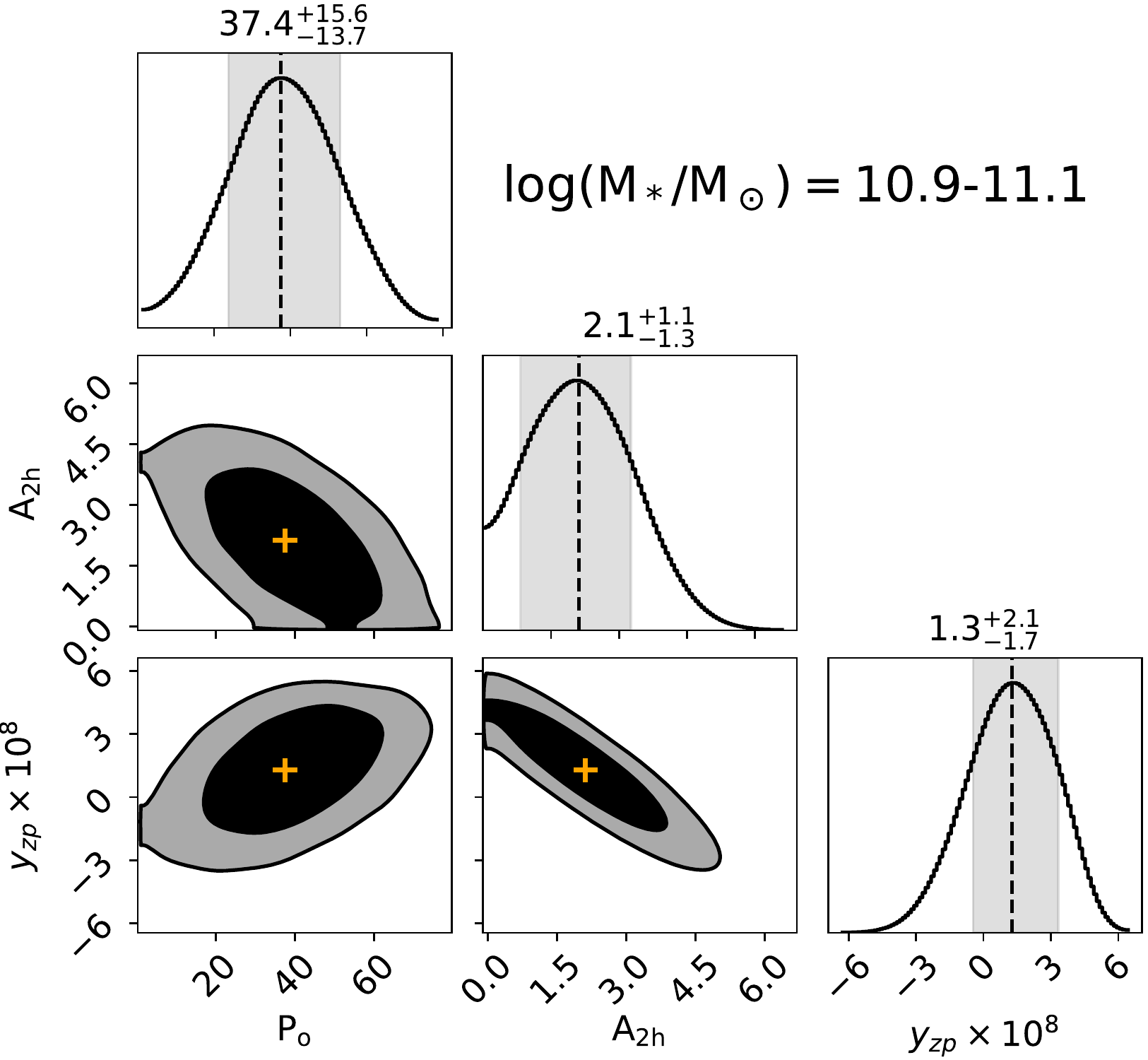}
    \includegraphics[width=0.15\textwidth]{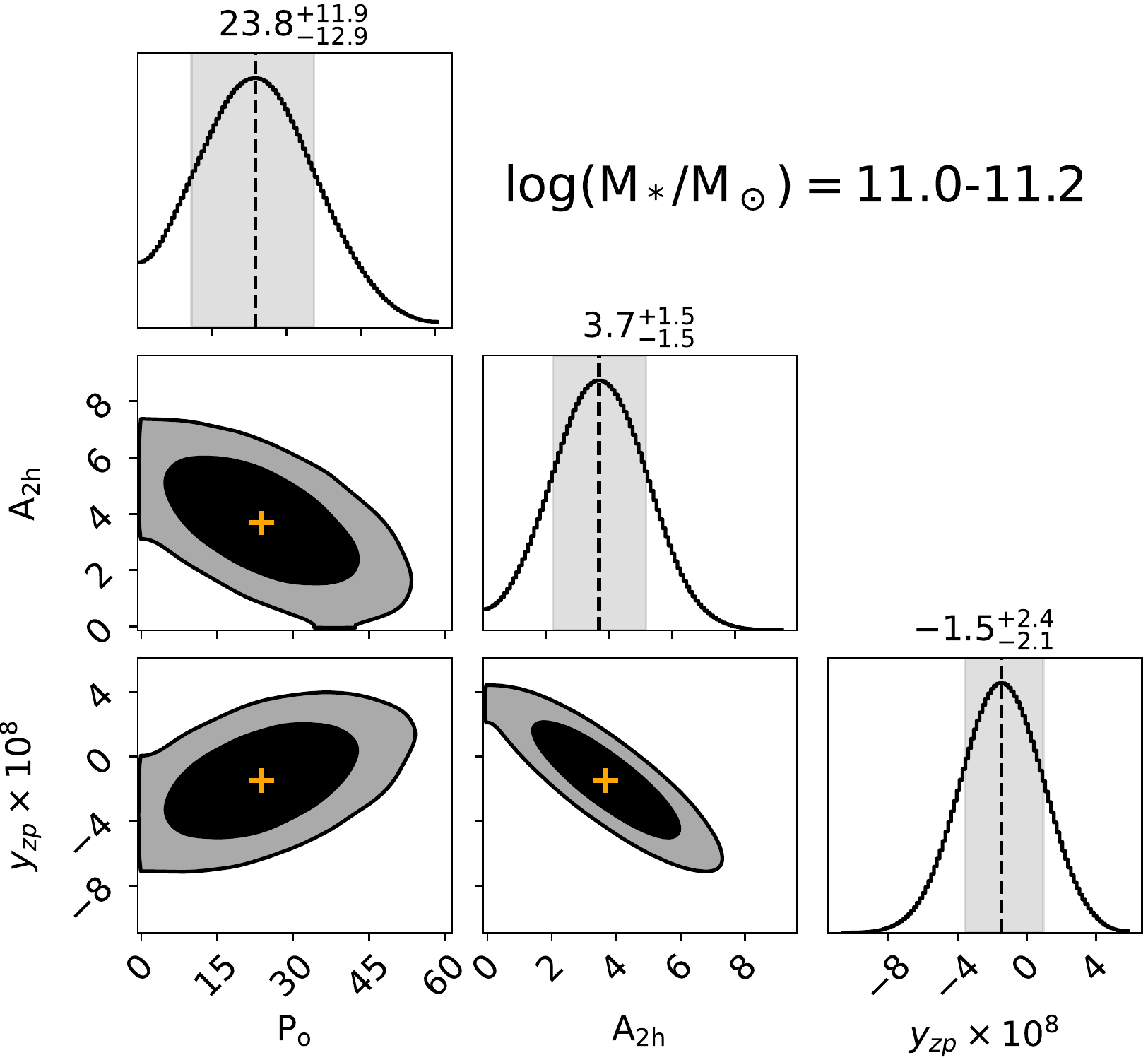}
    \includegraphics[width=0.15\textwidth]{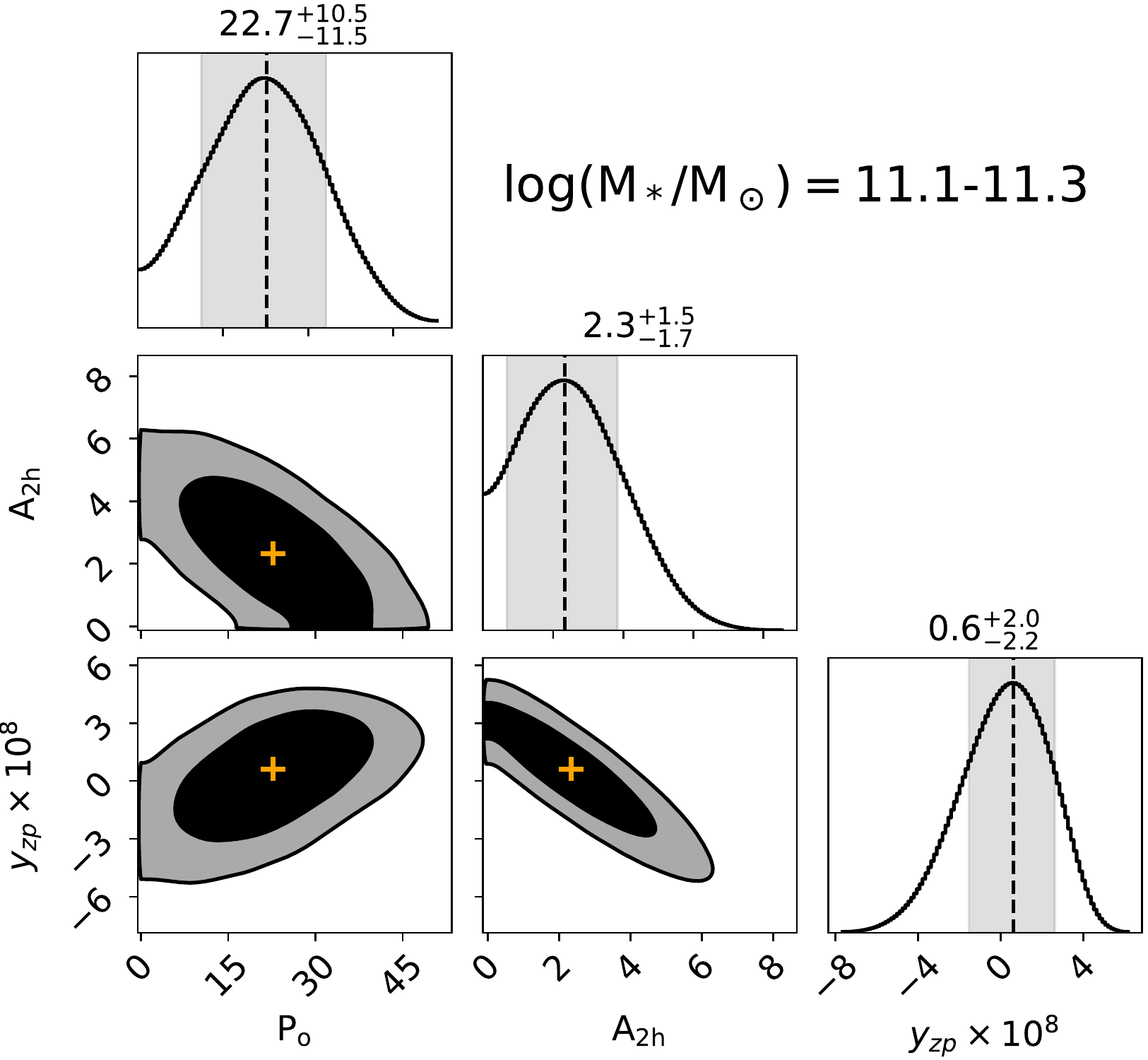}

    \includegraphics[width=0.15\textwidth]{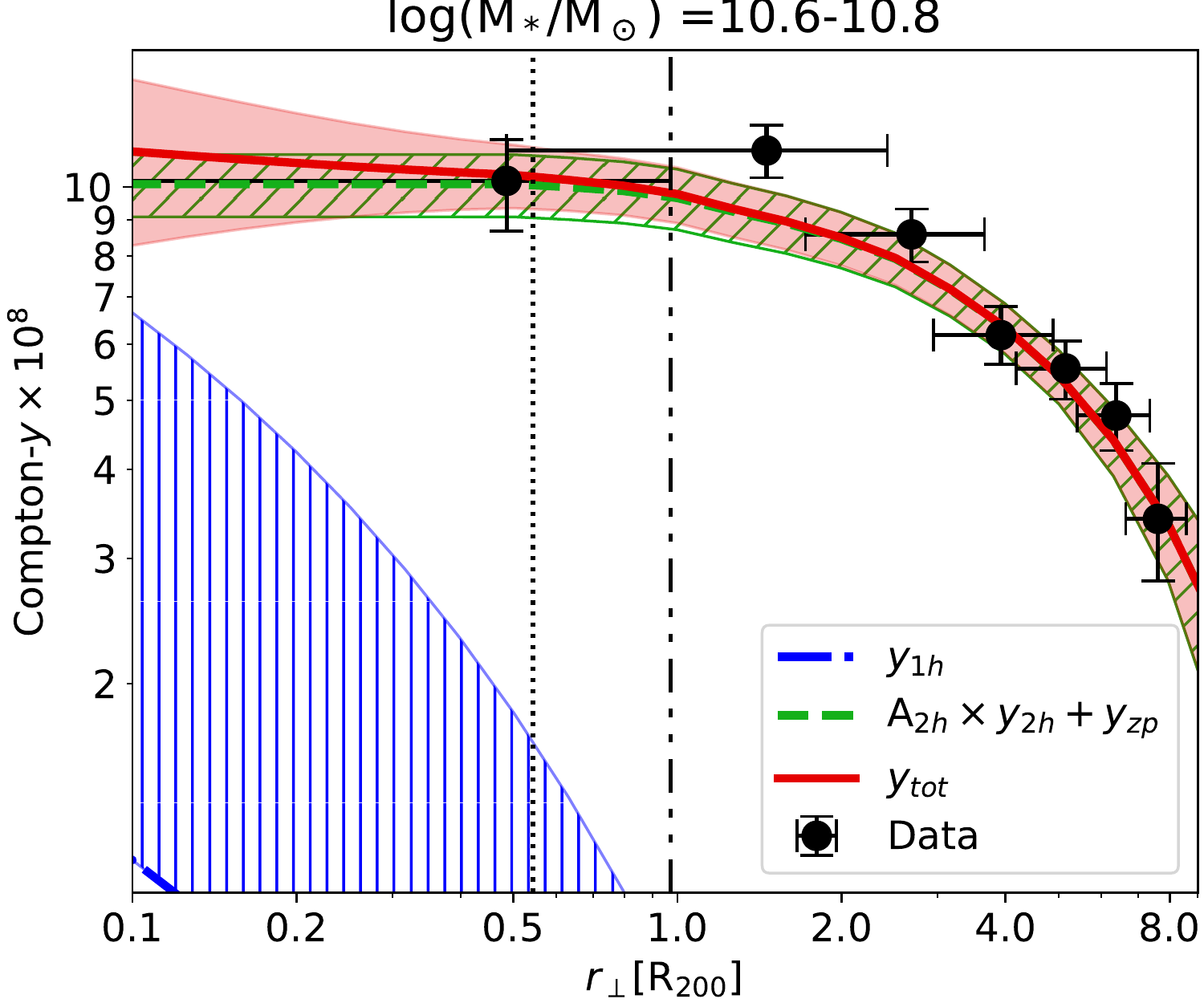}
    \includegraphics[width=0.15\textwidth]{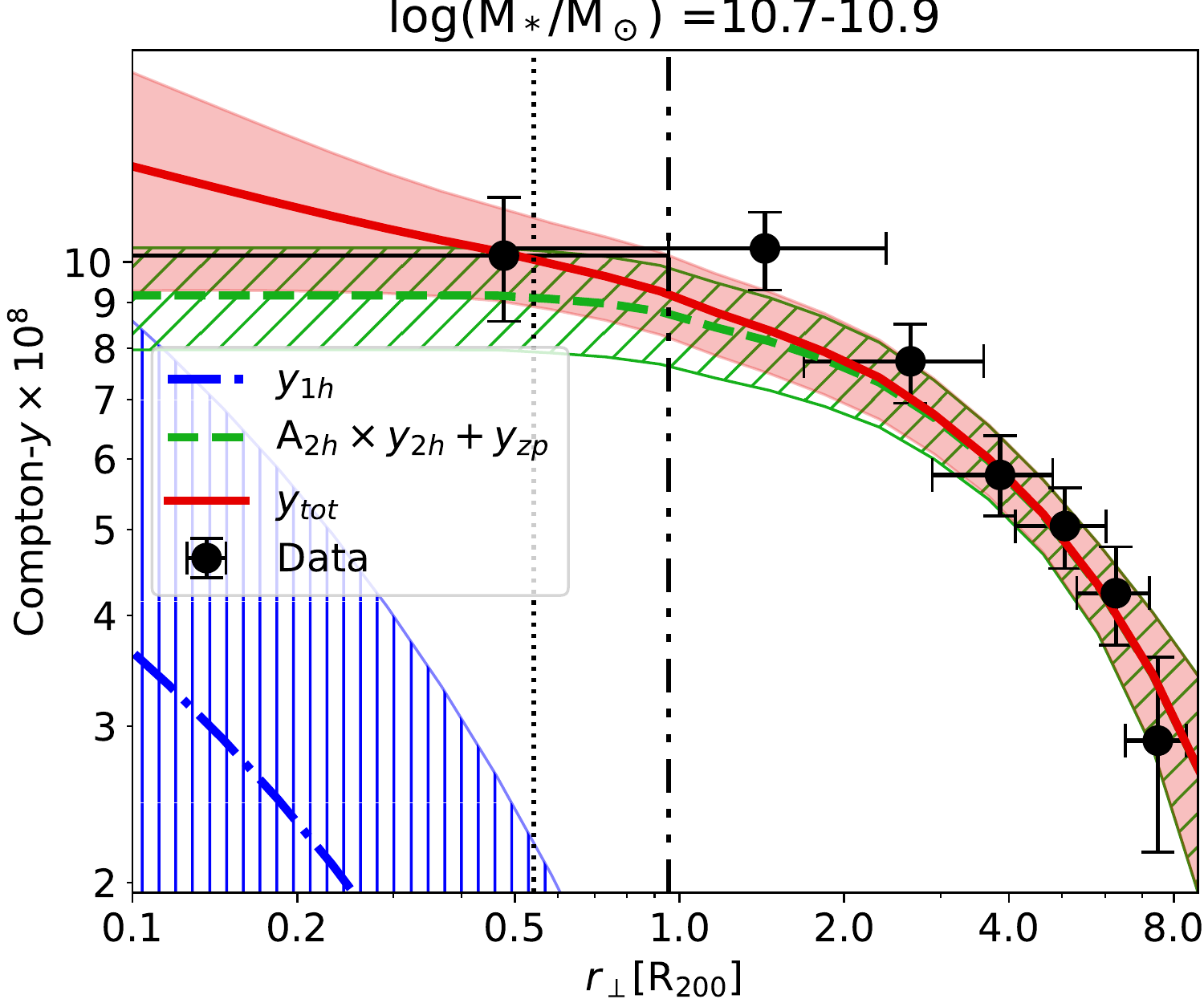}
    \includegraphics[width=0.15\textwidth]{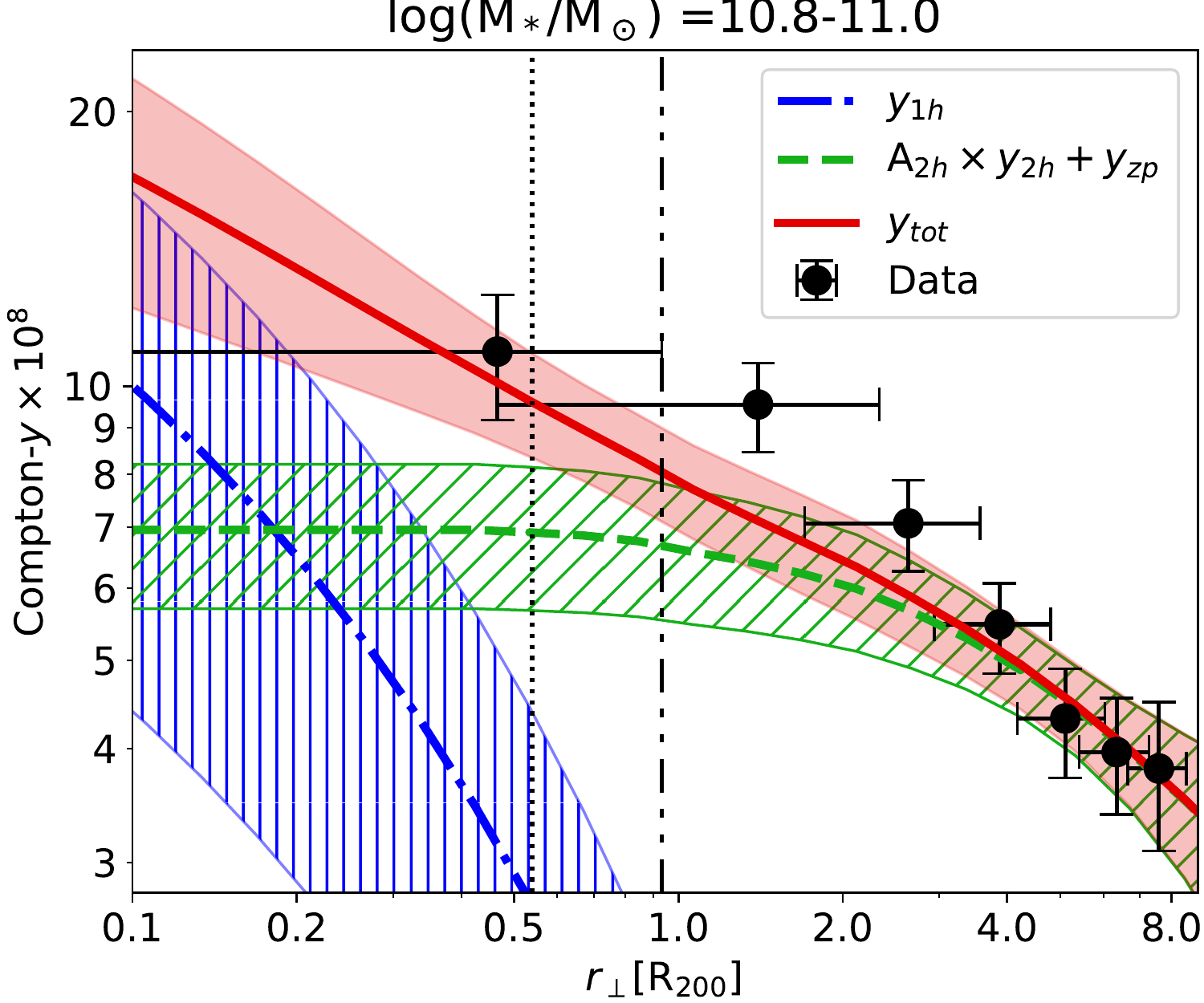}
    \includegraphics[width=0.15\textwidth]{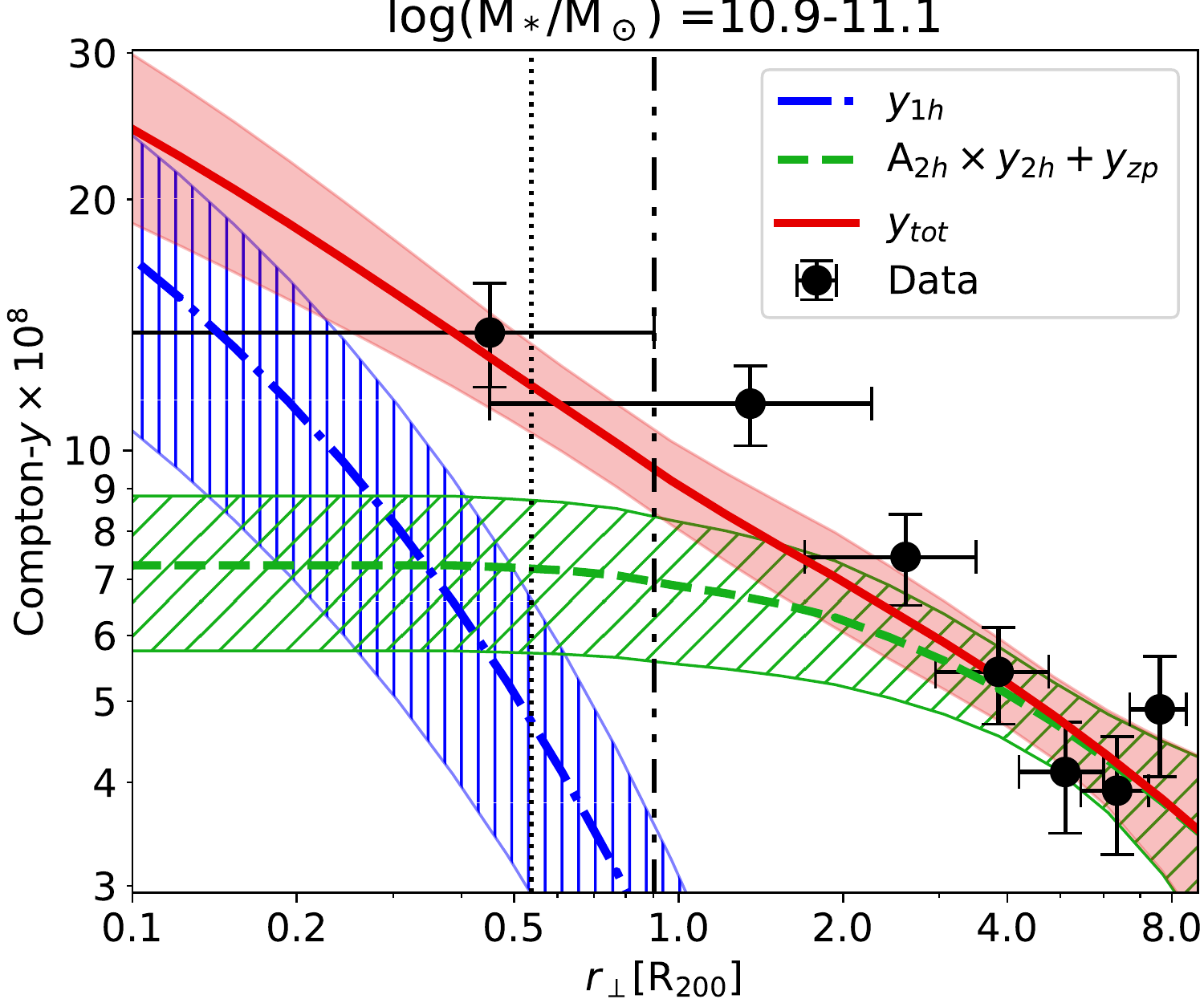}
    \includegraphics[width=0.15\textwidth]{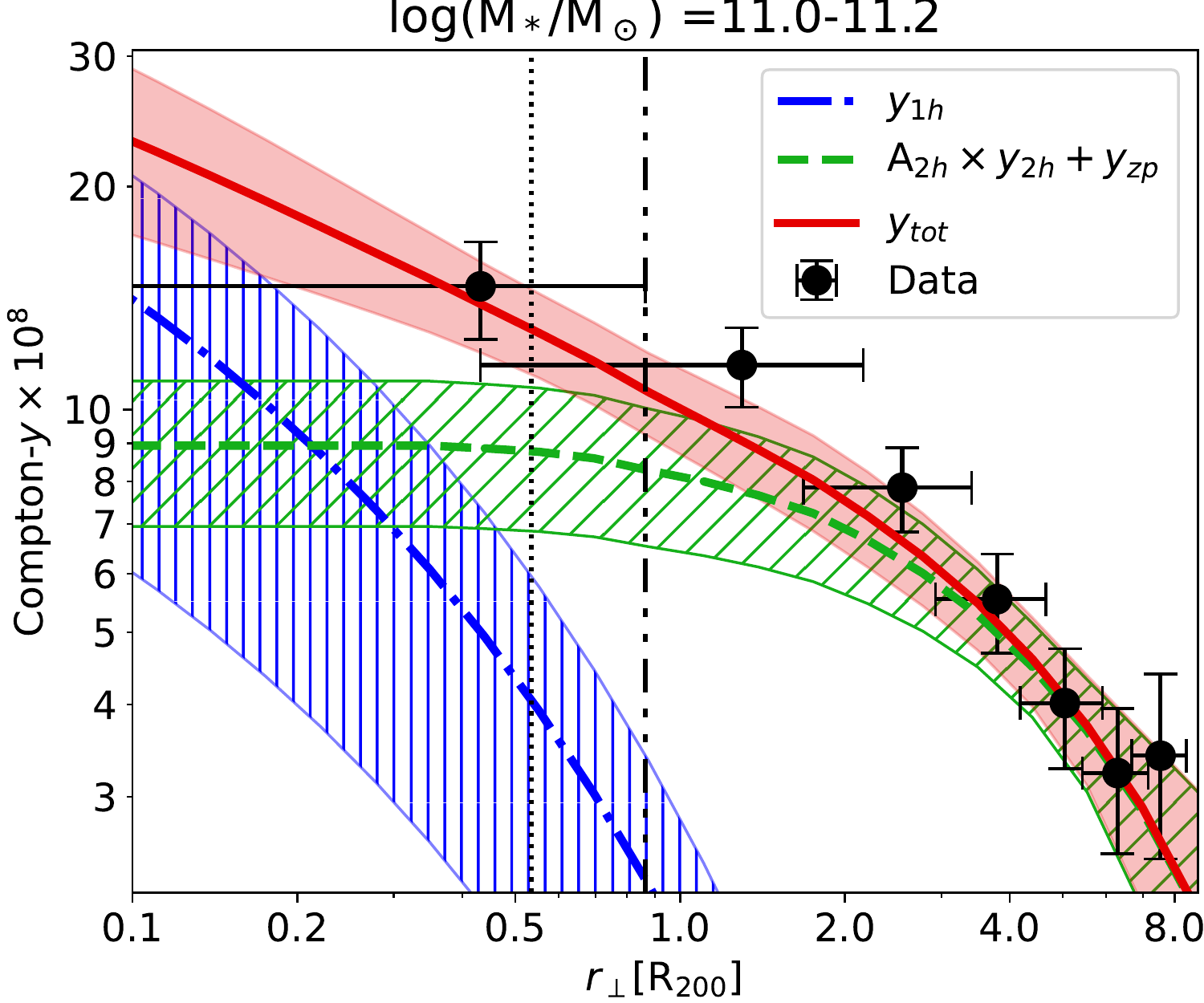}
    \includegraphics[width=0.15\textwidth]{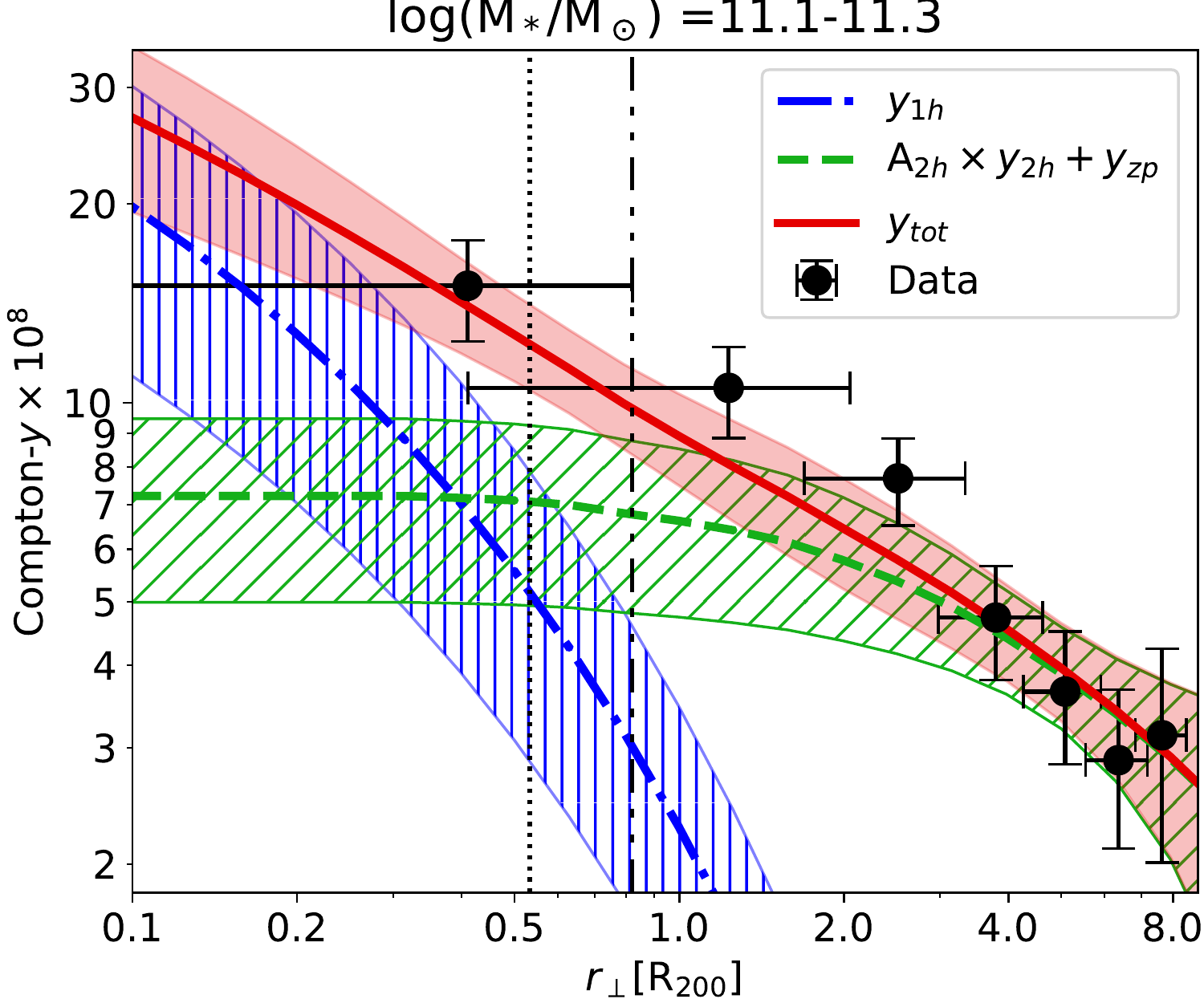}

    \includegraphics[width=0.22\textwidth]{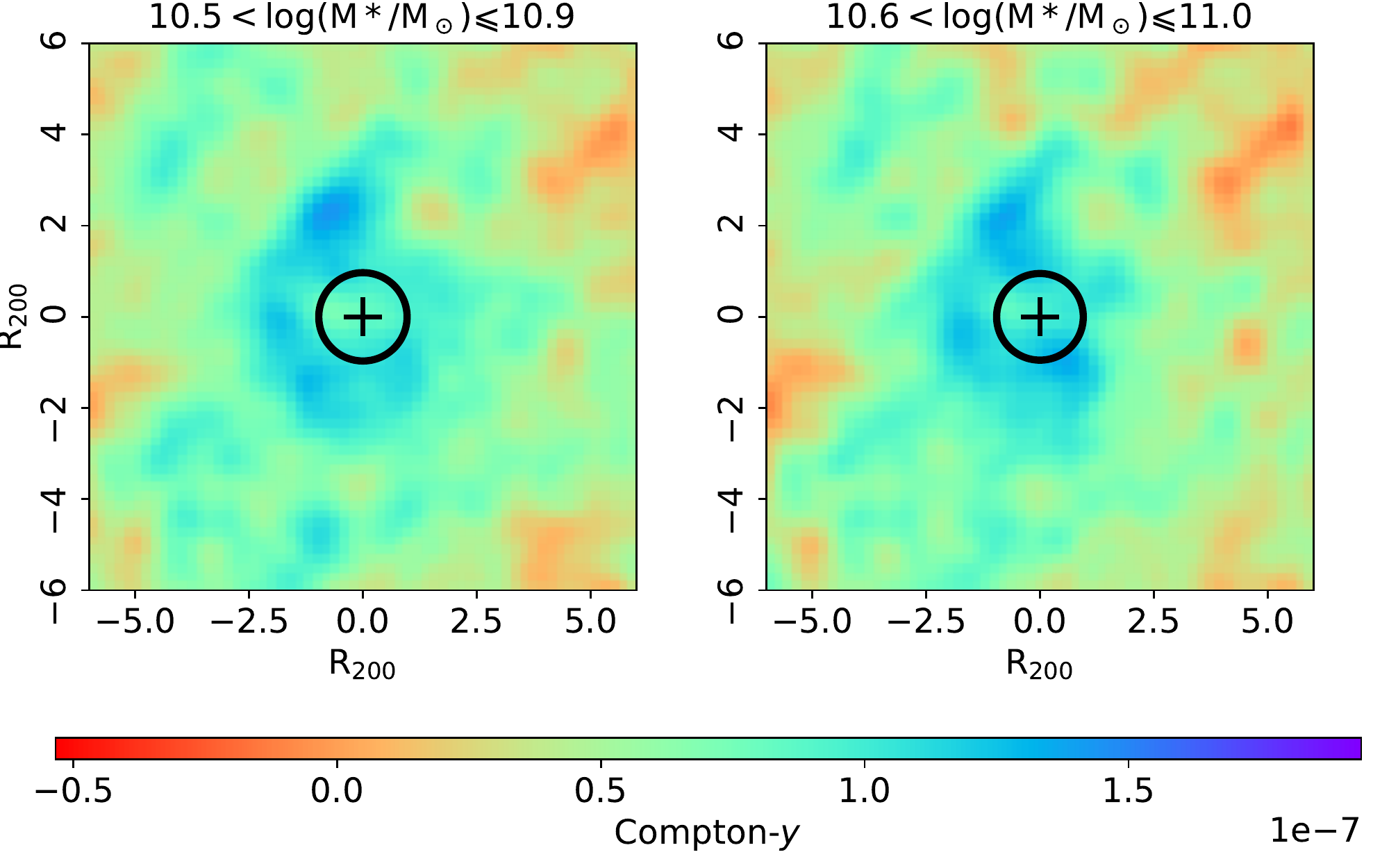}
    \includegraphics[width=0.15\textwidth]{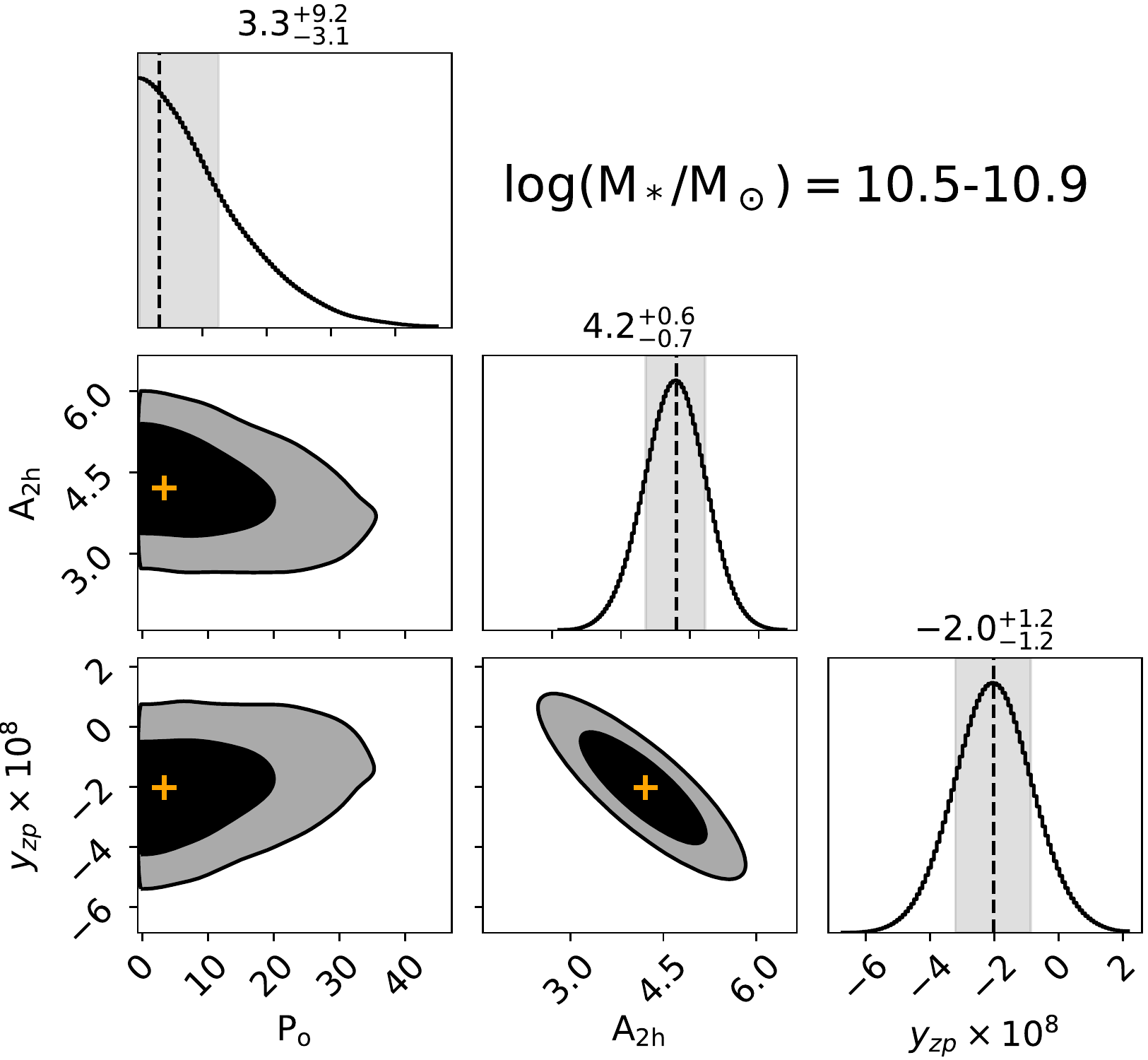}
    \includegraphics[width=0.15\textwidth]{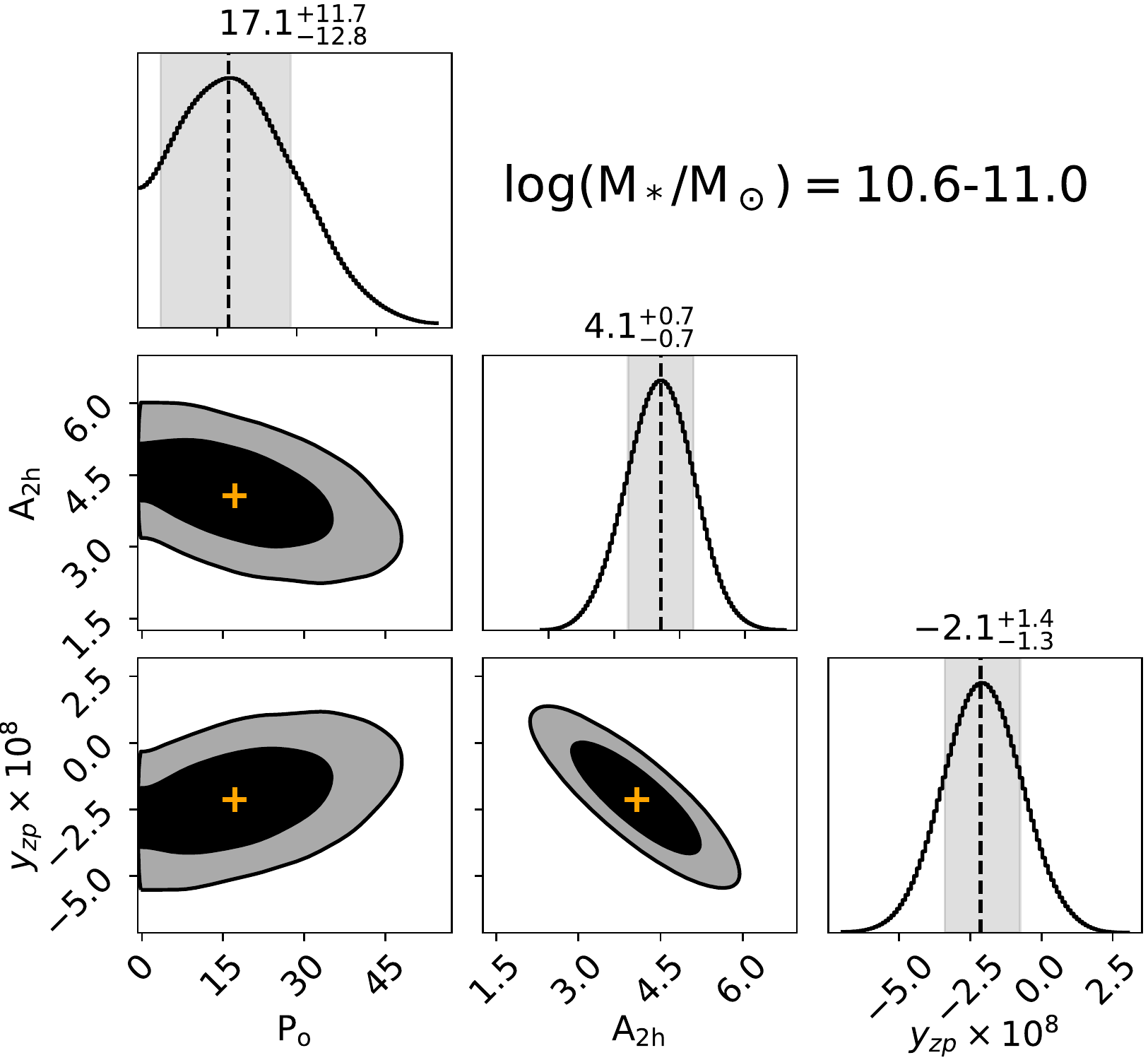}
    \includegraphics[width=0.15\textwidth]{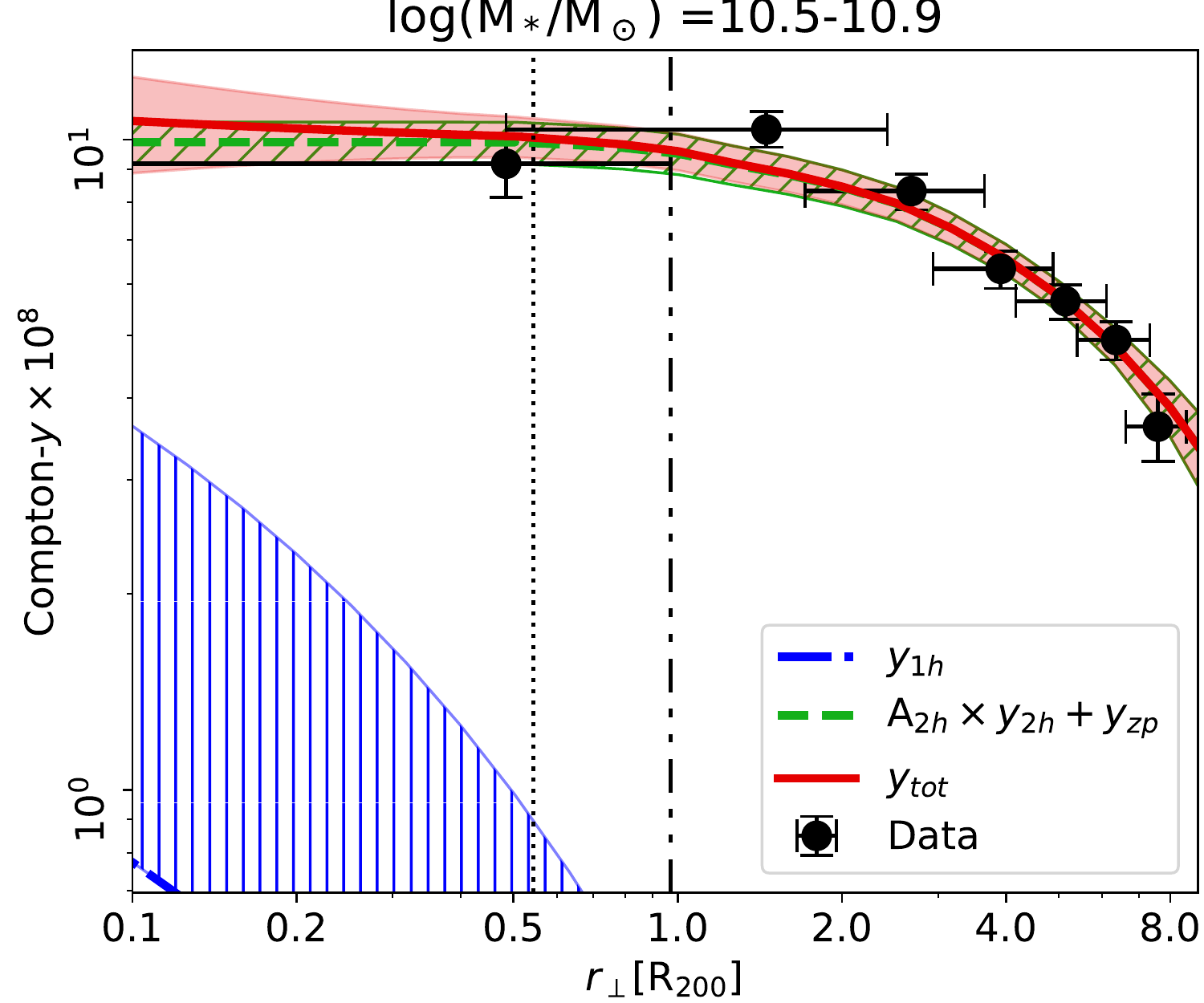}
    \includegraphics[width=0.15\textwidth]{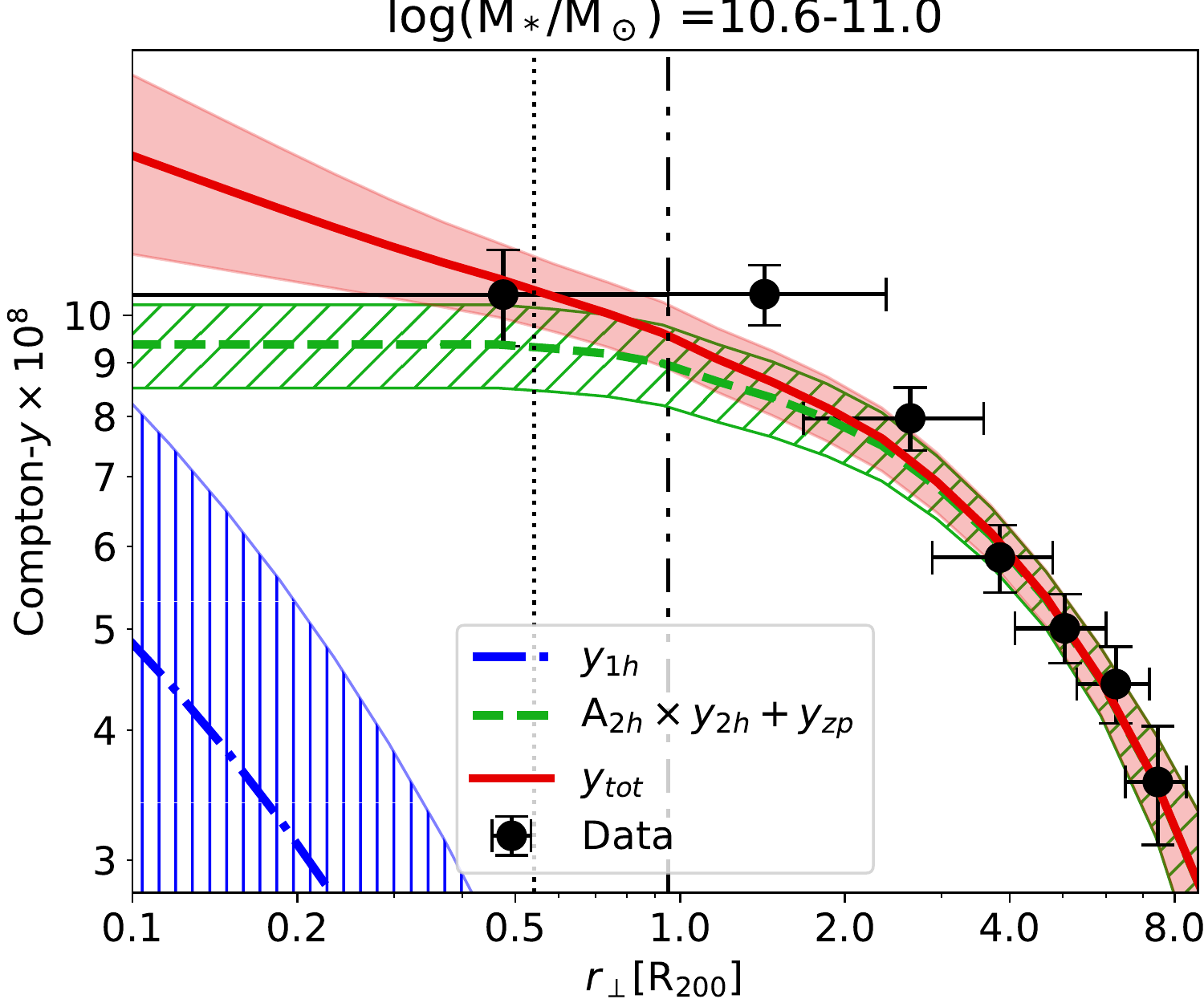}
    \caption{Stacked Compton-$y$ maps, posterior probability distributions of $\rm P_o$, $\rm A_{2h}$, and $y_{zp}$ obtained by fitting the GNFW pressure model to our tSZ measurements, and differential radial profile of Compton-$y$. $\Delta \rm M_\star = 0.3$\,dex (top three rows), 0.2\,dex (third, fourth, and fifth rows from top), and 0.4\,dex (bottom row). The range of stellar mass is mentioned in the title of each plot. See the top left, top right, and bottom left panel of Figure\,\ref{fig:bestfit} for more description. }
    \label{fig:result_others}
\end{figure*}

We show the stacked Compton-$y$ maps, the posterior distribution of the model parameters, and the differential radial profiles of Compton-$y$ along with the best-fitted model and its components in Figure\,\ref{fig:result_others}. An excess in the $\approx\rm 1-2\times R_{200}$ region compared to the region within $\rm R_{200}$ is visible in most of the $y$-maps stacked between $\rm M_\star=10^{10.4}$\msun~and $\rm M_\star=10^{11}$\msun, for different $\Delta \rm M_\star$. It leads to a flattening of the differential profile at a small radius, or even a deficit of Compton-$y$ in the innermost radial bin compared to the adjacent radial bin (visible in M$_\star$ = 10$^{10.4-10.7}$\msun), which we refer to as ``central dip". We can visually identify such deviation from radially declining Compton-$y$ only from differential profiles, not cumulative profiles. The ``central dip" could be an artifact due to the incomplete treatment of dust at the center of these galaxies, or the residual effect of fainter radio sources. Otherwise, it could also happen due to the galactic feedback snowplowing the CGM out of the halo. 
The ``central dip" leads to an underestimated normalization of the pressure profile with weak constraints at the respective stellar mass bins, but it does not rule out the possibility of the \cite{Arnaud2010} pressure profile altogether. We will address the diversity in the shape of the radial profiles and their physical implication in a companion paper with advanced modeling.  

\section{Central vs satellite galaxies}
\begin{figure}[h!]
    \centering
    \includegraphics[width=0.475\textwidth]{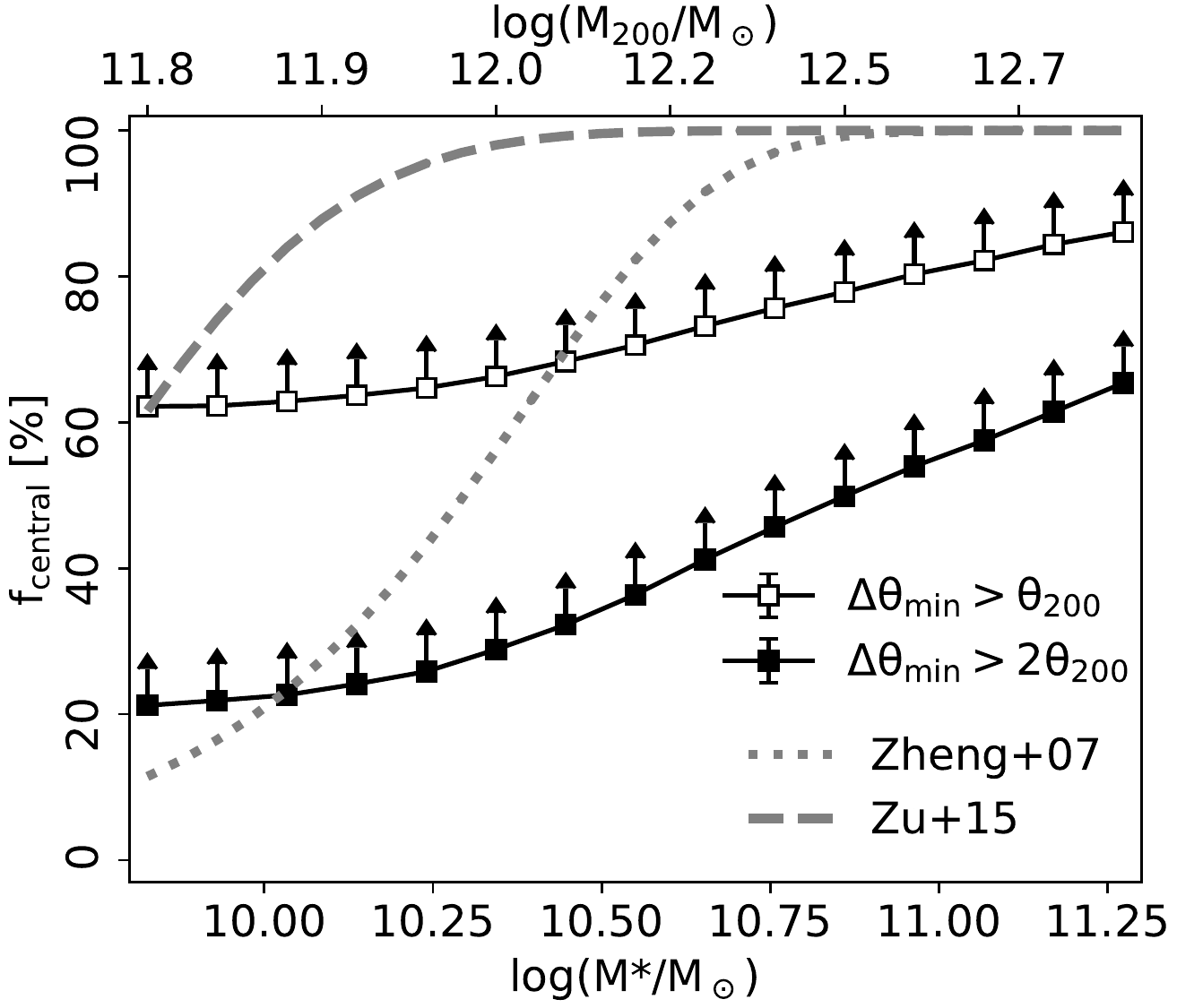}
    \caption{The fraction of central galaxies, $f_{central}$ as a function of the stellar mass (bottom x-axis) and virial mass (top x-axis). The black filled and unfilled squares are our calculated $f_{central}$ in the WISE$\times$SCOSPZ galaxy sample for the requirement  of minimum angular distance from more massive galaxies of $\Delta \theta_{min} > 2\theta_{200}$ and $\Delta \theta_{min} >\theta_{200}$, respectively. We show the speculated $f_{central}$ for different halo occupation statistics with the dotted \citep{Zheng2007} and dashed \citep{Zu2015} curves.}
    \label{fig:fcentral}
\end{figure}
The central and satellite galaxies evolve in different environments, therefore they could have different SHMR. The interpretation of our measured tSZ effect, and the quantities derived from it, would not be straightforward if our galaxy sample is dominated by satellite galaxies. In the following, we discuss the identification of central galaxies in our sample and the credibility of our analysis in the context of potential contamination from satellites. 

We restrict the calculation to the galaxy pairs that are at similar redshifts, i.e., whose photometric redshifts differ by less than their combined uncertainty in $z_{\rm ph}$: $|z_1 - z_2| < \sqrt{(\sigma_{z_1}^2+\sigma_{z_2}^2)}$, where $\sigma_z$ is the redshift uncertainty. By comparing the $z_{\rm ph}$ with the corresponding spectroscopic redshifts from multiple catalogs, \cite{Bilicki2016} estimated the overall accuracy in $z_{\rm ph}$ to be 0.035$\times$(1+$z_{\rm ph}$). We adopt it as an estimator of $\sigma_z$. If a galaxy is within 2$\times\theta_{200}$ of a more massive galaxy, we identify it as a satellite of the more massive galaxy. We find that $\approx 37\%$ of our galaxy sample is central galaxies according to this condition.

To test the validity of our calculation of the fraction of central galaxies, $f_{central}$, we compare it with the previous observational estimates or theoretical predictions of $f_{central}$. The halo occupation fraction of a central galaxy depends on the host halo mass. We calculate the $f_{central}$ as a function of stellar mass and virial mass using the occupation statistics prescription of \cite{Zheng2007} and \cite{Zu2015}, and show it in Figure\,\ref{fig:fcentral}. 

Because the uncertainty in photometric redshift is large, the redshift difference between galaxies, and hence, the number of galaxies to be considered as potential satellites is large too. Therefore, our estimate of $f_{central}$ is only a lower limit. The requirement of minimum angular distance, $\rm \Delta \theta_{min}$, between two central galaxies of 2$\times\theta_{200}$ could be aggressive as well. In fact, a lenient requirement of $\rm \Delta \theta_{min} > \theta_{200}$ increases the overall $f_{central}$ to $\approx 71\%$. Given the size of our galaxy sample, the stacked signal over these central galaxies would be too weak to put a constraint on. Therefore, instead of re-stacking over the central galaxies, we compare the existing prescriptions of $f_{central}$ with our estimate of $f_{central}$, and we try to understand how much our stacking results could be affected by satellite galaxies, if at all.   

The exact dependence of $f_{central}$ on mass is not of prime importance in our context. At a given stellar mass, if the speculated value of $f_{central}$ (gray lines in Figure \ref{fig:fcentral}) is larger than the $f_{central}$ of our galaxy sample (black squares), it would imply that our $f_{central}$ is not underestimated, and hence, reliable. The exact measure of satellite galaxy contamination in our galaxy population is model-dependent and inconclusive. 
In the most conservative prescription of $f_{central}$ \citep{Zheng2007}, the galaxy population is dominated by central galaxies at $> 10^{10.6}$\msun, i.e., $f_{central}>$80\%. Therefore, in the mass range where we can constrain the tSZ effect, it is safe to assume the central galaxy properties in the interpretation of the results. 


\end{document}